\documentclass[aps,prd,nofootinbib,floatfix,superscriptaddress,noeprint,longbibliography,11pt]{revtex4-1}
\usepackage{amsmath}
\usepackage{amsfonts}
\usepackage{amssymb}
\usepackage{verbatim}
\usepackage{mathtools}
\usepackage{color}
\usepackage[all]{xy}
\usepackage{setspace}
\usepackage{array}
\usepackage{amsthm}
\usepackage{enumitem}
\usepackage{bbm}
\usepackage{bbold}
\usepackage{graphicx}
\usepackage{subfigure}

\usepackage[english]{babel}
\usepackage[utf8]{inputenc}
\usepackage[T1]{fontenc}
\usepackage{mathrsfs}
\usepackage{bm}
\usepackage{enumerate}
\usepackage[dvipsnames]{xcolor}
\usepackage{booktabs}
\usepackage{multirow}

\usepackage[colorlinks,
linkcolor=BrickRed,
citecolor=MidnightBlue,
urlcolor=MidnightBlue,
bookmarks=true,
bookmarksopen=true,
bookmarksnumbered=true,
]{hyperref}

\newcounter{jvcc}
\newcounter{amg}

\newcounter{amgg}
\newcounter{ls}

\newcounter{yjcc}

\newcommand{\Tr}{{\rm Tr}}

\newcommand{\eref}[1]{(\ref{#1})}
\newcommand{\nn}{\nonumber}

\newcommand{\be}{\begin{eqnarray}}
\newcommand{\ee}{\end{eqnarray}}

\newcommand{\bmat}{\left ( \begin{array}{cc} }
	\newcommand{\emat}{\end{array} \right ) }

\renewcommand\epsilon\varepsilon
\renewcommand\phi\varphi

\DeclareMathOperator\erf{erf}
\DeclareMathOperator\erfc{erfc}

\renewcommand{\i}{\mathrm{i}}

\renewcommand{\mod}{\,\mathrm{mod}\,}

\newcommand{\brho}{\bar \rho}

\graphicspath{{figures/}}

\begin{document}

\title{Universality and its limits in non-Hermitian many-body quantum chaos using the Sachdev-Ye-Kitaev model}

\author{Antonio M. Garc\'\i a-Garc\'\i a}
\email{amgg@sjtu.edu.cn}
\affiliation{Shanghai Center for Complex Physics,
	School of Physics and Astronomy, Shanghai Jiao Tong
	University, Shanghai 200240, China}

\author{Lucas S\'a}
\email{lucas.seara.sa@tecnico.ulisboa.pt}
\affiliation{CeFEMA, Instituto Superior T\'ecnico, Universidade de Lisboa, Av.\ Rovisco Pais, 1049-001 Lisboa, Portugal}

\author{Jacobus J. M. Verbaarschot}
\email{jacobus.verbaarschot@stonybrook.edu}
\affiliation{Department of Physics and Astronomy, Stony Brook University, Stony Brook, New York 11794, USA}

\date{\today}

\begin{abstract}
\vspace{+0.5cm}
Spectral rigidity in Hermitian quantum chaotic systems signals the presence of dynamical universal features at timescales that can be much shorter than the Heisenberg time. We study the analog of this timescale in many-body non-Hermitian quantum chaos by a detailed analysis of long-range spectral correlators. For that purpose, we investigate the number variance and the spectral form factor of a non-Hermitian $q$-body Sachdev-Ye-Kitaev (nHSYK) model, which describes $N$ fermions in zero spatial dimensions. After an analytical and numerical analysis of these spectral observables for non-Hermitian random matrices, and a careful unfolding, we find good agreement with the nHSYK model for $q > 2$ starting at a timescale that decreases sharply with $q$.
The source of deviation from universality, identified analytically, is ensemble fluctuations not related to the quantum dynamics. 
For fixed $q$ and large enough $N$, these fluctuations become dominant up until after the Heisenberg time, so that the spectral form factor is no longer useful for the study of quantum chaos. 
In all cases, our results point to a weakened or vanishing spectral rigidity that effectively delays the observation of full quantum ergodicity.
We also show that the number variance displays nonstationary spectral correlations for both the nHSYK model and random matrices. This nonstationarity, also not related to the quantum dynamics, points to intrinsic limitations of these observables to describe the quantum chaotic motion. On the other hand, we introduce the local spectral form factor, which is shown to be stationary and not affected by collective fluctuations, and propose it as an effective diagnostic of non-Hermitian quantum chaos.
For $q = 2$, we find saturation to Poisson statistics at a timescale of $\log D$, compared to a scale of $\sqrt D$ for $ q>2$, with $D $ the total number of states.
\end{abstract}

\maketitle

\clearpage

{
\setstretch{1.25}
\hypersetup{linkcolor=black}
\tableofcontents
}

\clearpage

\section{Introduction}
The study of quantum chaotic dynamics attracts a great deal of interest in different fields because of its robust universal features. For sufficiently long timescales, the evolution of very different quantum systems is qualitatively similar provided that the dynamics is quantum chaotic. By contrast, the quantum dynamics of integrable systems is very sensitive to the details of the Hamiltonian.
A central result in the theory of quantum chaos is the Bohigas-Giannoni-Schmit (BGS) conjecture~\cite{bohigas1984} that states that spectral correlations of a quantum chaotic system are given by random matrix theory (RMT). In the context of single-body quantum mechanics, the conjecture has received strong analytic
support~\cite{berry1985semiclassical,richter2002,muller:2004nb} by using periodic orbit theory techniques. The BGS conjecture has been very influential because the spectrum of the Hamiltonian (or another relevant operator) is one of the least expensive quantities to obtain numerically even for quantum many-body systems. Therefore, a relatively straightforward spectral analysis is sufficient to determine the quantum chaotic nature of the motion for sufficiently long times of the order of the Heisenberg time---the timescale related to the (inverse) mean level spacing.
 
However, the agreement with RMT extends in many cases to substantially shorter times due to so-called spectral rigidity---directly related to the power-law tails of the two-level correlation function---which is responsible for the slow logarithmic growth of the number variance or the ramp of the spectral form factor.
For sufficiently short times, level statistics of realistic quantum chaotic Hamiltonians deviate from the random matrix prediction. The timescale that marks the onset of these deviations and delimits the region of universal quantum chaotic dynamics---related to the so-called dip or correlation hole~\cite{leviandier1986,wilkie1991,alhassid1992,torres2018} of the connected spectral form factor, or to power-law deviations of the logarithmic growth of the number variance---depends on details of the dynamics. For disordered systems, where it is called the Thouless time~\cite{braun1995}, it is related to the typical diffusion time needed for a single particle to cross the sample. However, also in the context of disordered systems, this timescale is sometimes determined by ensemble fluctuations not directly related to the type of motion~\cite{brody1981,flores:2000ew}. 

So far, the discussion has been restricted to Hermitian quantum systems.
A natural question to ask is to what extent these ideas and results are applicable to non-Hermitian quantum chaotic many-body Hamiltonians. 
The main goal of this paper is to address this question. 
We investigate long-range spectral correlations such as the number variance and the spectral form factor, which probe shorter timescales of the dynamical evolution of non-Hermitian systems.

The theory of non-Hermitian random matrices is well developed for some universality classes corresponding to the so-called Ginibre ensembles \cite{ginibre1965}. However, no equivalent of the BGS conjecture is known, and so the relation between dynamics and level statistics is less clear than in the Hermitian case.
There are also technical problems: correlations of complex eigenvalues are weakened, and the necessary unfolding of eigenvalues
may be problematic~\cite{markum1999PRL,akemann2019} when the
eigenvalue distribution is not radially symmetric.
Unfolding problems have been ameliorated in the last years
for short-range observables with the introduction of spectral observables such as the adjacent gap ratios \cite{atas2016} for complex spectra \cite{sa2020} that do not require unfolding. They have already been applied in a variety of non-Hermitian systems: phase transitions in many-body Liouvillians~\cite{rubio2021,yusipov2022,hamazaki2022,prasad2022}, non-Hermitian Anderson localization~\cite{huang2020a,luo2021,luo2021a,yan2022}, nonunitary open quantum circuits~\cite{sa2021,prosen2021}, two-color QCD at imaginary chiral chemical potential~\cite{kanazawa2021}, and, more recently, the Sachdev-Ye-Kitaev (SYK) model~\cite{garcia2022,garcia2022a,sa2022PRR}. 

Long-range spectral correlators such as the number variance~\cite{jancovici1981,fyodorov1997,lebowitz1999,garcia2002,lacroix2019,huang2020} or spectral form factor (SFF)~\cite{fyodorov1997,chan2021} have already been investigated in the context of non-Hermitian systems but there are still problems with the unfolding procedure, which is necessary, especially for the SFF, for a correct dynamical interpretation of the results. Moreover, the role of spectral rigidity, if any, and the determination of the timescale that signals deviations from universality are still poorly understood in quantum chaotic non-Hermitian systems. 
We aim to shed light on this problem by computing these spectral observables for the non-Hermitian SYK model~\cite{garcia2021,pengfei2021,garcia2022a,garcia2021b}. This model is a natural building block of Euclidean~\cite{garcia2021,garcia2022a} and Keldysh ~\cite{garcia-garcia2022keldysh} wormholes, models for Lindbladian dissipation~\cite{sa2022PRR,kulkarni2022PRB}, and entanglement dynamics~\cite{liu2021SciPost}.

The SYK model, describing $N$ fermions with infinite-range interactions in zero spatial dimensions, was introduced more than 50 years ago \cite{french1970,french1971,bohigas1971,bohigas1971a,mon1975} in the context of nuclear physics as a toy model
for nuclei. Later, it played an important role in the development of so-called many-body quantum chaos \cite{benet2001,benet2003,kota2014,kota2001} and also in the description of certain aspects of spin liquids \cite{sachdev1993}. The revival of interest in the SYK model is motivated by its role in quantum gravity as a toy model for holography \cite{kitaev2015,maldacena2016,sachdev2010} and also due to the use of Majorana fermions, proposed by Kitaev \cite{kitaev2015}, that simplifies the model allowing analytical calculations for some region of the
parameters. For instance, it was possible \cite{kitaev2015,maldacena2016} to demonstrate that the SYK model saturates a universal bound \cite{maldacena2015} on the exponential growth in time of certain out-of-time correlation functions that probe quantum chaos at short timescales of the order of the Ehrenfest time---the time for which quantum effects start to become relevant.
It has also been shown that the SYK model, both Hermitian and non-Hermitian, is quantum chaotic with spectral correlations well described by RMT~\cite{garcia2016,garcia2017,cotler2016,garcia2022}. By tuning $q$ and $N$, the SYK model can also reproduce several of the different universality classes, controlled by the global symmetries of the system, in which a many-body quantum chaotic system can relax to ergodicity~\cite{you2016,garcia2018a,li2017,kanazawa2017,sun2020,garcia2021d,garcia2022}. It is therefore a natural choice for the problems we will be addressing. 

We initiate our analysis in Sec.~\ref{sec:nHRMT} with a description of long-range spectral correlations of non-Hermitian random matrices belonging to the Ginibre ensemble \cite{ginibre1965}. We will focus on the number variance and the spectral form factor of the real parts of the eigenvalues, which has recently been proposed as a measure of quantum chaos in non-Hermitian systems~\cite{chan2021}.
Some of the analytic results of this section were already derived in Refs.~\cite{fyodorov1997,fyodorov1998,fyodorov2003,chan2021}, but we present them here from a unified viewpoint. Then, in Sec.~\ref{sec:nHSYK} we apply the same tools to investigate the emergence of random-matrix universality, and its limits, in the non-Hermitian SYK model.

\section{Spectral form factor and number variance of non-Hermitian random matrix models}
\label{sec:nHRMT}

For real spectra, some widely used long-range correlators are the number variance, the $\Delta_3$ statistic, and the spectral form factor \cite{mehta2004}. In this section, we study analog statistics for non-Hermitian spectra of the Ginibre unitary ensemble (GinUE) of random matrices 
\cite{ginibre1965}.

\subsection{The Ginibre ensemble}

Throughout this paper, we frequently refer to the Ginibre ensemble. Here, we collect some well-known results that we use below. The Ginibre ensemble is the ensemble of
$D\times D$ complex random matrices $H$ with probability distribution given by \cite{ginibre1965}
\be
P(H) \sim e^{-D\,\Tr H^\dagger H}.
\ee
With this normalization, in the limit of large $D$, the eigenvalues $z_i$ of $H$ are distributed uniformly inside the complex unit disk. The real parts $E_i=\mathrm{Re} z_i$ of the eigenvalues, therefore, follow the semicircular distribution
\be
\bar \rho(E) = \frac {2D}\pi\sqrt{1-E^2},
\label{rhosc}
\ee
normalized as 
\begin{equation}
\int d E\, \bar \rho(E)=D.
\end{equation}
 
At finite $D$, the connected two-point correlation function of the eigenvalues is given by~\cite{ginibre1965,mehta2004}
\begin{equation}
\rho_{2c}(z_1,z_2)=K(z_1,z_1)\delta^2(z_1-z_2) - |  K(z_1,z_2)|^2,
\label{eq:correlator_kernel}
\end{equation}
with the kernel given by:
\be
\label{eq:kernel}
K(z_1,z_2) =\frac D\pi e^{-\frac D2 (|z_1|^2 +|z_2|^2)}\sum_{k=0}^{D-1}\frac{(D z_1 z_2^*)^k}{k!}.
\ee
In the large-$D$ limit, this simplifies to~\cite{ginibre1965,mehta2004}
\be
\label{eq:2pt-Ginibre}
  \rho_{2c}(z_1,z_2) = \bar \rho(z_1) \delta^2(z_1-z_2)-
  \bar \rho(z_1) \bar \rho(z_2) e^{-D|z_1-z_2|^2}.
\ee
  
\subsection{Unfolding}
\label{sec:unfolding}

Depending on the long-range correlator, we may have to unfold the eigenvalues (i.e., reparametrize them such that the spectral density is constant in the new variables) in order to obtain universal results. Since the number variance is defined as the
variance of the number of eigenvalues in a fixed interval, there is
no need for unfolding, although it is convenient to do so if it is
calculated by spectral averaging or a combination of spectral averaging
and ensemble averaging. The spectral form factor is an observable that
involves the entire spectrum so in this case unfolding is essential for
making quantitative comparisons between different systems. Since the contribution of eigenvalue pairs with large spacings
	is suppressed by large phase oscillations, the main contribution to the
	spectral form factor is due to eigenvalues that are close. Therefore, it
	is possible to give a local definition of the spectral form factor that only
	includes the eigenvalues on a scale where the average spectral density
	is approximately constant, see Sec.~\ref{sec:local-form} below.
Because our goal is to identify universal features of the quantum dynamics, we only consider connected two-point correlators in the analysis of the spectral correlations.

For the Ginibre ensemble, there is no need to unfold the spectrum because,
well away from the spectral edge, the spectral density is constant and a rescaling is enough. However, for the nHSYK model, to be studied later,
the average spectral density is not constant and nonuniversal as it
is determined by the details of the phase space and the dynamics.
Therefore, we first provide a detailed explanation of the unfolding of complex spectra. For simplicity, we focus on the case of radially symmetric spectra. 

For a radially symmetric spectrum, $\{z_k\}_{k=1,\dots,D}$, the spectral density $\bar\rho(z,z^*)$ satisfies
\be
\bar \rho(z,z^*) d^2 z = \bar \rho(|z|)\,|z|\, d|z|\, d ({\rm arg}(z)).
\ee
Because of unfolding ambiguities, we only unfold the eigenvalues in this case\footnote{There is still an ambiguity. For example, we could have unfolded the eigenvalues $z_k$ so that the density of
	${\rm Re}(z_k e^{i\theta})$ (i.e., the projection of the eigenvalues along some axis with angle $\theta$) becomes constant. However, since analytical results are only available for
	constant density inside the unit disk we unfold the eigenvalues this way.}
and we only have to 
reparametrize the absolute value of the eigenvalues.
The unfolding is performed using the
average radial spectral density, which is a smooth function 
and therefore does not affect
local statistics. If the average radial density is given by $\bar \rho(r)$, the unfolded eigenvalues, $z_k^{\rm unf} $ in terms of the original eigenvalues are given by
\be
z_k^{\rm unf} = \left(\int_0^{|z_k|} 2\pi r dr \bar \rho(r) \right )^{1/2}\frac{z_k}{|z_k|}.
\ee
As a check of this transformation, we can take the flat density $\bar \rho(r) =1/\pi$ which results
in $z_k^{\rm unf} = z_k$.
The unfolded eigenvalues have constant density inside the unit disk.
Below, we will see that the spectral density of the nHSYK model for $q\ge 6$ is almost constant,
and unfolding can be approximated by 
just rescaling the eigenvalues to the unit disk by
\be
z_k \to  \frac{z_k \pi \bar \rho{(0)}}D.
\ee

When the eigenvalue density is not strictly isotropic, generically the spectrum is still locally isotropic. In that case, unfolding
ambiguities can be resolved by requiring that local spectral isotropy is preserved.

\subsection{Number variance}

\subsubsection{Complex eigenvalues}

The best-known spectral correlator to probe the existence of spectral
rigidity is the number variance, defined as the variance of the number of eigenvalues in
a spectral window containing a fixed number of eigenvalues on average. Let that window be a compact domain ${\cal D}$ inside the unit disk,
containing $n$ eigenvalues. In terms of the connected two-point correlation function, the number variance reads
\begin{equation}\label{eq:nv_def}
\Sigma^2(n)=\int_\mathcal{D} d^2 z_1 d^2 z_2 \rho_{2c}(z_1,z_2).
\end{equation}

For spectra of Hermitian quantum chaotic systems and Hermitian random matrices,
the growth of the number variance increases only logarithmically
with the number of eigenvalues, while it grows linearly in the case of a nondegenerate uncorrelated spectrum typical of
an integrable system. For non-Hermitian systems, the spectral rigidity takes a different form because
the spectrum is now two-dimensional. One can imagine the $D$ eigenvalues in the unit disk as small disks
with area $\sim 1/D$.
For the compact domain ${\cal D}$ containing $n$ eigenvalues, only the eigenvalues near the edge can diffuse in and out of the disk.
There are order of $\sqrt n$ eigenvalues near the surface. Therefore, the
number variance (i.e., the variance of the number of eigenvalues inside ${\cal D}$) behaves as $\Sigma^2(n) \sim \sqrt{n}$ for sufficiently large $n \gg \sqrt D$, while for $n \sim O(\sqrt D)$ all eigenvalues
can move in and out resulting in Poisson statistics with $\Sigma^2(n) =n$.
It is clear that the coefficient of $\sqrt n$ depends on the length of the
perimeter of ${\cal D}$ for $n$ eigenvalue inside this domain. Below, we
will see that for a disk \cite{lacroix2019} we have that $\Sigma^2(n) = \sqrt{ n/\pi}$,
while for a square we find $\Sigma^2(n) =2 \sqrt{ n}/\pi$, which behaves as the
ratio of the perimeters for the same area.

Next, we calculate the number variance
of the Ginibre ensemble
for two simple geometries, the
rectangle and the disk. Using Eq.~(\ref{eq:2pt-Ginibre}), the number variance~(\ref{eq:nv_def}) for the rectangle with sides $2a$ and $2b$ and centered at the origin is given by
\be
\label{eq:nv_rect_ini}
\Sigma^2(n) = n - \frac{D^2}{\pi^2}\int_{-a}^a\int_{-a}^a d x_1 d x_2 \int_{-b}^b
\int_{-b}^b d y_1 dy_2 e^{-D(x_1-x_2)^2 -D(y_1-y_2)^2},
\ee
with
\be
n = \frac D\pi 4ab.
\label{n-area}
\ee
The second term in Eq.~(\ref{eq:nv_rect_ini}) factorizes into $F(a) F(b)$ with
\be
F(a) &=& 
\frac D\pi \int_{-a}^a\int_{-a}^a d x_1 d x_2  e^{-D(x_1-x_2)^2} \nn\\
 &=& 
\frac 1\pi\left( e^{-4a^2D}-1\right ) + \frac {2a\sqrt D}{\sqrt \pi}
\erf(2a\sqrt D),
\ee
where $\erf$ stands for the error function.
Using Eq.~\eref{n-area}, this result can be expressed in terms of $n$ and
the aspect ratio of the rectangle, $\alpha\equiv b/a$,
\be
F(a) = \frac 1\pi \left( e^{-n\pi/\alpha}-1\right )
+\sqrt{\frac n\alpha }
\erf\left( \sqrt{\frac{n\pi}\alpha} \right ),
\ee
which has a well-defined large-$D$ limit. $F(b)$ is obtained by replacing $\alpha\to1/\alpha$.
If the rectangle contains more than a few eigenvalues, this result is
well approximated by
\be
F(a) \sim \sqrt{\frac{n}\alpha} -\frac 1\pi.
\ee
This gives the number variance
\be
\Sigma^2(n) =\frac 1\pi \left( \sqrt \alpha + \frac 1{\sqrt\alpha} \right )
\sqrt n  - \frac 1{\pi^2}.
\ee
This approximation is already accurate to four digits for $n=2$ and becomes rapidly more accurate for larger values of $n$.

A second interesting case is a rectangle with $b$ independent of $n$ and $ a \sim b$.
We again have that $4ab D/\pi =n$, but now $4 a^2 D = n^2\pi^2 /(4D b^2)$
instead of $n\pi /\alpha$.
In this case,
\be
F(a) = \frac 1\pi\left(e^{-n^2\pi^2/4 D b^2} -1\right )+\frac {n \pi}{2 b \sqrt {\pi D}}
{\rm erf}(n\pi/2b\sqrt D). 
\ee
If $b$ is fixed we have in the large-$D$ limit that
\be
F(b) \approx \frac{2 b \sqrt D}{\sqrt \pi} -\frac 1\pi.
\ee
The second term is subleading and will be neglected.
This gives the number variance
\be
\Sigma^2(n) &=&n-\frac{2 b \sqrt D}{\sqrt \pi}\left [ \frac 1\pi\left(e^{-n^2\pi^2/4 D b^2}-1 \right )+\frac {n \pi}{2 b \sqrt {\pi D}}
  {\rm erf}\left(\frac{n\pi}{2b\sqrt D}\right) \right ]\nn \\
&=&
\frac{2 b \sqrt D}{\sqrt \pi} \frac 1\pi\left(1-e^{-n^2\pi^2/4 D b^2} \right )
+n\, {\rm erfc}\left(\frac{n\pi}{2b\sqrt D}\right).
\label{double-scaling-0}
\ee
Note that this result for the number variance is nonuniversal: it depends on
$D$ and $b$. A stable double scaling limit is obtained by taking the large-$D$ limit at fixed $b\sqrt D$.

We now turn to the evaluation of the number variance~(\ref{eq:nv_def}) for a disk of radius $R$ centered at zero. This can
be done most simply by using the exact finite-$D$ expression for the connected two-point
correlation function of the Ginibre ensemble written in terms of the kernel, Eq.~(\ref{eq:correlator_kernel}). Inserting it into Eq.~(\ref{eq:nv_def}) we obtain
\be
\Sigma^2(n) = \int_{|z_1|<R} d^2 z_1 K(z_1,z_1) - \int_{|z_1|<R} d^2 z_1\int_{|z_2|<R} d^2 z_2
|  K(z_1,z_2)|^2,
\ee
with the kernel given by Eq.~(\ref{eq:kernel}).
Using polar coordinates, the integrals can be simplified to
\be
\Sigma^2(n) &=& 2D \int_0^R d r_1 \sum_{k=0}^{D-1}
\frac 1{k!}r_1 ^{2k+1}D^k e^{-D r_1^2}
- 4 D^2 \int_0^R d r_1  \int_0^R d r_2 \sum_{k=0}^{D-1}
\frac 1{(k!)^2}(r_1 r_2)^{2k+1}D^{2k} e^{-D(r_1^2+r_2^2)}\nn\\
&=& \sum_{k=0}^{D-1} \left (1-\frac{\Gamma(k+1,R^2D)}{\Gamma(k+1)} \right ) - \sum_{k=0}^{D-1} \left (1-\frac{\Gamma(k+1,R^2D)}{\Gamma(k+1)} \right )^2
\nn \\
&=&\sum_{k=0}^{D-1} \frac{\Gamma(k+1,R^2D)}{\Gamma(k+1)}-\left (\frac{\Gamma(k+1,R^2D)}{\Gamma(k+1)} \right )^2,
\label{numvar-cir}
\ee
which was first obtained in Ref.~\cite{lacroix2019}. In the large-$D$ limit, we have
\be
D R^2 =n,
\ee
which allows us to take the large-$D$ limit of Eq.~\eref{numvar-cir}.
Using the asymptotic expansion of the incomplete Gamma function, the number variance for finite but large $n$ is approximated by
\be
\Sigma^2(n) &=& \frac 14 \sum_{k=0}^\infty  \erfc\left (  \frac {k-n}{\sqrt {2n}} \right )
\erfc\left (  \frac {n-k}{\sqrt {2n} }\right )\nn\\
&\approx & \frac{\sqrt{2n}}4 \int_{-\infty}^\infty dx \erfc(x) \erfc(-x) \nn\\
&=& \frac {\sqrt n }{\sqrt \pi}
\ee

\subsubsection{Real parts of complex eigenvalues}

In this paper, we also consider level correlations
of the projections of the eigenvalues on a line through the origin. If the spectrum is radially symmetric, these
are just the level correlations of the real parts of the eigenvalues.
Again only eigenvalues near the surface can diffuse in and out of the area $|{\rm Re}[z]|< n/ D$ which contains $O(n)$ eigenvalues with $2\sqrt D$ eigenvalues near the surface. The number variance saturates therefore at $\sqrt D$
for sufficiently large $D$. When $n < O(\sqrt D)$ all eigenvalues can diffuse in and out of the segment, giving Poisson
statistics with $\Sigma^2(n) =n$. The number variance
of the real parts of the eigenvalues
therefore behaves roughly as
\be
\Sigma^2(n) = n \theta(2\sqrt{D}  -n) + 2\sqrt D \theta(n-2\sqrt{D} ),
\label{numgin}
\ee
where we expect a rounding to occur when $ n \approx 2\sqrt D$. 

The number variance of the real parts of the eigenvalues (or the eigenvalues projected on a ray through the origin) is not universal. It depends sensitively on the geometry of the eigenvalue
support. For example, for the elliptic Ginibre ensemble supported on an ellipse with $a_x$ and $a_y$ as long and short semiaxis ($a_xa_y=1$), order $2 a_y\sqrt N$ eigenvalues can diffuse in and out of a 
segment. More generally, the number variance of the real parts depends on the vertical dimension of the eigenvalue support. The geometry of the eigenvalue support thus determines the fraction of eigenvalue pairs with projections that
are much closer than their spacing in the complex plane and are, therefore, essentially uncorrelated. We refer
to this as the 
Poisson admixture. For the same reason, we expect that the
number variance of the projected eigenvalues is not stationary. Closer to the edge,
there is less Poisson admixture resulting in a smaller value of the number variance.

The number variance of the real parts of
the eigenvalues at energy $E$ can be obtained from the number variance in a rectangle with fixed side $b$, Eq.~(\ref{double-scaling-0}). In this case, $b$ is equal to the range of the imaginary parts, i.e., $b  =\sqrt {1-E^2} =\pi\bar \rho(E)/2D$ with $\bar \rho(E)$ the density of the real parts of the eigenvalues
$\bar \rho(E) =(2D/\pi) \sqrt{1-E^2}$ [see Eq.~\eref{rhosc}]. Recall that $\bar \rho(E)$ is normalized as $\int \bar\rho(E) d E = D$.
This results in the number variance
\be
\Sigma^2(n)&=& \frac {\bar \rho(E)}{\sqrt{\pi D}}
\left(1- e^{-\frac{n^2 D}{\bar \rho^2(E)}} \right )
+  n\,  \erfc\left (\frac{n\sqrt D}{\bar \rho(E)} \right),
\label{sig-local}
\ee
showing that the number variance
\be\label{eq:univSigma}
\Sigma^2(n)=  \frac{\bar \rho (E)}{\sqrt D} \,g\!\left(\frac{n\sqrt D}{\bar \rho(E)}\right)
\label{double-scaling}
\ee
is determined by a universal function $g$.

We shall employ these expressions in the comparison with the number variance of the nHSYK model in Sec.~\ref{sec:nHSYK}.

\subsection{Spectral form factor}
For real spectra, the Fourier transform of the two-point correlation function, called the spectral form factor, although non-self-averaging~\cite{prange1997}, is also a popular probe of quantum chaos. The logarithmic growth in energy of the number variance, which signals spectral rigidity, translates into linear growth in time of the spectral form factor for intermediate times much larger than the Ehrenfest time but smaller than the Heisenberg time.

The connected spectral form factor of $D$ (real) eigenvalues $x_k$ is defined by
\be
\label{eq:SFF_def}
K_c(t) =
\frac 1D \left\langle \sum_{kl} e^{it(x_k-x_l)}\right\rangle_c
=
\frac 1D \left\langle \sum_{kl} e^{it(x_k-x_l)}\right\rangle-\frac1D \left|\left\langle \sum_{k} e^{itx_k}\right\rangle \right|^2  .
\ee
The normalization is such that
\be
\lim_{t\to \infty} K_c(t) = 1.
\ee
The second term in Eq.~(\ref{eq:SFF_def}) is proportional to the nonuniversal disconnected spectral form factor:
\begin{equation}
\label{eq:dsff_dis}
K_{\rm dis}(t)
=
\left | \left \langle
\sum_{k} e^{it x_k}
\right \rangle  \right|^{2}.
\end{equation}
Only $K_c(t)$ provides direct information on the quantum dynamics.

A spectral form factor for complex eigenvalues was first
introduced in Ref.~\cite{fyodorov1997},
\be
K_c(t,s_1,s_2)&=&\frac 1D\left \langle \sum_{k,l} e^{\frac i2 z_k (t+s_1)+\frac i2 z_k^*(t-s_1)}
e^{-\frac i2 z_l (t+s_2)-\frac i2 z_l^*(t-s_2)} \right \rangle_c
\nn \\
&=&
\int dx_1 dx_2 dy_1 dy_2\, e^{i(x_1-x_2)t + iy_1 s_1 -i y_2 s_2}
\rho_{2c}(z_1,z_2).
\label{kfyod}
\ee
with $z_1 = x_1+i y_1$ and $z_2 = x_2+i y_2$. 
For $s_1=s_2=0$, this becomes the
spectral form factor of the real parts of the eigenvalues.
In Ref.~\cite{fyodorov1997},
the spectral form factor was calculated for the weak 
non-Hermiticity limit of the elliptic Ginibre ensemble but, as stated by the authors,
their calculations are also valid in the case of interest here, termed the 
strong non-Hermiticity limit. They found: 
\be
K_c(t) \equiv K_c(t,0,0) =1 -e^{-\frac{ t^2}{4D}},
\label{form-an}
\ee
in the normalization where the support of the $D$ eigenvalues is the unit disk.
A derivation of this result from the finite-$D$ Ginibre kernel
is given in Appendix~\ref{app:DSFF_GinUE}, see also Ref.~\cite{chan2021,shivam2022}. 
The spectral density of the real parts
of the eigenvalues of the Ginibre ensemble is given by the semicircle distribution, Eq.~(\ref{rhosc}).
We note that unfolding the eigenvalues from semicircular to constant density can almost entirely be absorbed by rescaling the argument of the exponential in Eq.~\eref{form-an} by $1.18$.

More generally, setting $t=\tau \cos\theta$, $s_1=-i \tau \sin\theta$, and $s_2=i\tau\sin\theta$ in Eq.~\eref{kfyod}, we can define the spectral form factor of the eigenvalues projected onto the direction defined by the angle $\theta$:
\be
K_c(\tau, \theta) 
&=& \frac1D
\left \langle
\left| \sum_{k} e^{i\tau {\rm Re}(e^{i\theta} z_k)} \right|^{2}
\right \rangle_{c}.
\label{form}
\ee
If the spectral properties are axially symmetric, this spectral form factor does not depend
on the angle $\theta$ onto which the eigenvalues are projected.
By averaging over $\theta$ it is possible to increase the statistics of the
form factor \cite{chan2021}.
The spectral form factor of the projected eigenvalues was recently proposed~\cite{chan2021} as a measure of quantum chaos in dissipative systems, see also Refs.~\cite{shivam2022,ghosh2022}. 
It was dubbed the dissipative spectral form factor~\cite{chan2021} not to be confused with other closely related quantities: the dissipative form factor introduced in Ref.~\cite{Can:2019znk} and the open-system spectral form factor put forward in Refs.~\cite{xu2021,cornelius2022}.

The spectral form factor \eref{form} measures the same long-range spectral correlations as the number variance of the real parts and therefore
depends sensitively on the geometry of the spectrum. This can be seen explicitly from the
relation between the number variance and the spectral form factor~\cite{leviandier1986}
\be
\Sigma^2(n) = \frac {n^2}{2\pi} \int_{-\infty}^\infty d\tau\, K_c(\tau)\,
\frac {\sin^2(n\tau/2)}{(n\tau/2)^2},
\label{sigform}
\ee        
which is valid if the spectral form factor is calculated for unfolded eigenvalues. 
Equation~(\ref{sigform}) shows that the number variance and the spectral form factor are complementary observables. In the normalization with $D$ eigenvalues in the complex unit circle, the spectral form factor
for $\tau > T$ determines the number variance for $n < D/(2T)$. Likewise, the spectral form factor
for $\tau < T$ mostly contributes to the number variance for $n > D/(2T)$. This
does not imply, however, that the asymptotic behavior of the number variance is given by the
small-$\tau$ behavior of the spectral form factor.
In particular,
the spectral form factor for $\tau \ll \sqrt D$ increases quadratically
\be
\label{eq:SFF_small_t}
K_c(\tau) = \frac 1{2D} \tau^2.
\ee
This region contributes to the number variance for $n > \sqrt N /2$
but it does not determine the large-$n$ saturation value of the number variance.

Finally, we note that the limits $D	\to\infty$ and $\tau\to0$ do not commute (see Appendix~\ref{app:DSFF_GinUE} for details). Indeed, the coefficient of $\tau^2$ in the expansion of $1-\exp(-\tau^2/4D)=\tau^2/4D+O(\tau^4)$, obtained by taking the large-$D$ limit first, is not equal to the coefficient obtained by taking the small-$\tau$ limit first. The latter is given by $K_{c, {\rm exact}}''(0)/2$, where $K_{c, {\rm exact}}(\tau)$ is the exact finite-$D$ result for the Ginibre ensemble, Eq.~\eref{kcgin}. Its Taylor expansion to second order in $\tau$ is given by $K_{c, {\rm exact}}(\tau) =\tau^2/2D +O(\tau^4)$. It is actually a factor of 2 larger than the result obtained by taking the large-$D$ limit first
and is equal to the perturbative coefficient
\be
\frac 1D \left\langle \sum_k {\rm Re} (z_k)  \sum_l {\rm Re} (z_l)\right\rangle_c,
\ee
which can be easily checked numerically, for example, for an ensemble of 1000 of
$100 \times 100$ Ginibre matrices.
For a real spectrum belonging to the Gaussian Unitary Ensemble (GUE), we also find that for sufficiently small $\tau$, the spectral form factor $K(\tau) \sim \tau^2$ with the perturbative prefactor given by the analogous expression.

\subsection{Local spectral form factor}
\label{sec:local-form}

\subsubsection{Real eigenvalues}

Let us again start by considering the case of real spectra. Except for times much shorter than the inverse mean level spacing, the main contribution to
the spectral form factor (\ref{eq:SFF_def}) comes from eigenvalue pairs that are sufficiently close. That is, we can define a local connected spectral form factor
\be
\label{kc-loc}
K_c(x,t) &=& \frac 1{\brho(x) |\pi(x)|}\left\langle \sum_{x_k,x_l\in \pi(x)} e^{it(x_k-x_l)}\right\rangle_c\nn \\
&=& \frac 1{\brho(x) |\pi(x)|}\left\langle \sum_{x_k,x_l\in \pi(x)} e^{it \Delta(x) \frac{(x_k-x_l)}{\Delta(x)}}\right\rangle_c,
\ee
where $\Delta(x)$ is the average spacing of the eigenvalues at $x$, the level
density $\brho(x) = 1/\Delta(x)$, and $|\pi(x)|$ is the length of
an interval located at $x$ that satisfies $ \Delta(x) \ll |\pi(x)| \ll D \Delta(x)$. We have, for now, chosen a hard cutoff to enforce locality. Other choices are possible, however, and below we will find a Gaussian cutoff more useful.

Locally, unfolding is just rescaling the eigenvalues by the local level spacing $\Delta(x)$. We thus have
\be
K_c^{\rm unf}(x,t) = K_c(x,t \brho(x)).
\label{local-unf}
\ee
Integrating over the entire spectrum gives
\be
\label{eq:global_local_SFF}
K_c^{\rm unf}(t) = \int dx \brho(x) K_c(x,t \brho(x)).
\ee
We can also invert the relation \eref{local-unf}:
\be
K_c(x, t) = K_c^{\rm unf}(x, t/ \brho(x)).
\ee

Using Eq.~(\ref{sigform}), the spectral form factor is related to the number variance by
\cite{leviandier1986}
\be
\Sigma^2(n) &=& \frac {n^2}{2\pi} \int_{-\infty}^\infty dt K_c^{\rm unf}(t)
\frac {\sin^2(nt/2)}{(nt/2)^2}\nn\\
&=&\int dx \brho(x)\frac {n^2}{2\pi} \int_{-\infty}^\infty dt K_c^{\rm unf}(x,t)
\frac {\sin^2(nt/2)}{(nt/2)^2}\nn\\
&\equiv&\int dx \brho(x) \Sigma^2(x,n),
\label{sigform1}
\ee        
where $\Sigma^2(x,n)$ is the local number variance.
Using Eq.~(\ref{local-unf}), it evaluates to
\be
\Sigma^2(x,n) =\frac {n^2}{2\pi} \int_{-\infty}^\infty dt \frac 1{|\pi(x)| \brho(x)}
\sum_{x_k,x_l \in \pi(x)}\left \langle e^{it(x_k-x_l)/\Delta x} \right \rangle_c \frac {\sin^2(nt/2)}{(nt/2)^2}.
\ee
The integral over $t$ can be evaluated analytically resulting in the number variance
\be
\Sigma^2(x,n)&=&  \frac 1{\brho(x) |\pi(x)|}
\left\langle
\sum_{x_k,x_l\in \pi(x)} \frac {\Delta(x)}D  \left(n-\frac{|x_k-x_l|}{\Delta(x)}\right)
\,\theta\left(n-\frac{|x_k-x_l|}{\Delta(x)}\right)
\right \rangle_c-n ^2 \nn\\
&=&\int_{\pi(x)}d \bar x\, \frac{\brho(\bar x)}{\brho(\bar x)|\pi(x)|}
\left \langle
\left(\sum_{|x_k-\bar x|< n\Delta}1 \right)
\left(\sum_{|x_l-\bar x|< n\Delta}1 \right)\right \rangle_c
-n^2.
\ee
This shows that the local number variance can be obtained by integrating the local spectral form factor.
The spectral form factor of the entire spectrum is thus related to the spectral
average of the number variance. This is not an issue if the number
variance is stationary (i.e., independent of the point $x$) but, as we will see below, the number variance
of the real parts of the eigenvalues of both the Ginibre ensemble and the nHSYK model are not
stationary. On the other hand, the local spectral form factor turns out to be stationary.

The local spectral form factor \eref{kc-loc} shows strong oscillations
resulting from the Fourier modes of the hard cutoff. These oscillations can
be eliminated by introducing a smooth cutoff. For a Gaussian cutoff, we
obtain the local spectral form factor \cite{Gharibyan:2018jrp}
\be
K_c^{\rm loc} (\bar x, t ) =\frac 1{\cal N} \int d x_1 dx_2 \rho_{2c}(x_1,x_2)
e^{it(x_1-x_2)-\frac{(x_1-\bar x)^2+(x_2-\bar x)^2}{2w^2}},
\label{kt-loc}
\ee
where ${\cal N}$ is a normalization factor chosen such that $K_c^{\rm loc} (\bar x,t ) $ asymptotes to 1 for large $t$ and $w$ is the width of the cutoff. The large-$t$ behavior 
determined by the contribution of the self-correlations is given by
\be
\int d x_1 dx_2  \brho(x) \delta(x_1-x_2)
e^{it(x_1-x_2)-\frac{(x_1-\bar x)^2+(x_2-\bar x)^2}{2w^2}}
=
\int d x   \brho(x)
e^{-\frac{(x-\bar x)^2}{w^2}}.
  \ee
  This results in the normalization factor
  \be
  \label{eq:norm_local}
  {\cal N} = \int d x   \brho(x)
  e^{-\frac{(x-\bar x)^2}{w^2}}.
  \ee

\subsubsection{Real parts of complex eigenvalues}

So far, we have only considered the local spectral form factor of real eigenvalues. We now turn to the local spectral form factor of the real parts of complex eigenvalues, i.e., the local counterpart of Eq.~(\ref{kfyod}). A straightforward generalization of Eq.~(\ref{kt-loc}) yields
\be
K_c^{\rm loc} (\bar x, \tau ) =\frac 1{\cal N} \int d^2 z_1 d^2 z_2 \rho_{2c}(z_1,z_2)
e^{i\tau(x_1-x_2)-\frac{(x_1-\bar x)^2+(x_2-\bar x)^2}{2w^2}},
\label{kt-loc-real-parts}
\ee
where $z_j=x_j+iy_j$.

We now show that this local spectral form factor for eigenvalues unfolded to constant density inside the unit disk is stationary.
    The non-Hermitian two-point correlator is given by
    \be
    \rho_{2c}(z_1,z_2) = \brho (z_1) \delta^2(z_1-z_2) - \bar \rho^2(z_c)
    R_{\rm univ}(|z_1-z_2|^2  \brho(z_c)),
    \ee
    where $ R_{\rm univ}$ is the universal two-point correlator and $z_c=(z_1+z_2)/2$.
    Changing to variables
    \be
    z_1 =z_c +\delta z/2,\qquad  z_2 =z_c -\delta z/2,
    \ee
    the universal contribution to the spectral form factor in Eq.~(\ref{kt-loc-real-parts}) can be written as
    \be
-   \frac 1{\cal N} \int d^2 z_c  \brho^2(z_c) e^{-(x_c-\bar x)^2/w^2}  \int d^2\delta z  e^{i \tau\delta x- \frac 14 \delta x^2/w^2}
    R_{\rm univ}(|z_1-z_2|^2  \brho(z_c)).
    \ee
     For eigenvalues unfolded to the complex unit disk, this becomes
 \be
 -  \frac 2{\cal N}   \left (\frac D\pi \right )^2 \int_{-1}^{1} d x_c \sqrt{1-x_c^2} e^{-(x_c-\bar x)^2/w^2}   \int d^2\delta z e^{i \tau\delta x - \frac 14 \delta x^2/w^2}
    R_{\rm univ}(|z_1-z_2|^2\brho(z_c)).
    \ee
    For the Ginibre universality class with universal two-point correlator~(\ref{eq:2pt-Ginibre}), the second
    integral can be worked out:
\be
   \frac 1{\cal N}  \int d^2\delta z e^{i \tau\delta x- \frac 14 \delta x^2/w^2} 
   R_{\rm univ}(|z_1-z_2|\bar \rho(z_c))
    &=&
   \frac 1{\cal N}    \int d^2\delta z  e^{i \tau\delta x - \frac 14 \delta x^2/w^2} 
      e^{-D(\delta x^2+\delta y^2 )}\nn\\
    &=&
    \frac{1}{\mathcal{N}} \frac{\pi}{\sqrt{ D(D +1/4w^2)} } e^{-\tau^2/(4D+1/w^2)}.
\ee
Collecting all terms and using that the normalization factor is equal to [see Eq.~(\ref{eq:norm_local})]
      \be
         {\cal N} =\frac D\pi   \int d x_c \sqrt{1-x_c^2} e^{-(x_c-\bar x)^2/w^2}.
      \ee
We find that, for eigenvalues unfolded to a constant density inside the unit disk and in the limit $w^2\gg 1/D$, the local spectral form factor of the real parts of the eigenvalues is given by:
         \be
         K_c^{\rm loc}(\bar x, \tau)= 1 - e^{-\tau^2/(4D)}.
         \ee
This shows that the spectral form factor is stationary. In contrast, the global spectral form factor, given by Eq.~(\ref{eq:global_local_SFF}), is the integral of a nonstationary quantity due to the semicircular distribution of the projected eigenvalues multiplying the local spectral form factor.

\subsubsection{Relation to the local number variance}

It is possible to calculate the number variance of the real parts of the eigenvalues from the spectral
form factor
using the relation \eref{sigform} which is valid for unfolded
eigenvalues \cite{fyodorov1997}.
If the local eigenvalue density is $\bar\rho(E)$ (normalized to $\int \bar\rho(E) dE= D$), the spectral form factor unfolded to the unit density is given by
[see Eq.~\eref{local-unf}]
\be
K_c^{\rm unf}(E,\tau) = 1- e^{- \tau^2 \rho(E)^2/ 4D }.
\ee
The resulting integral can be performed analytically
\cite{fyodorov1997} leading to
\be
\Sigma^2(n,D) = n \, {\rm erfc}\left(\frac{n \sqrt D}{\brho(E)} \right )
+\frac {\brho(E)}{\sqrt{\pi D}}\left ( 1 - e^{-Dn^2/\brho^2(E)}
\right ) .
\label{nv-ann}
\ee
This is in agreement with the previous result \eref{sig-local} obtained from the number variance of a rectangular
geometry with vertical side $b = \sqrt {1-E^2}$. 
If we take the limit $D\to \infty$ at fixed $n$, we get Poisson statistics:
\be
\lim_{D\to \infty} \Sigma^2(n,D) = n.
\ee
However, the expression for the number variance also has
a nontrivial double scaling limit
\be
\lim_{D\to \infty} \frac{\sqrt{D}}{\rho(E)} \Sigma^2 \left(\frac{n \rho(E)}{\sqrt{D}},D\right)
=n\, {\rm erfc}\!\left(n \right) + \frac 1{\sqrt{\pi}}
\left( 1-e^{-n^2}\right ).
\label{nv-annds}
\ee
It is tempting to interpret
the existence of this scaling limit as a signature
of quantum
chaos in non-Hermitian systems. However, a similar scaling behavior
has been observed for the number variance of integrable
systems at a finite distance above the ground state
\cite{berry1985semiclassical,seligman1984quantum,seligman1985spectral,verbaarschot1987higher}.

\begin{figure}[t!]
	\centerline{\includegraphics[width=8cm]{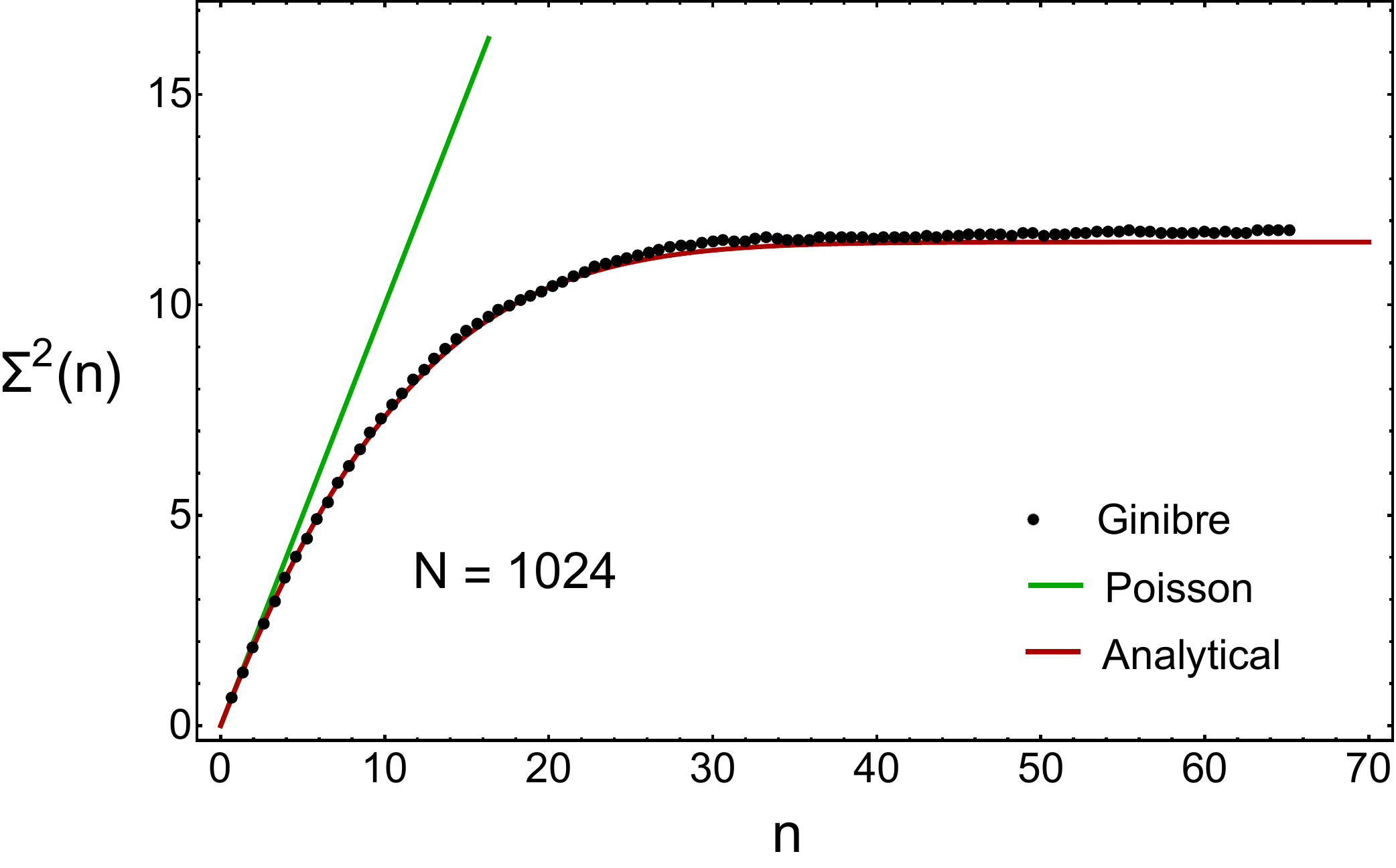}}
	\caption{Number variance of the real parts of the eigenvalues in an interval located symmetrically about zero. 
		The dotted line corresponds to the numerical result for an ensemble of 8192 $1024\times 1024$ random matrices belonging to the GinUE (class A). The agreement with the analytical expression \eref{nv-ann}
		(red curve) is excellent. We also show the result for Poisson statistics (green line).}
	\label{fig:nv-gin}
\end{figure}

As a check of the analytical result \eref{nv-ann}, we compare
in Fig.~\ref{fig:nv-gin} this expression to the numerically calculated
number variance at the center of the spectrum
for an ensemble of $8192$ realizations of $1024 \times 1024$
complex Ginibre matrices.
The normalization is such that the support of the eigenvalues is the complex unit disk but this does not affect the number variance. The discrepancy between the analytical and numerical results for large $D$ is due to finite-size corrections. There is also a correction due to the semicircular shape of the spectral density, but
because $ n\ll D$, this correction is much smaller and can be neglected.

\subsection{Spectral form factor for universality classes AI$^\dagger$ and AII$^\dagger$}

So far, we have restricted our discussion to the spectral form factor of the GinUE (also known as symmetry class A). However, there exist two other classes of bulk level repulsion, depending on the behavior of the Hamiltonian under transposition~\cite{hamazaki2020,kanazawa2021,garcia2022} besides class A, which has no transposition symmetry. If there exists an antiunitary operator $\mathcal{C}_+$ such that $\mathcal{C}_+H^\dagger\mathcal{C}_+^{-1}=H$, then $H$ belong to class AI$^\dagger$ if $\mathcal{C}_+^2=+1$ and to class AII$^\dagger$ if $\mathcal{C}_+^2=-1$. In contrast, if no such operator exists, $H$ belongs to class A. As in the Hermitian case, where the spectral form factors of the GOE, GUE, and GSE distinguish the increasing degrees of level repulsion, so do the spectral form factors of classes AI$^\dagger$, A, and AII$^\dagger$.

\begin{figure}[t!]
	\centerline{\includegraphics[width=0.9\textwidth]{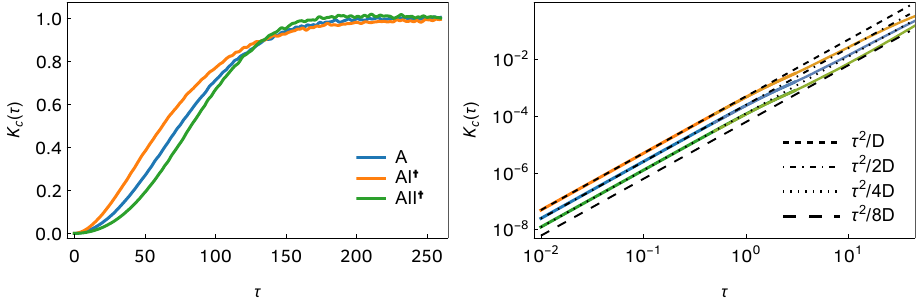}}
	\caption{The spectral form factor of the real parts of the eigenvalues of non-Hermitian matrices from the three bulk universality classes, A, AI$^\dagger$, and AII$^\dagger$. The solid curves correspond to the numerically obtained results for an ensemble average of $10^4$ $2048\times2048$ matrices drawn from the respective RMT ensemble. The right panel shows a magnification of the region close to the origin, where the spectral form factor shows quadratic growth in time (dashed lines) with a coefficient that decreases by a factor of 2 going
		from the perturbative to the nonperturbative domain.}
	\label{fig:A_AId_AIId}
\end{figure}

The spectral form factor for the other two universal bulk statistics,
AI$^\dagger$ and AII$^\dagger$, is much less understood and no analytical results are available. Some numerical results were presented in Ref.~\cite{ghosh2022} for class AI$^\dagger$, while the spectral form factor for class AII$^\dagger$ has not been investigated before.
We obtained them numerically and plot them in Fig.~\ref{fig:A_AId_AIId}. As for class A, for classes AI$^\dagger$ and AII$^\dagger$ there is also an early quadratic growth, albeit with a different prefactor---for class AI$^\dagger$ it is twice the prefactor of A, while for class AII$^\dagger$ it is half.
Note that the prefactor for class A, $1/(2D)$, is in agreement with our previous considerations, see Eq.~(\ref{eq:SFF_small_t}). As can be seen from Fig.~\ref{fig:A_AId_AIId} (right), in all three cases, the prefactor of the $\tau^2$ dependence decreases by a factor of 2 going from the perturbative to the nonperturbative domain [see the discussion below
Eq. \eref{eq:SFF_small_t} for an explanation of this anomaly for class A].
The approach to the late-time plateau is slower for AI$^\dagger$ than for A, while it is faster for AII$^\dagger$.
Contrary to the Hermitian case, there is no nonanalyticity in the spectral form factor around the Heisenberg time.

The spectral form factor is a bulk observable, namely, it is defined as a sum over all eigenvalues,
so in principle it can identify only three different classes A, AI$^\dagger$, and AII$^\dagger$.
However, we will show in the next section, that for ensembles with chiral symmetry, the number variance calculated for a symmetric interval around zero is a factor of 2 different with respect to the Ginibre ensembles, so that it can be employed to identify non-Hermitian systems with chiral symmetry.

\subsection{Spectral form factor for Poisson statistics}

In this section, we calculate the spectral form factor for spectra with
Poisson statistics (i.e., 2d uncorrelated points) unfolded to constant density inside the complex unit disk.
The connected two-point correlation function is given by
\be
\rho^P_c(z_1,z_2) = \bar \rho (z_1) \delta^2(z_1-z_2)
- \frac 1D \bar \rho(z_1)\bar \rho(z_2).
\ee
The connected spectral form factor is given by
\be\label{eq:Poisson}
K_c^P(\tau)&=& \frac 1{D} \int d^2z_1 d^2z_2 \rho^P_c(z_1,z_2) e^{i\tau(x_1-x_2)}\nn\\
&=&1- \frac 1{D^2} \left |\int d^2z_1\brho(z_1) e^{i\tau x_1} \right |^2\nn\\
&=&1-  \left |2\int_{-1}^1  dx_1 \sqrt{1-x_1^2} e^{i\tau x_1} \right |^2\nn\\
&=& 1 -  4\left(\frac{J_1(\tau)}{\tau}\right )^2.
  \ee

\begin{figure}[t!]
	\centerline{\includegraphics[width=8cm]{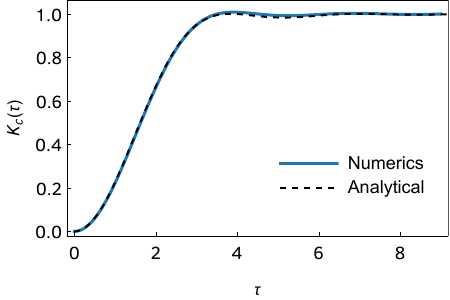}}
	\caption{The spectral form factor of the real part of uncorrelated random variables uniformly distributed on the unit disk. The full blue line corresponds to the numerically obtained spectral form factor for $10^4$ sets of $2048$ independent random complex numbers with a flat distribution on the unit disk, while the dashed black line gives the analytical prediction of Eq.~(\ref{eq:Poisson}). There is a perfect agreement between the two.}
	\label{fig:dsff_Poisson}
\end{figure}

In Fig.~\ref{fig:dsff_Poisson}, we show a numerical verification of the result of Eq.~(\ref{eq:Poisson}), finding perfect agreement. We will see in the next section that although the $q=2$ nHSYK model is integrable, its spectral form factor has more structure than plain Poisson statistics of completely uncorrelated random variables.
  
\section{Long-range spectral correlations of the Sachdev-Ye-Kitaev model with complex couplings}
\label{sec:nHSYK}

We now probe the dynamics of the non-Hermitian SYK (nHSYK) model for timescales shorter than the Heisenberg time by a detailed comparison of the unfolded spectral form factor and the number variance with the random matrix predictions worked out in the previous section. The nHSYK Hamiltonian is defined as~\cite{kitaev2015,maldacena2016} 
\begin{align}\label{hami}
	H \, &= \, \sum_{i_1<i_2<\cdots<i_q}^{N} (J_{i_1i_2\cdots i_q}+i M_{i_1i_2\cdots i_q}) \, \psi_{i_1} \, \psi_{i_2} \, \cdots \, \psi_{i_q},
\end{align}
where $N$ and $q$ are integers ($N$ is taken to be even), $J_{i_1\cdots i_q}$ and $ M_{i_1\cdots i_q}$ are real Gaussian random variables with zero mean and variance 
\be
\sigma^2 =\frac 1{6(2N)^{q-1}},
\ee
and $\psi_i$ are Majorana fermions satisfying $\{ \psi_{i}, \psi_{j} \} = 2\delta_{ij}$. To be precise, we note that for odd $q$, $H$ corresponds to a supercharge (not Hamiltonian) operator.

\begin{table}[]
	\caption{Classification of the nHSYK Hamiltonian into non-Hermitian bulk universality classes for all $q$ and even $N$. Note that this does not correspond to the full symmetry classification, which is richer and goes beyond bulk level statistics~\cite{garcia2022}.}
	\label{tab:classification}
	\begin{tabular}{@{}l cccc@{}}
		\toprule
		$N\,\mathrm{mod}\,8$   & 0              & 2               & 4               & 6               \\ \midrule
		$q\,\mathrm{mod}\,4=0$ & AI$^\dagger$   & A               & AII$^\dagger$   & A               \\
		$q\,\mathrm{mod}\,4=1$ & AI$^\dagger$   & AI$^\dagger$    & AII$^\dagger$   & AII$^\dagger$   \\
		$q\,\mathrm{mod}\,4=2$ & A              & A               & A               & A               \\
		$q\,\mathrm{mod}\,4=3$ & AI$^\dagger$   & AII$^\dagger$   & AII$^\dagger$   & AI$^\dagger$    \\ \bottomrule
	\end{tabular}
\end{table}

The symmetry classification of the nHSYK model was put forward in Ref.~\cite{garcia2022}. It was found that, depending on $q\mod4$ and $N\mod8$, it belongs to nine out of the 38 non-Hermitian symmetry classes. However, the bulk correlators we are employing here only capture the local level repulsion, i.e., only distinguish the bulk universality classes A, AI$^\dagger$, and AII$^\dagger$. The bulk universality classes for different $q$ and $N$~\cite{kanazawa2021,garcia2022} are tabulated in Table~\ref{tab:classification}.

We obtain the spectrum by exact diagonalization techniques. We carry out an ensemble average to suppress statistical fluctuations, reaching at least $10^5$ eigenvalues for a given $q$ and $N$. 
Since the spectrum is radially symmetric, the necessary unfolding is carried out as explained in Sec.~\ref{sec:unfolding}. Depending on $q$, we shall employ polynomials of different degrees to approximate $\bar \rho(r)$. For $q=4$, the radial spectral density is well approximated
by a fourth-order even polynomial, while for $q=2$ it is close to a Gaussian, and
for $q=6$ the radial spectral density is almost constant so that unfolding
is basically a rescaling of the eigenvalues. For $q=3$, the radial spectral density can be unfolded by an eighth-order even polynomial. 

In what follows, we will compare the spectral form factor and number variance of the nHSYK model with the random matrix prediction in the corresponding universality class (Table~\ref{tab:classification}). We start our analysis with the spectral form factor.

\subsection{Spectral form factor of the nHSYK model}
We carry out the numerical evaluation of the 
ensemble-averaged spectral form factor corresponding to
the real parts of the eigenvalues, $K_c(\tau, \theta)$,
for various values
of $N$ and $q$.
As before, the spectral form factor is normalized such that it asymptotes to 1 for large times.
Since the spectrum is rotationally invariant, we additionally average the spectral form factor
over $10$ values of $\theta_j = \pi j/5, j =1,\ldots, 10$.

\subsubsection{Ginibre universality for $q=4$}

\begin{figure}[t!]
	\centerline{  \includegraphics[width=8cm]{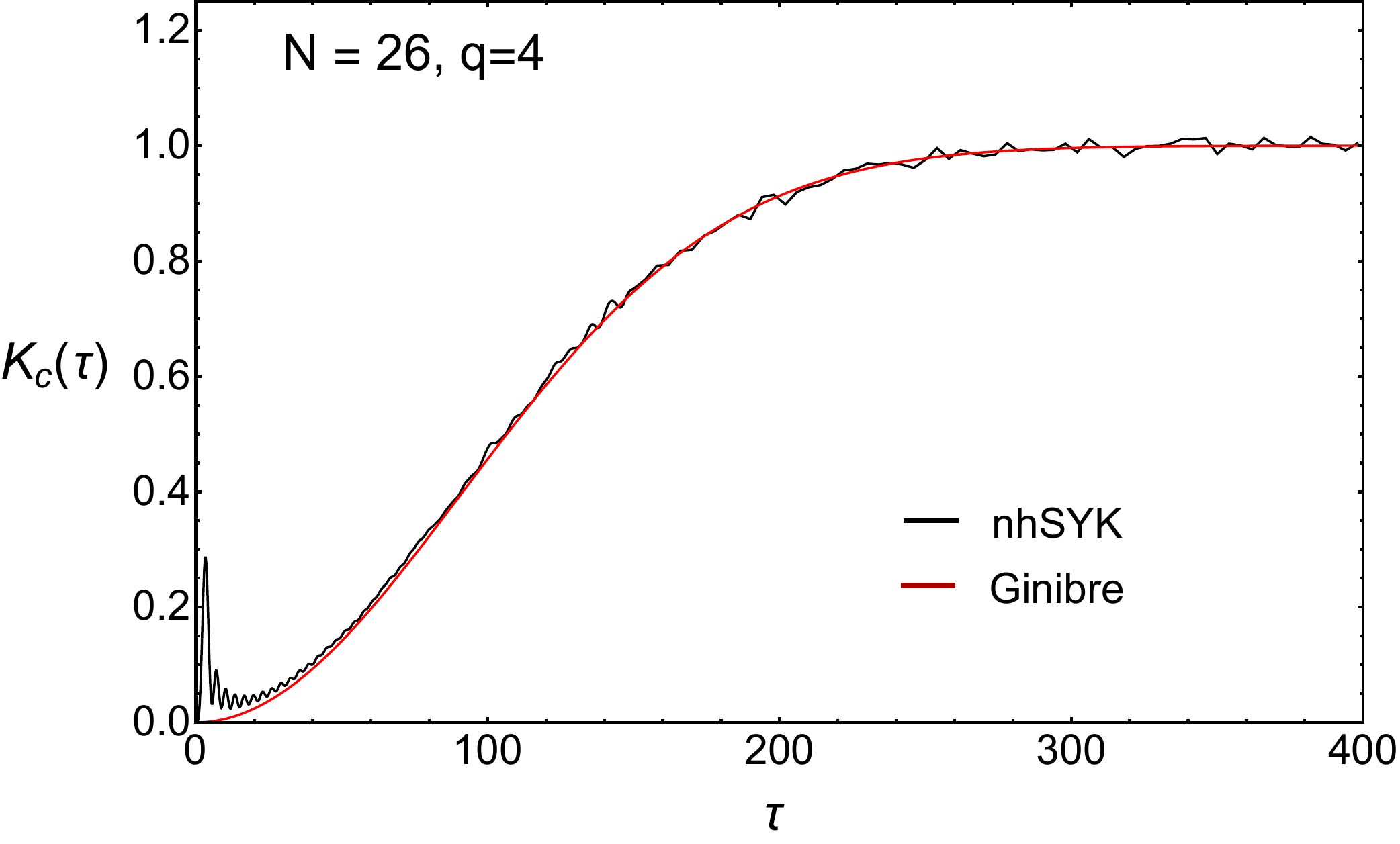}
		\includegraphics[width=8cm]{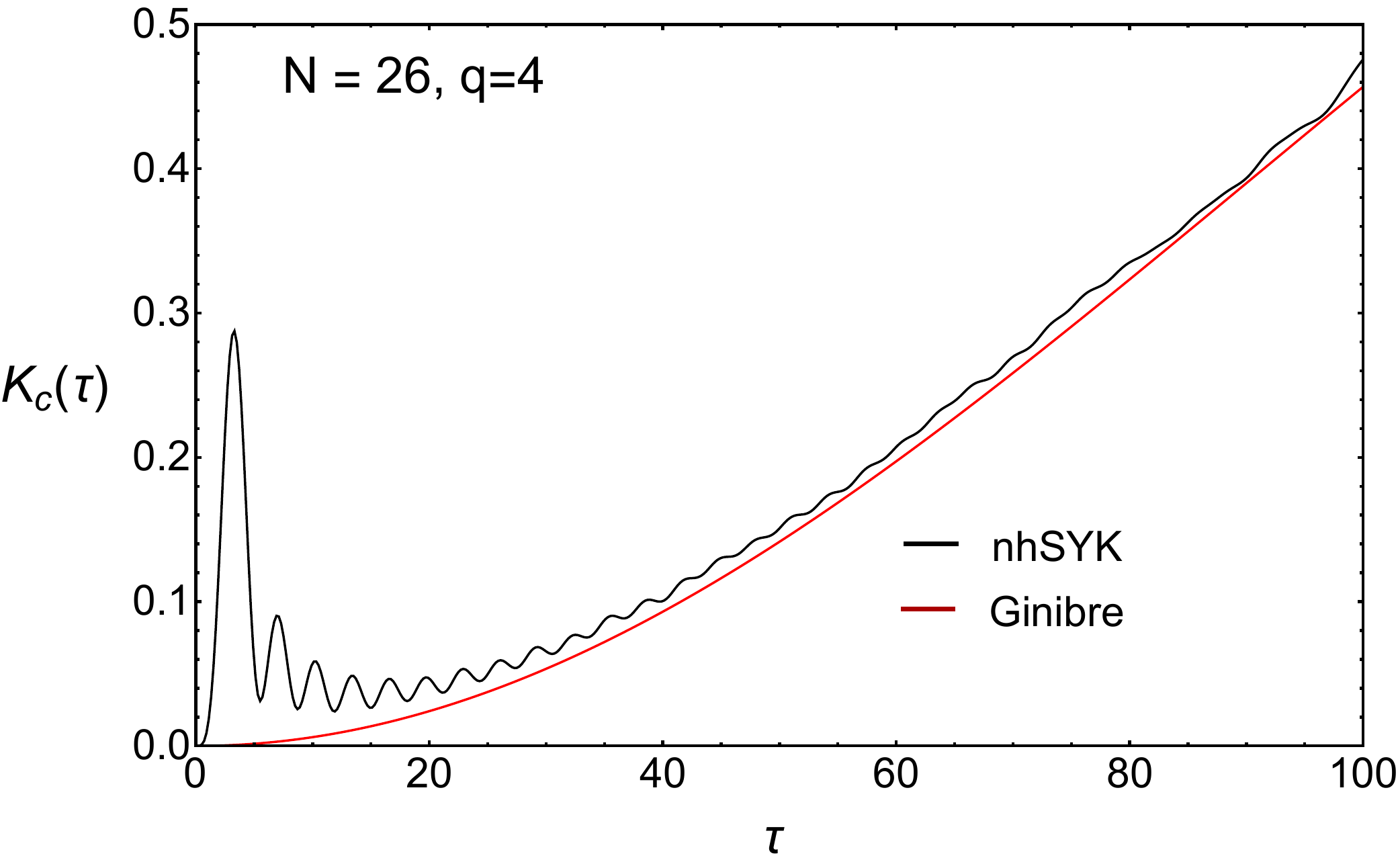}}
	\centerline{  \includegraphics[width=8cm]{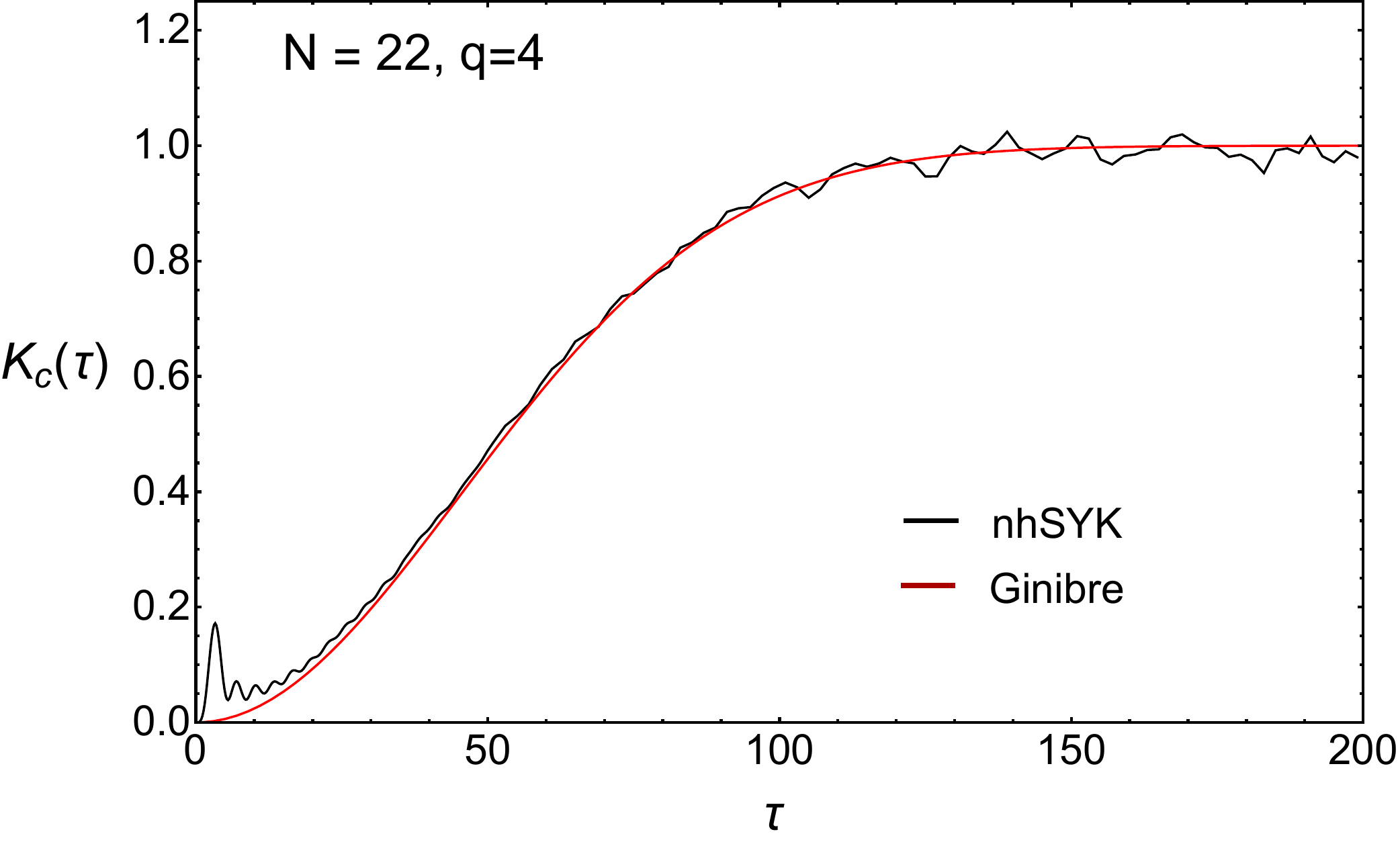}
		\includegraphics[width=8cm]{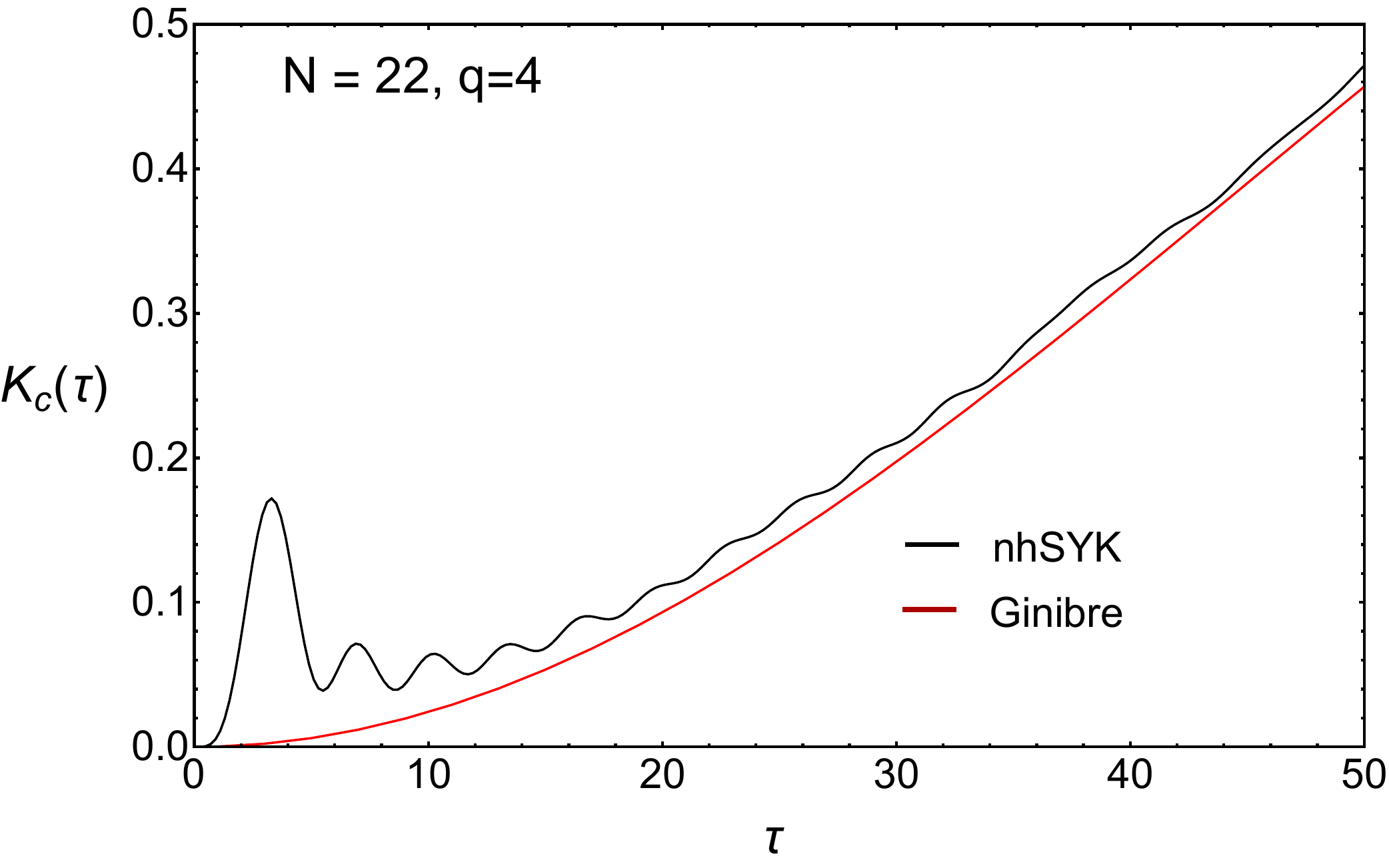}}
	\caption{The connected part of the spectral form factor of the real parts of the unfolded eigenvalues, Eq.~(\ref{form}), normalized by the number of eigenvalues $D\sim2^{N/2-1}$, of the $q = 4$ nHSYK model for $N =26$ (upper) and $N = 22$ (lower). The results are compared to the analytical prediction for the GinUE (red curves) with the same number of eigenvalues. The right panels, which are a magnification of the left panels, show that the spectral form factor of the nHSYK model differs from the spectral form factor of the GinUE up to a scale of about $\tau \sim \sqrt D$.
	\label{fig:kt22-26}
	}
\end{figure}

\begin{figure}[t!]
	\centerline{\includegraphics[width=8cm]{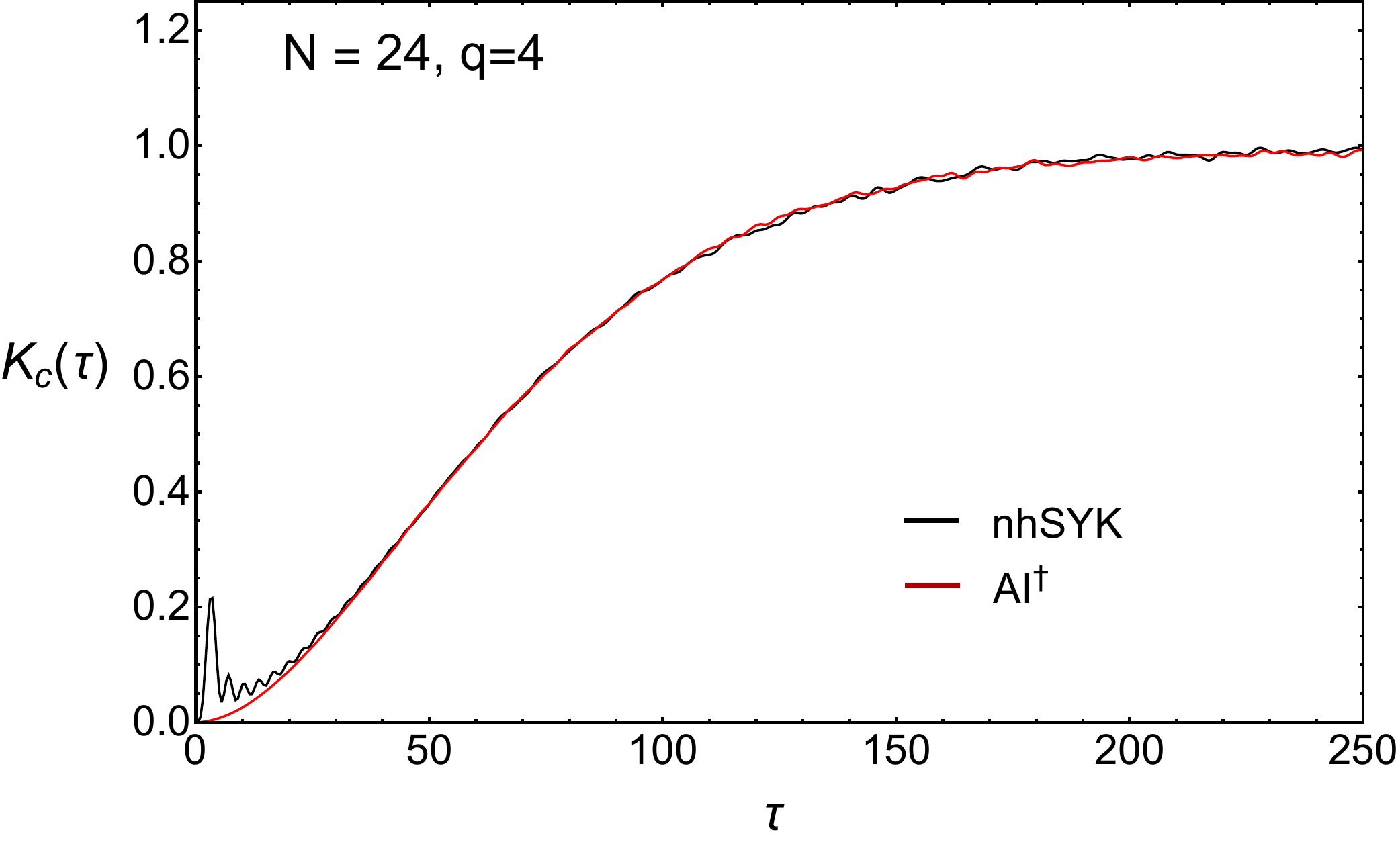}
		\includegraphics[width=8cm]{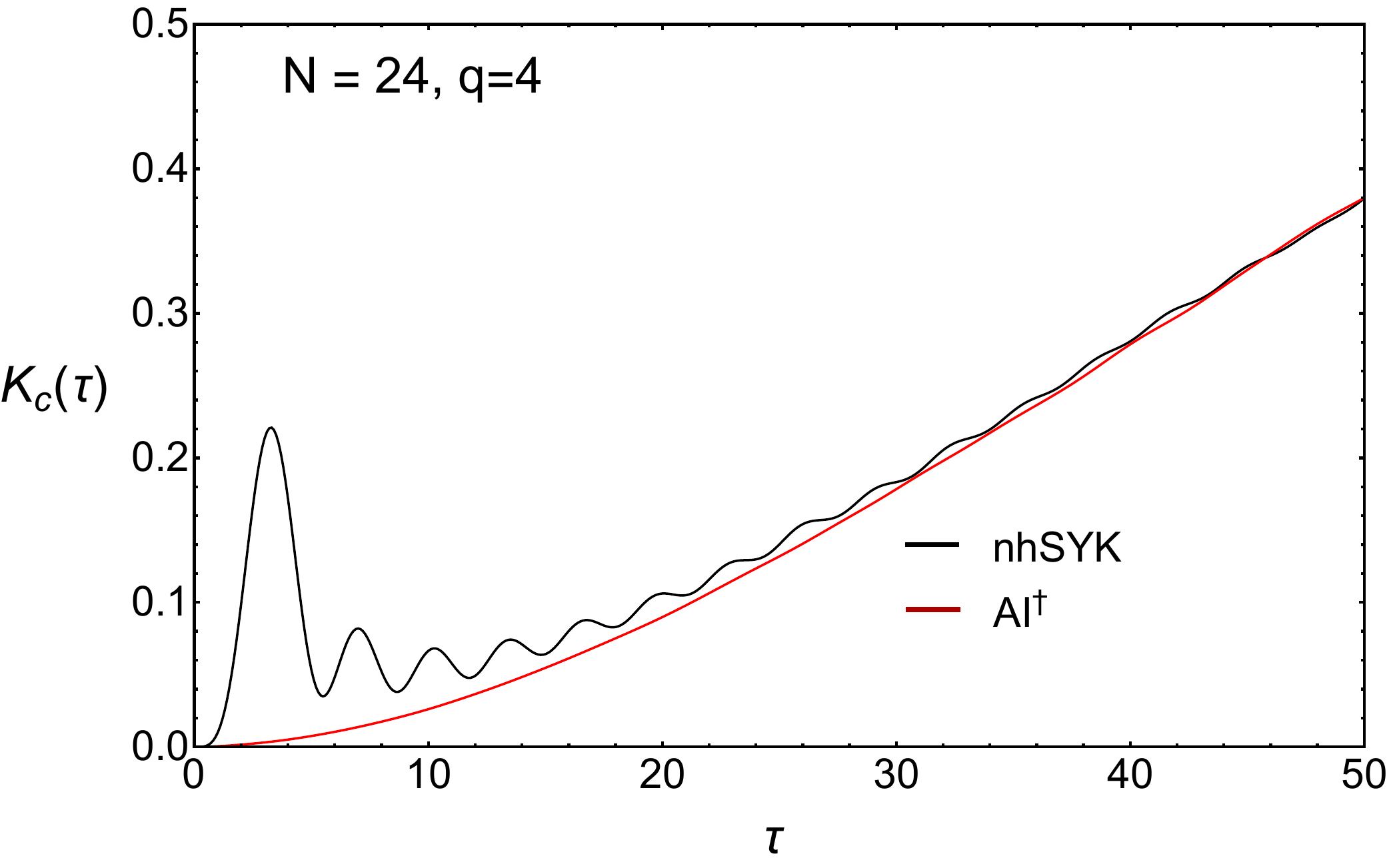}}
	\centerline{\includegraphics[width=8cm]{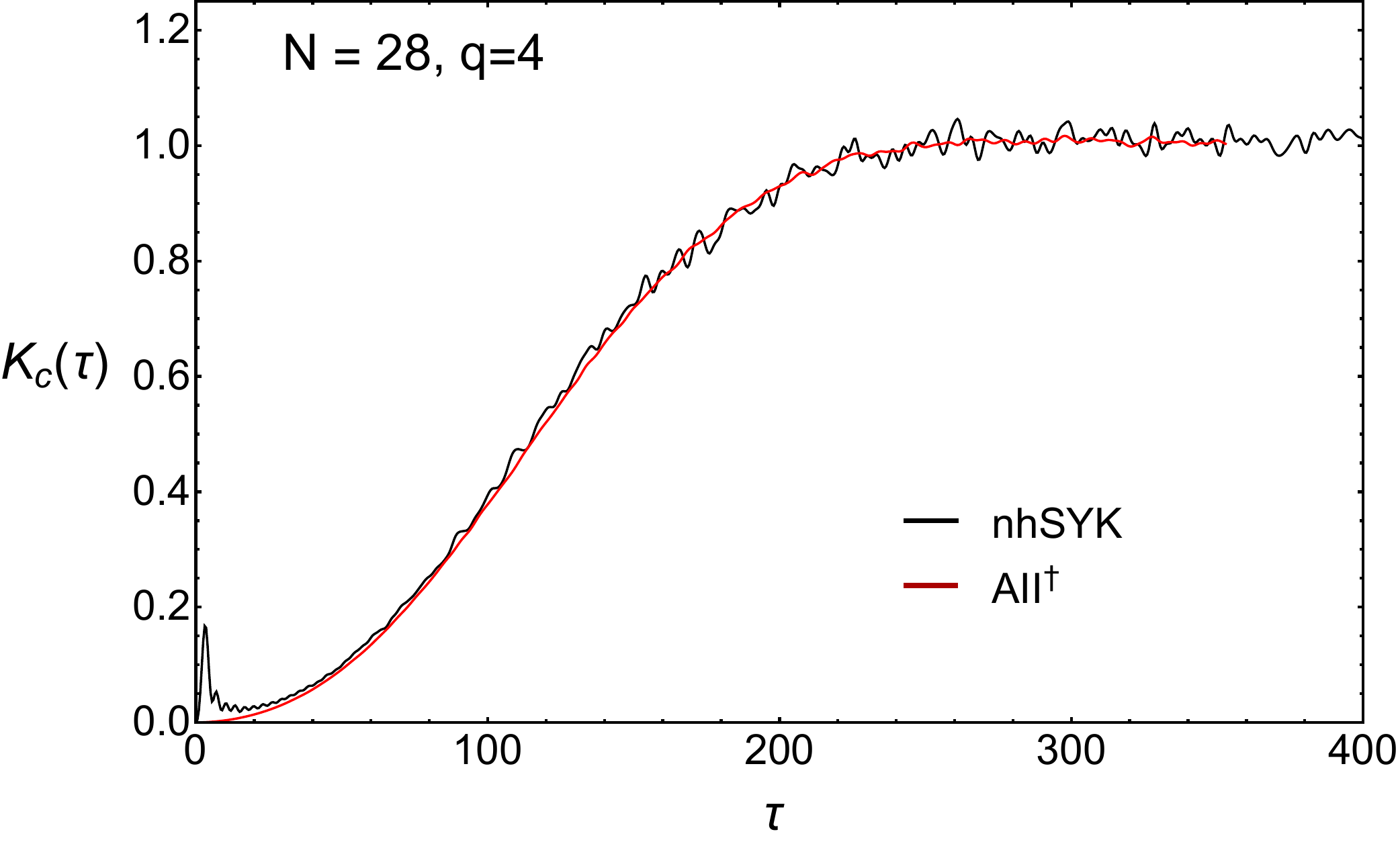}
		\includegraphics[width=8cm]{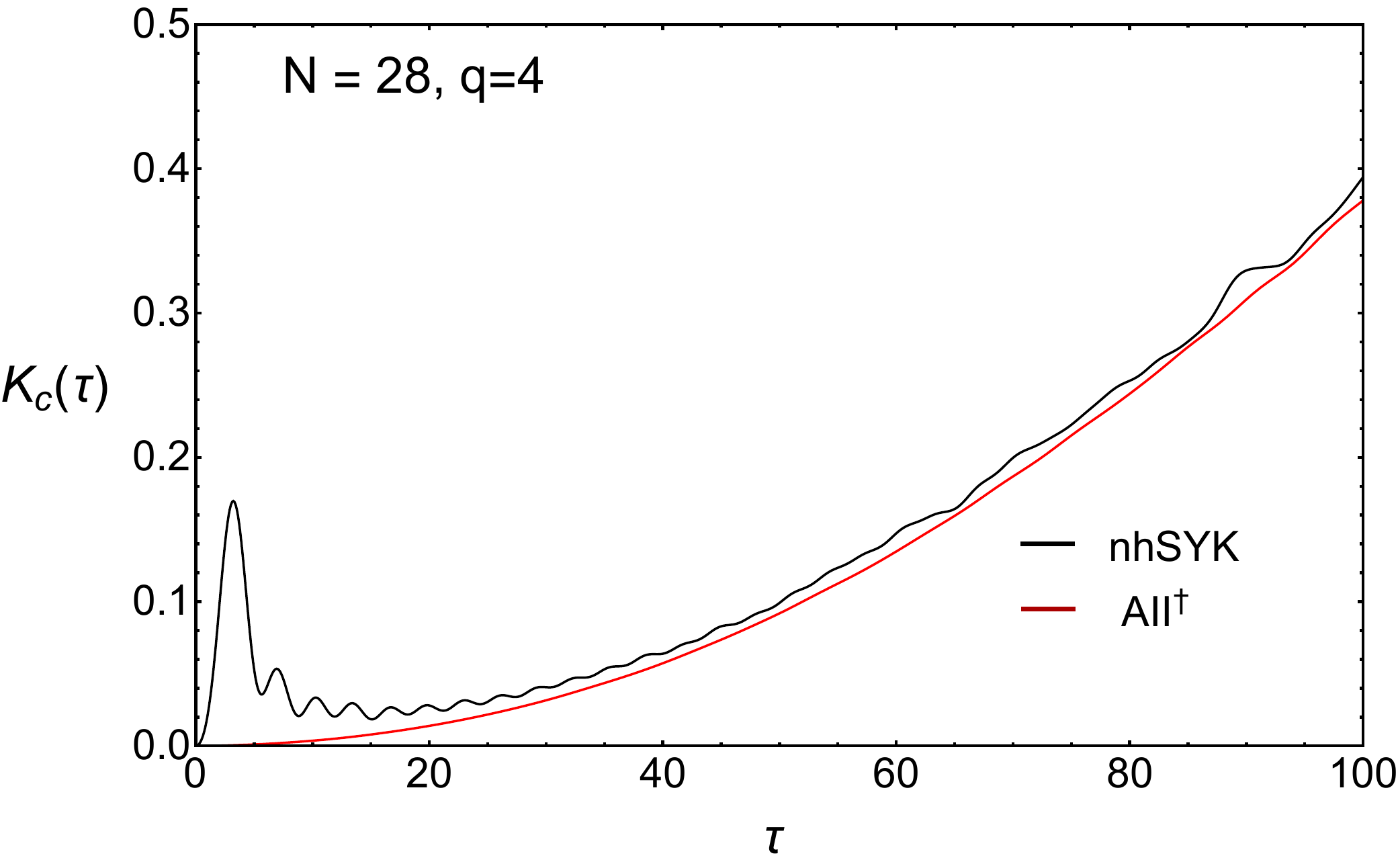}}
	\caption{The connected part of the spectral form factor of the real parts of the unfolded eigenvalues, Eq.~(\ref{form}), normalized by the number
		of eigenvalues $D\sim2^{N/2-1}$, of the $q = 4$ nHSYK model
		for $N=24$ (upper) and $N =28$ (lower). The red curves are the random matrix prediction which was obtained numerically for the AI$^\dagger$ and AII$^\dagger$
		universality classes. The right plots are a magnification of the left ones. We find excellent quantitative agreement without any fitting up to relatively small $\tau \sim \sqrt{D}$, of the order of the correlation hole, which signals the timescale for which the quantum chaotic dynamics is universal.
	}
	\label{fig:dsffq4}
\end{figure}

In Fig.~\ref{fig:kt22-26}, we depict the results for the $q=4$ nHSYK Hamiltonian for
$N=22$ and $N=26$ (black curves)
and compare them to the analytical
result \eref{form-an} for the Ginibre ensemble (red curves).
We find agreement with the random matrix prediction for $\tau > \sqrt D$ but the two results differ for smaller times.
This is fully consistent with the results of short-range correlations \cite{garcia2022} which
are insensitive to these deviations.
In Fig.~\ref{fig:dsffq4}, we show the results for $N=24$ and $N=28$. The agreement with the corresponding random matrix universal result is excellent for $\tau > \sqrt D$. In these two cases, which are in
the universality class of AI$^\dagger$ $(N=24)$ and AII$^\dagger$ $(N=28)$, no
analytical formula is available and the random matrix result was obtained
numerically for $D =2048$. The spectral form factor for other values of
$D$ can be obtained by using the scaling
relation
\be
K_c^{D_2} (\tau) = K_c^{D_1} \left(  \frac {\sqrt {D_1}}{\sqrt{D_2}} \tau\right ).
\ee

\subsubsection{The limits of universality: Collective scale fluctuations}

To better understand the short-time behavior of the spectral form factor we have enlarged the region close to the origin in the plots of the right column of Figs.~\ref{fig:kt22-26} and \ref{fig:dsffq4}. 
The local minimum of $K_c(\tau)$
for $\tau > 0$, usually termed correlation hole~\cite{leviandier1986,wilkie1991,alhassid1992,torres2018,schiulaz2019}, defines, for a real spectrum, the maximum timescale for which
the dynamics did not fully relax to the universal prediction of RMT.
In the Hermitian SYK model~\cite{garcia2018a}, it is determined by the collective fluctuations
of the spectrum that arise because the number of independent matrix elements ($\sim N^q$) is much smaller than the number of matrix elements of the Hamiltonian ($2^{N/2}$). The same mechanism is at
work in the non-Hermitian case, where we have the same mismatch in the number of matrix elements.

The oscillations for small times are mostly due to collective \emph{scale} fluctuations~\cite{Jia:2019orl}. They correspond to fluctuations in the overall scale of eigenvalues from one realization to the next: $x_n\to x_n(1+\xi)$, for all $n$, where $x_n$ are the real parts of the eigenvalues. $\xi$ is a random variable with zero mean and gives rise to the scale fluctuation of the
spectral density:
\be
\rho^\mathrm{scale}(x) =\frac 1{1+\xi}\,\bar \rho\left( \frac x{1+\xi} \right ),
\ee
where $\bar\rho$ is the ensemble-averaged spectral density. 
The connected two-point correlator for these scale fluctuations is given by
\be
\left\langle \rho^\mathrm{scale}(x)\rho^\mathrm{scale}(y)\right \rangle_c&=&
\left\langle\left[\bar\rho(x)(1-\xi)-\xi x \bar\rho'(x)\right]\left[
\bar\rho(y)(1-\xi)-\xi y\bar\rho'(y)\right]\right\rangle -\bar\rho(x) \bar \rho(y)\nn\\
&=&
\left[\bar\rho(x)+x\bar\rho'(x)\right]
\left[\bar\rho(y)+y\bar\rho'(y)\right]\langle \xi^2\rangle,
\ee
where the prime denotes the derivative, we have used $\langle\xi\rangle=0$, and we have dropped all terms of order $\langle\xi^4\rangle$ and above.
Recall that $\bar \rho(x) $ is normalized as $\int \rho(x) dx =D$.
This contributes to the spectral form factor as~\cite{Berkooz:2020fvm}
\be
\delta K_c(\tau) &=& \frac {\langle \xi^2\rangle}D \int dx \frac d{d x}[x\bar\rho(x)] e^{i \tau x}
\int dx \frac d{dx} [ x\bar \rho(x)] e^{ -i\tau x}\nn\\
&=& \frac {\langle \xi^2\rangle}D \tau^2
\left |\int dx\, x\,\bar \rho(x) e^{i \tau x}\right |^2 .
\label{kt-sc}
\ee

The spectral density of the real parts of
the eigenvalues is given by
\be
\bar \rho(x) = \frac 2\pi \frac D{E_0^2} \sqrt{E_0^2-x^2},
\ee
with $E_0 =1$.
The relevant Fourier integral is
\be
\frac 2\pi \frac D{E_0^2} \int_{-E_0}^{E_0} dx\, x \, e^{i \tau x} \sqrt{E_0^2-x^2}
= 2 i  D  \frac{J_2(E_0 \tau)}\tau,
\ee
where $J_2$ is a Bessel function, resulting in the contribution of the scale fluctuations to the spectral
form factor
\be
\label{delK}
\delta K_c(\tau) &=& 4\langle \xi^2\rangle  D [J_2(E_0 \tau)]^2,
\ee
which decreases as $1/\tau$ for large $\tau$. The analytical result for
total spectral form factor in class A including the scale fluctuations factor is then given by
\be
\label{eq:total_SFF_with_scale}
K_c(t) &=& 1-e^{-t^2/4D} + 4\langle \xi^2\rangle  D [J_2(E_0 t)]^2.
\ee

The variance $\langle\xi^2\rangle$ can be computed as
\be
\langle \xi^2 \rangle =\frac 14 \left(\frac {M_{2,2}}{M_2^2} -1\right ),
\label{xi}
\ee
with the moments defined as
\be
M_2 &=& \frac 1D \left \langle \sum_k E_k^2 \right\rangle,\nn\\
M_{2,2} &=& \left\langle \frac 1D \sum_k E_k^2\frac 1D \sum_k E_k^2 \right \rangle.
\ee
Here, $E_k$ are either the real eigenvalues or the real parts of the complex eigenvalues and the brackets $\langle \cdot\rangle$ denote ensemble averaging. For the Hermitian SYK model, the moments can be evaluated exactly and we find that~\cite{Jia:2019orl}
\be
\langle\xi^2\rangle=\frac 12 {N \choose  q}^{-1}.
\label{xi2}
\ee
For the real parts of the eigenvalues
of a non-Hermitian matrix with spectral density unfolded to constant density
inside the unit disk, $\langle \xi^2\rangle$ cannot be obtained
from traces of moments of the Hamiltonian, but its exact
numerical value can be obtained from the real parts of the eigenvalues
using the definition \eref{xi}. For our non-Hermitian SYK model, we find that it is approximately equal to
\be
\langle \xi^2\rangle \approx {N \choose q}^{-1}.
\ee

\begin{figure}[t!]
	\centerline{\includegraphics[width=8cm]{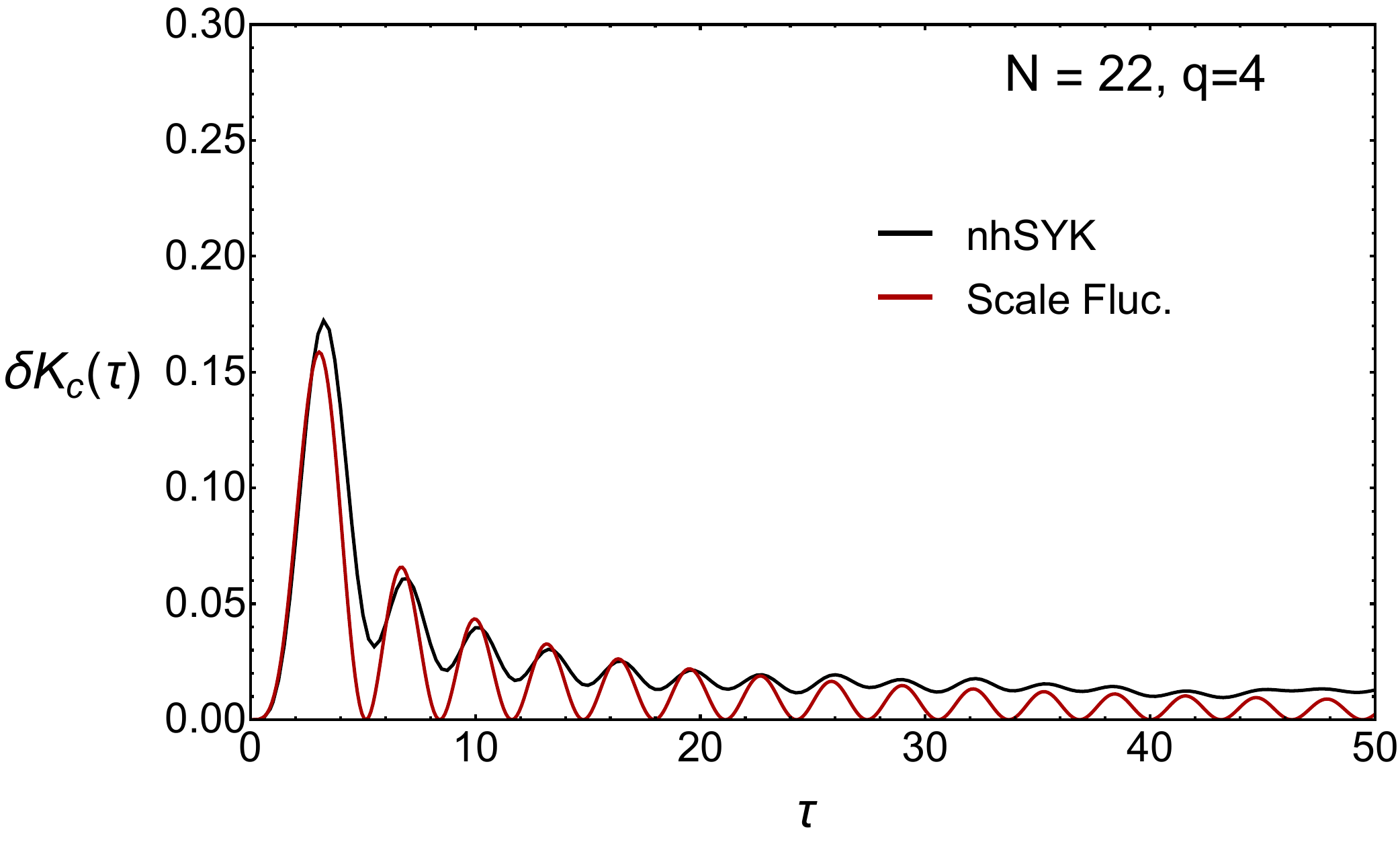}
		\includegraphics[width=8cm]{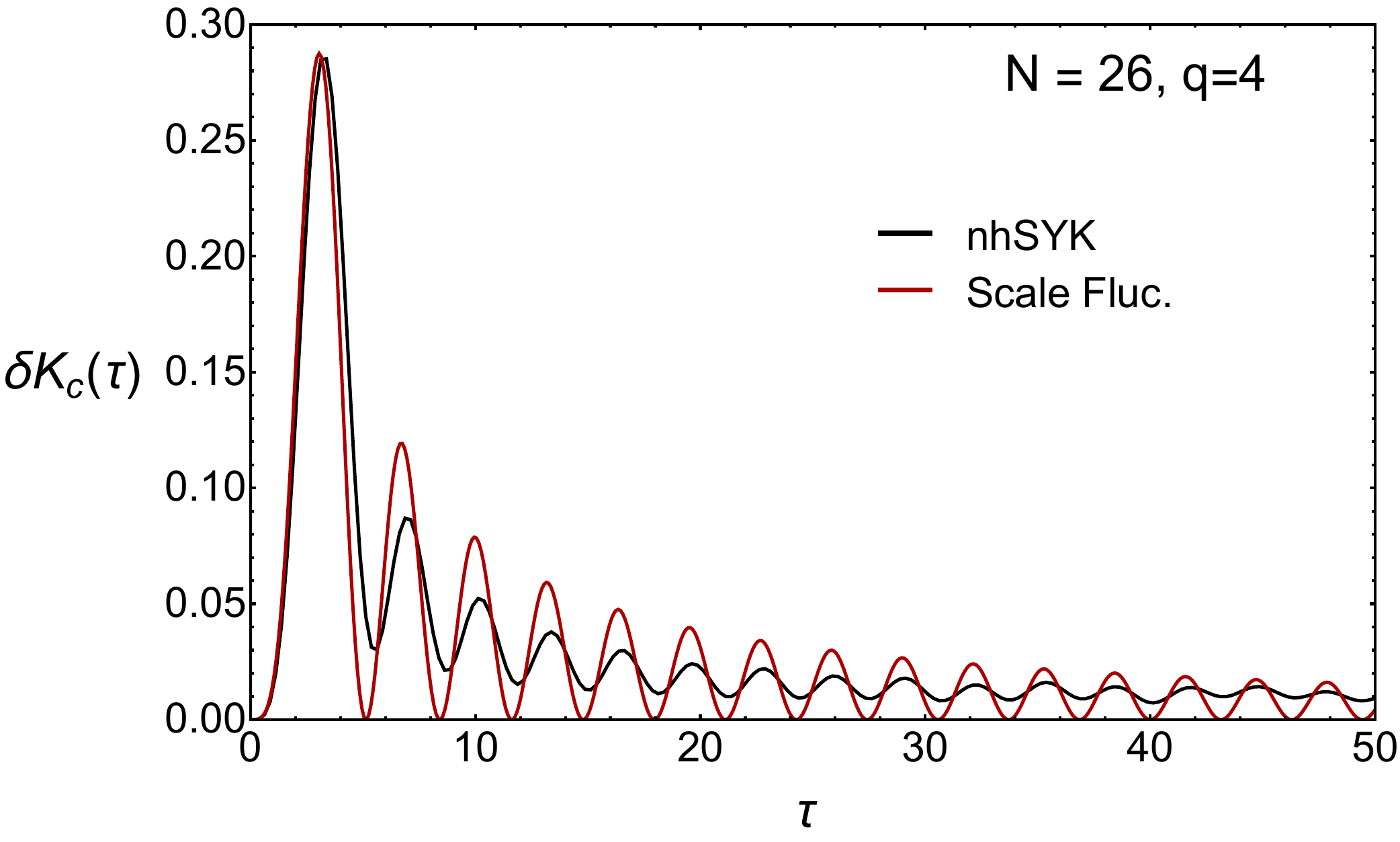}}
	\caption{The difference $\delta K_c$ between the connected spectral form factor of the $q=4$ nHSYK model and the
		GinUE (black curves) for $N=22$ (left) and $N=26$ (right) compared to
		the analytical result due to scale fluctuations, Eq.~(\ref{delK}) (red curves).
		\label{fig:ktdel}}
\end{figure}

\begin{figure}[t!]
	\centering
	\includegraphics[width=8cm]{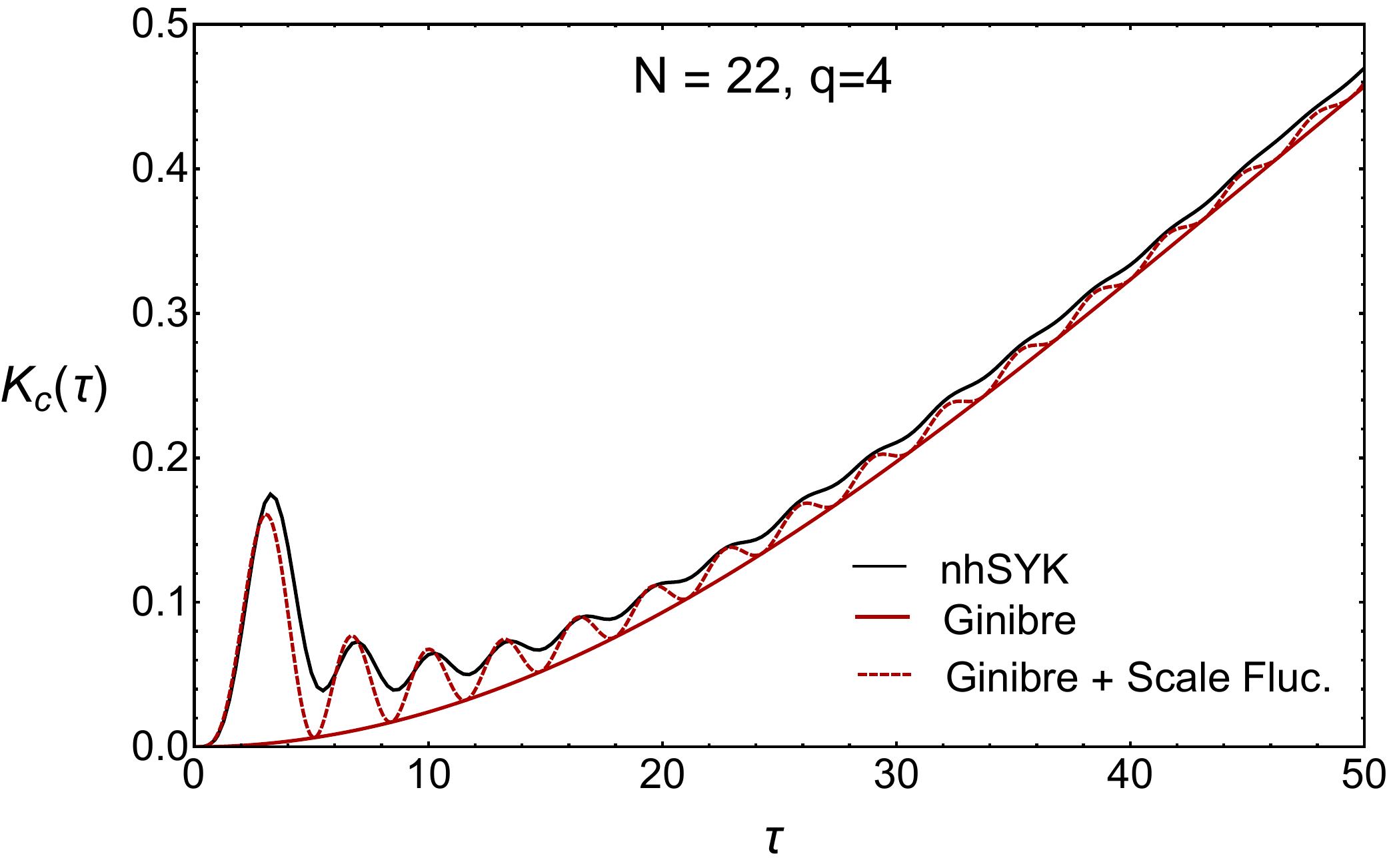}
	\includegraphics[width=8cm]{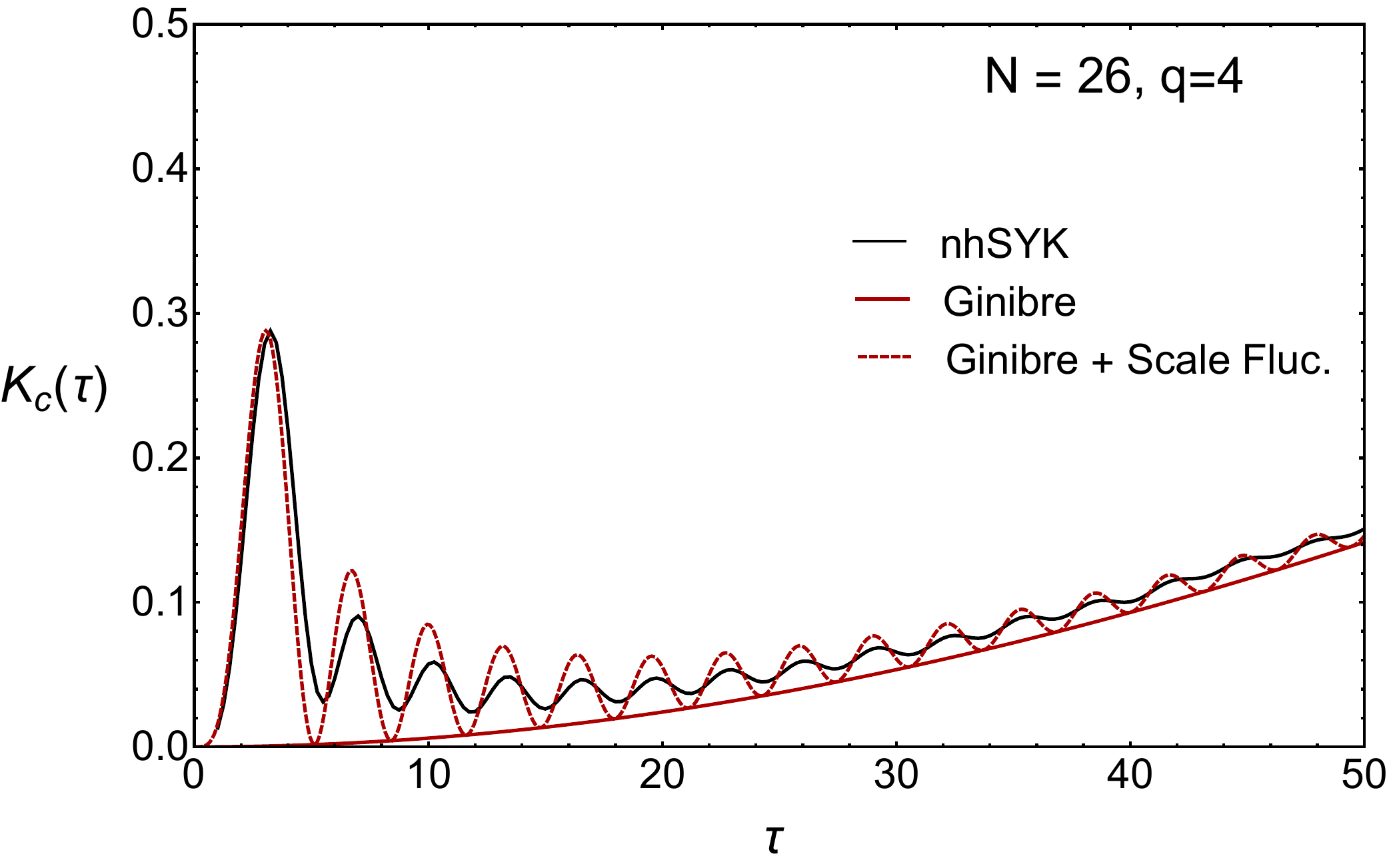}
	\caption{The spectral form factor including the collective scale fluctuations. The sum
		(dashed red curve)
		of the analytical results for the scale fluctuations and the spectral form factor
		of the Ginibre ensemble (red curve) explains the oscillations in the spectral
		form factor
		of the $q=4$ SYK model. Results are given for $N=22$ (left) and $N=26$ (right).}
	\label{fig:fluc}
\end{figure}

In Fig.~\ref{fig:ktdel}, we
show the difference between the spectral form factor
of the real parts of the eigenvalues of the nHSYK model for $q=4$ and the result for the Ginibre ensemble. The numerical results (black curves) for $N=22$ (left) and $N=26$ 
(right), which are both in the Ginibre universality class,
are compared to the analytical result \eref{delK} (red curves). 
In Fig.~\ref{fig:fluc}, we compare the full analytical spectral form factor, Eq.~(\ref{eq:total_SFF_with_scale}), with numerical results. 
Given that
higher multipole collective fluctuations also contribute to the difference,
the agreement
with the analytical result is better than expected, in particular for small times. We thus conclude that most of the oscillatory behavior is due to the lowest-order multipole, i.e., the scale fluctuations. We emphasize that the results for $\delta K_c(\tau)$ are obtained without
using fitting parameters.
Note that the period of the oscillations does not depend
on $N$ and is close to the period of the oscillations of $J_2^2(\tau)$. The amplitude
increases with $D$ and also varies as the amplitude of $J_2^2(\tau)$. 

The location of the correlation hole can be obtained by equating the two contributions to the spectral form factor: the universal Ginibre contribution, Eq.~(\ref{form-an}), and the collective fluctuations contribution, Eq.~(\ref{delK}). By replacing the oscillatory part of $J_2^2$
by its asymptotic average, the location of the correlation hole is thus given by the minimum of 
\be
4\langle \xi^2\rangle\frac 1\pi  \frac { D }{\tau} + 1-e^{- \frac {\tau^2}{4D}}.
\label{benv}
\ee
This condition cannot be solved analytically, but it gives the rough position of the correlation hole and can
be studied numerically for small values of $N$. For the cases studied in this paper belonging to class A, we
compare in Table \ref{tab1} the position of the correlation hole obtained from the figures for
the spectral form factor with the result given by the minimum of \eref{benv} and find
good agreement between the two.

The condition~(\ref{benv}) can be recast as
\be
\frac{1}{\pi}\langle\xi^2\rangle \sim \frac{\tau^3}{8D^2}e^{-\tau^2/4D}.
\label{scale_nH}
\ee
When $\langle \xi^2\rangle \gtrsim 1.29/\sqrt D $, this condition no longer has a solution for a real time $\tau$ and there is no correlation hole. In the case $q=4$, this occurs for $N\gtrsim 80$. Since the contribution of the scale fluctuation dominates the $\tau$-dependence of the spectral form factor
all the way up to the Heisenberg time, the spectral form factor of the real parts of the eigenvalues is no longer a useful measure for spectral fluctuations
due to quantum chaos. 

This result is to be contrasted with the Hermitian case, for which the location of the correlation hole is roughly determined by the condition,
\be
\langle \xi^2 \rangle \sim \frac{\tau^3}{D^2}.
\ee
Although $\langle \xi^2 \rangle$ is approximately the same as before, the Heisenberg time is now of order $D$ (instead of $\sqrt{D}$). As a consequence, there are real solutions $\tau$ for all values of $\langle\xi^2\rangle$. Furthermore, the correlation hole would
only be larger than the Heisenberg time if $\langle \xi^2\rangle\gtrsim D$, a condition that is never satisfied. We conclude that for the Hermitian SYK model, there is always a parametrically large separation between the timescale where collective fluctuations are relevant for the spectral form factor and the Heisenberg time, contrary to the spectral form factor of the real parts of the eigenvalues of the nHSYK model.

\begin{table}[t!]
	\caption{Comparison of the position of the correlation hole obtained from the figures of the spectral form factors for the cases belonging to class A (Figs.~\ref{fig:dsffq4} and \ref{fig:dsffq3q6}) with the estimate given by the minimum of Eq.~\eref{benv}.}
	\begin{tabular}{cccc}
		\toprule
		$\quad N \quad$ & $\quad q\quad $ & $\tau_{\rm hole} $ & $\tau_{\rm estimate}$ \\
		\midrule
		22 & 4  & 10 & 8 \\
		26 & 4 &  16 & 15\\
		22 & 6  &5  & 4 \\
		\bottomrule
	\end{tabular}
	\label{tab1}
\end{table}

Finally, we note that it is possible to eliminate the collective spectral fluctuations by unfolding
the spectrum realization by realization \cite{Gharibyan:2018jrp,Jia:2019orl}. Then
these oscillations do not show up in the spectral form factor.

\subsubsection{Dependence of nonuniversal features on $q$}

\begin{figure}[t!]
	\centerline{  \includegraphics[width=8cm]{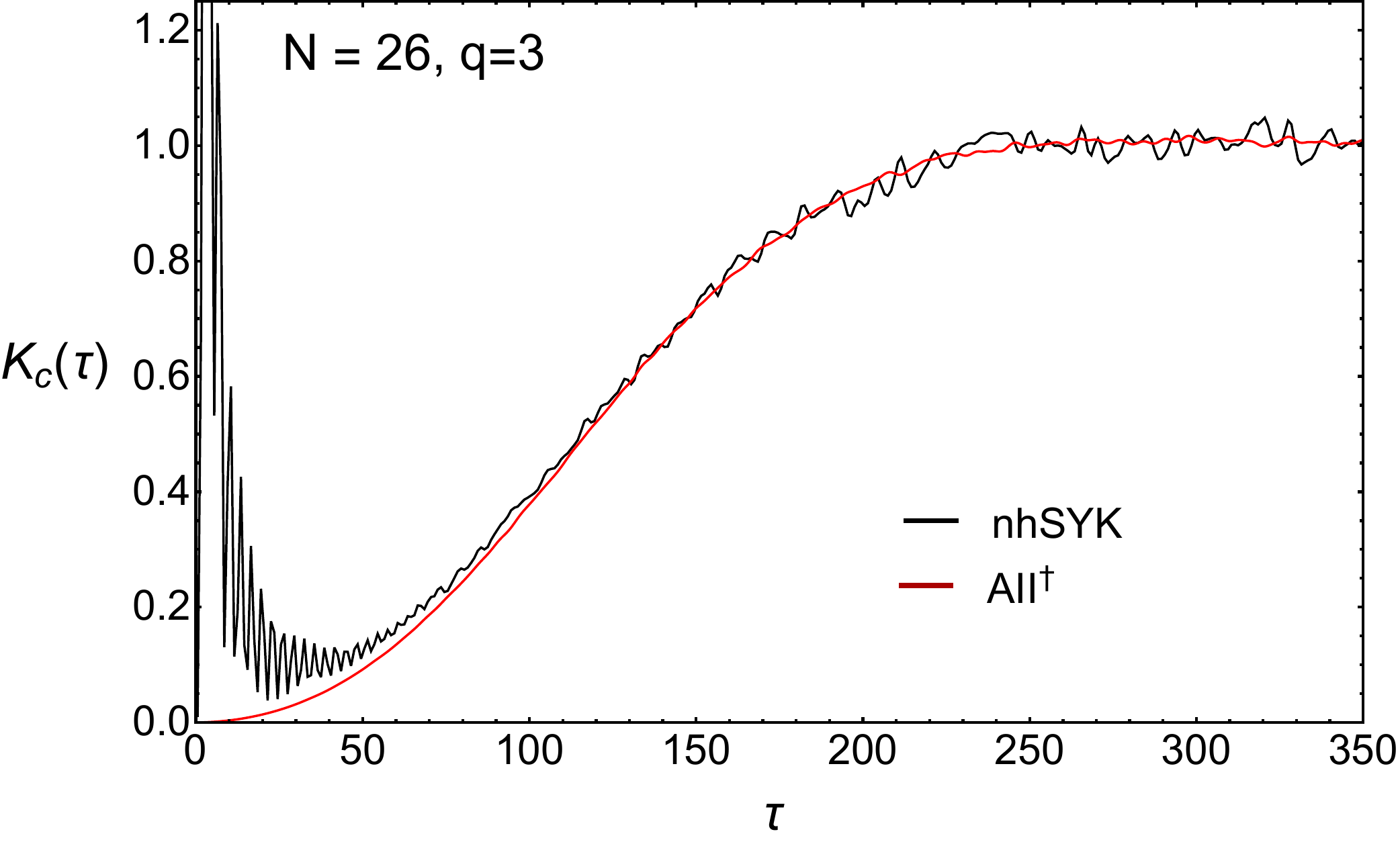}
		\includegraphics[width=8cm]{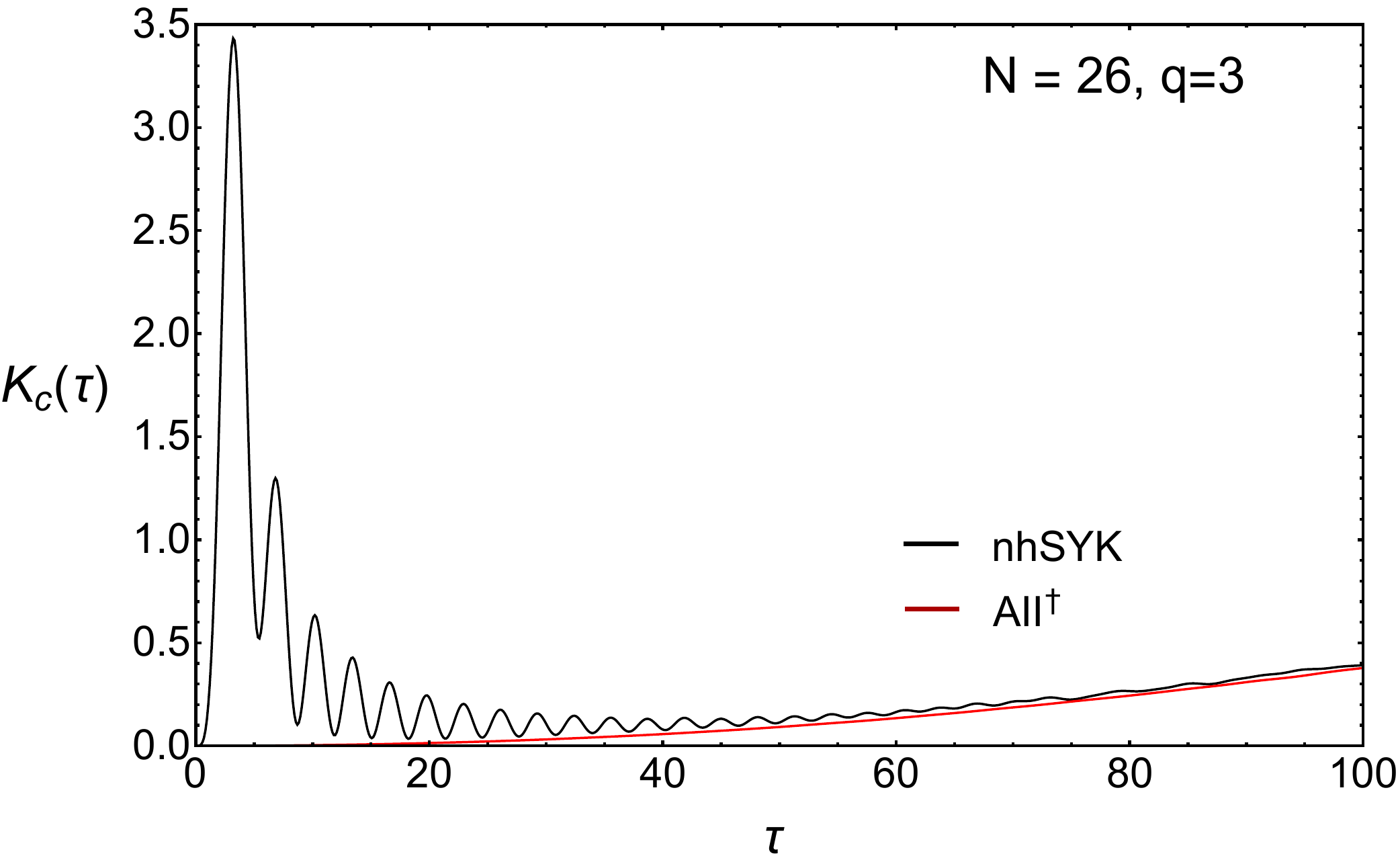}}
	\centerline{  \includegraphics[width=8cm]{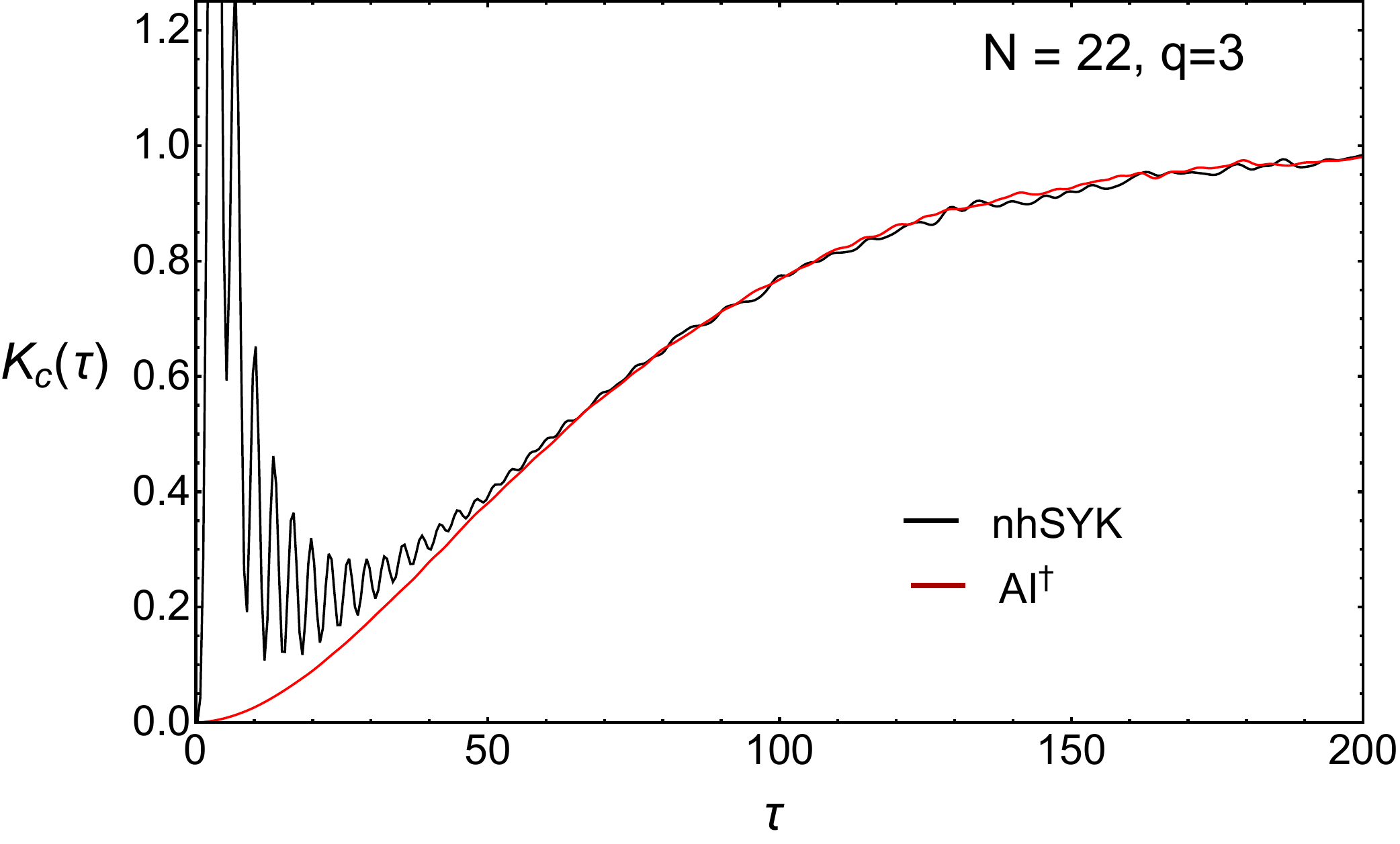}
		\includegraphics[width=8cm]{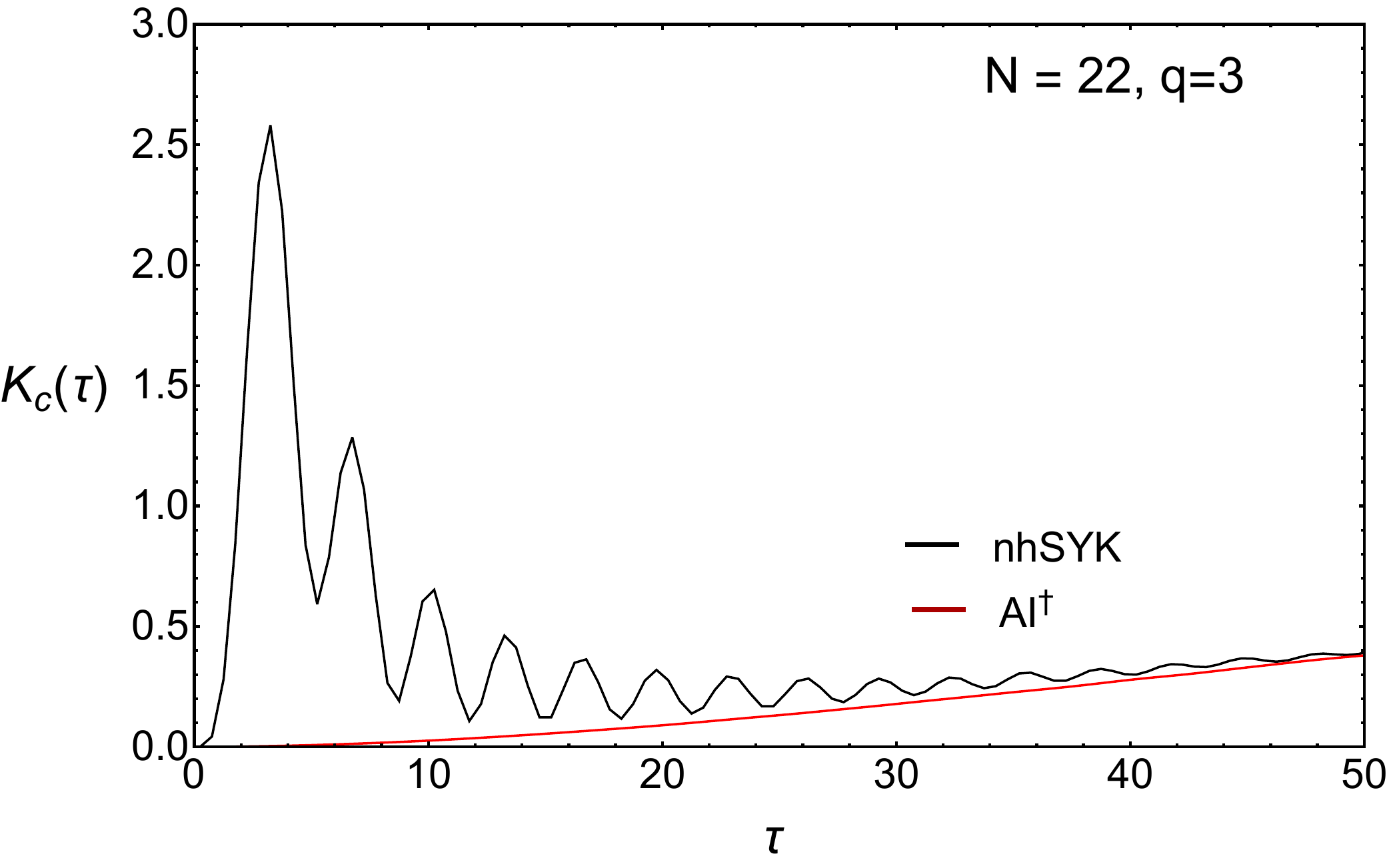}}
	\caption{The connected part of the spectral form factor of the real parts of the unfolded eigenvalues, Eq.~(\ref{form}), normalized by the number
		of eigenvalues $D\sim2^{N/2-1}$, of the $q = 3$ nHSYK model
		for $N=26$ (upper) and $N =22$ (lower). The results are compared
		to the AII$^\dagger$ ensemble for $N=26$ (red curves)
		and the AI$^\dagger$ ensemble for $N=22$ (red curves)
		scaled to the same number of eigenvalues. The
		right figures show the small-time behavior of the left figures in more detail. The spectral form factor
		of the nHSYK model differs from the spectral form factor of the AII$^\dagger$ ensemble only for $\tau \lesssim \sqrt D$. For $N = 26$ and $N = 22$, we employed $1279$ and $10000$ disorder realizations, respectively.
		\label{fig:kt26q3}
	}
\end{figure}

\begin{figure}[t!]
	\centerline{	\includegraphics[width=8cm]{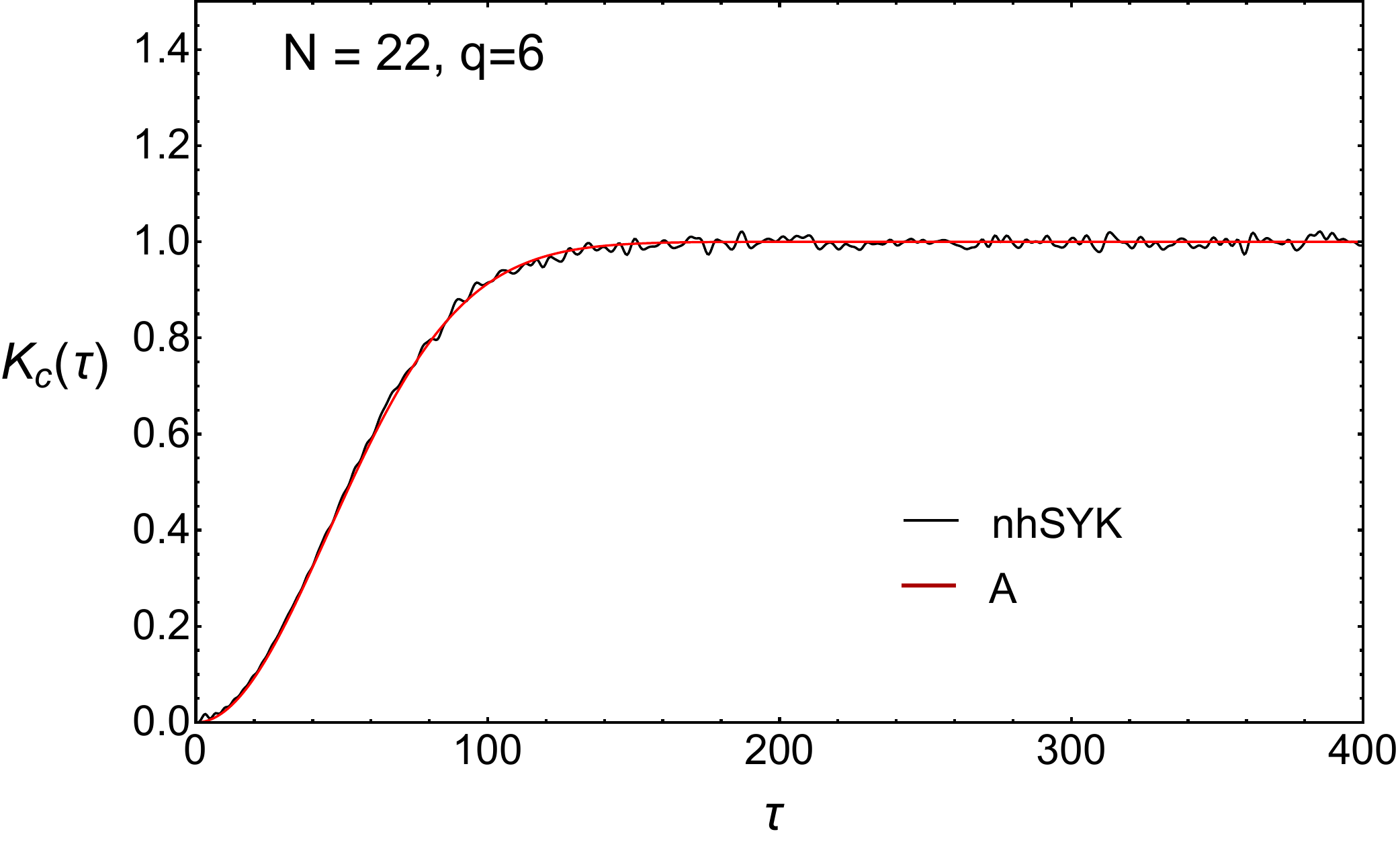}
		\includegraphics[width=8cm]{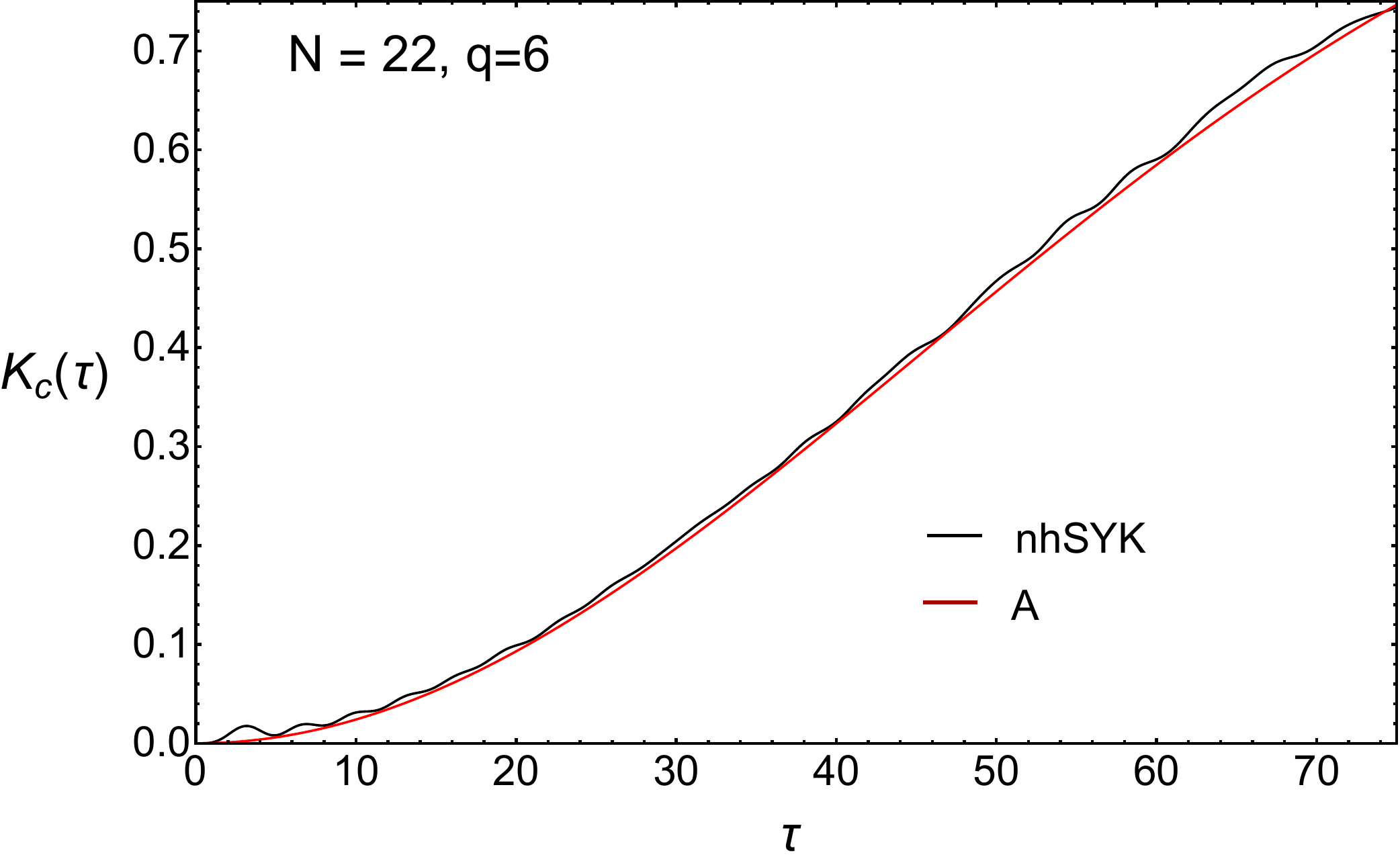}}
	\centerline{	\includegraphics[width=8cm]{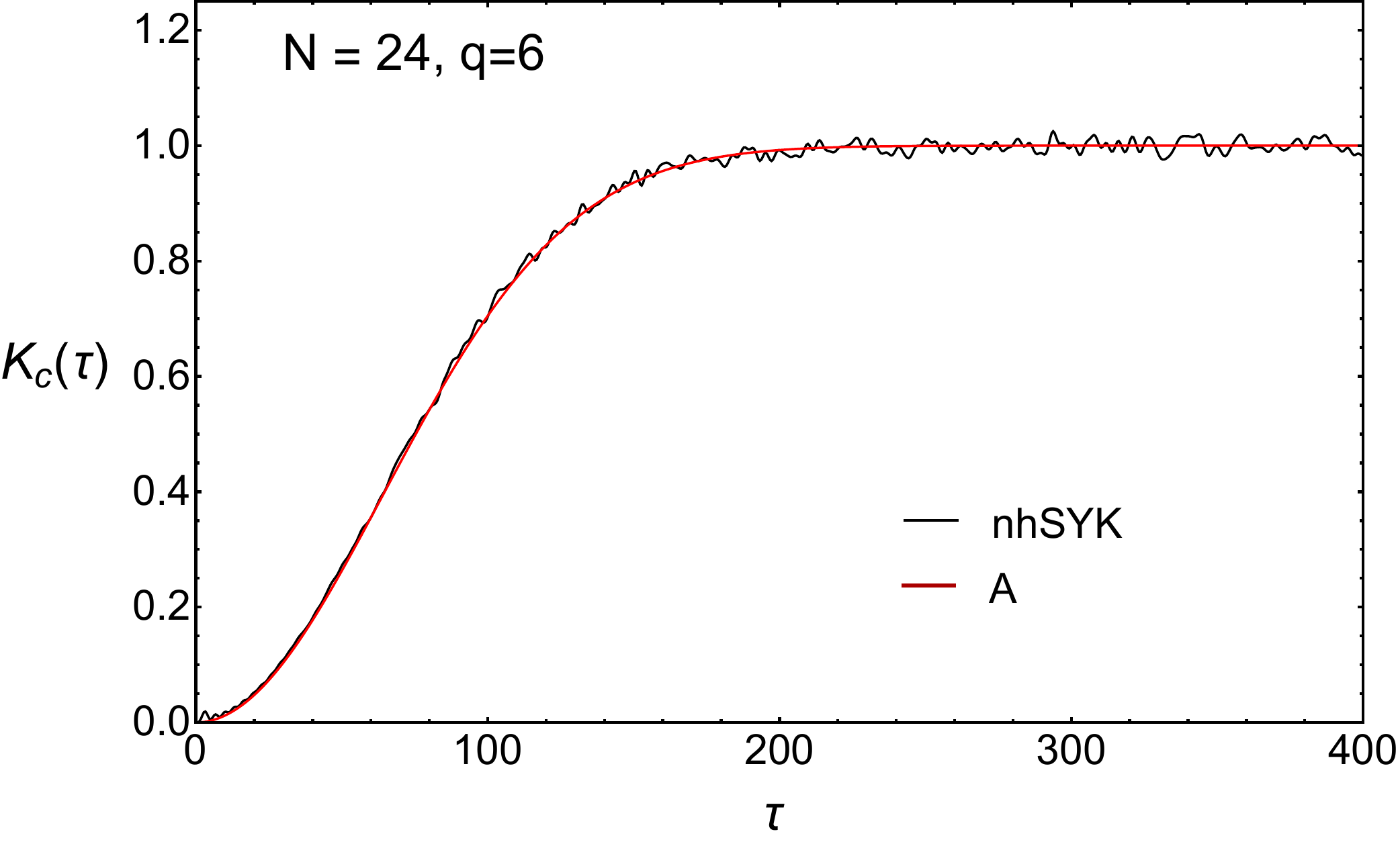}
		\includegraphics[width=8cm]{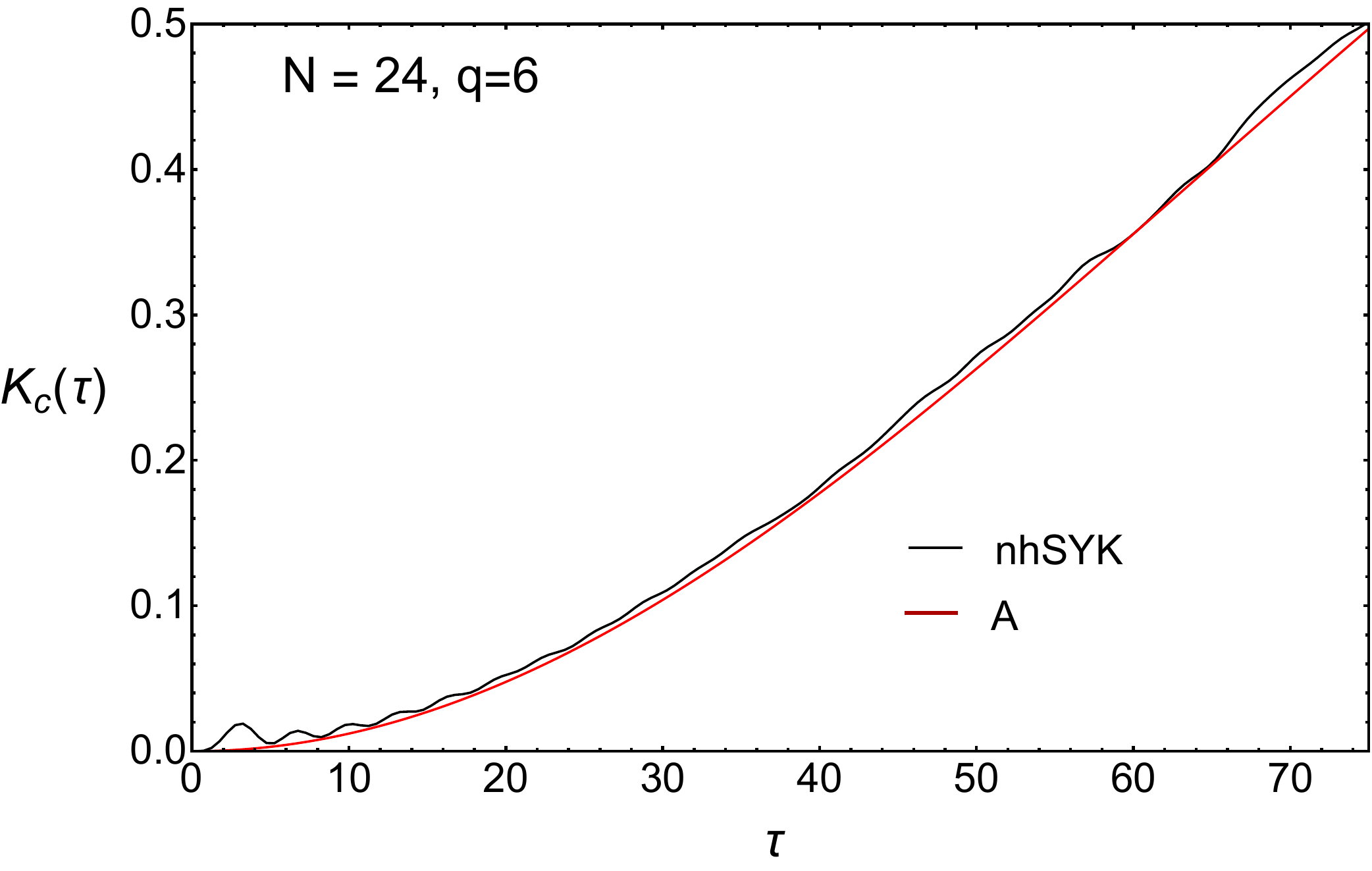}}
	\caption{The connected part of the spectral form factor of the real parts of the unfolded eigenvalues, Eq.~(\ref{form}), normalized by the number
		of eigenvalues $D\sim2^{N/2-1}$, of the $q = 6$ nHSYK model
		for $N=22$ (upper) and $N =24$ (lower).
		We find excellent quantitative agreement with the random matrix prediction for the GinUE.
		We note that this is a global observable and therefore the RMT prediction for class D is indeed identical to class A because the two only differ for eigenvalues around $E = 0$.
		A remarkable feature of the $q = 6$ results compared to smaller values of $q$ is that
		the correlation hole has almost disappeared.
	}
	\label{fig:dsffq3q6}
\end{figure}

Results for the $q = 3$ and $q=6$ nHSYK Hamiltonian, see Figs.~\ref{fig:kt26q3} and \ref{fig:dsffq3q6}, confirm
the picture obtained for $q=4$. Agreement with the random matrix predictions corresponding to the expected universality class is observed for $\tau > \sqrt D$.
The area below the small-time peak decreases markedly for increasing values of $q$. This is expected since a larger $q > 2$ brings the nHSYK Hamiltonian closer to a random matrix, as more entries of the Hamiltonian are nonzero. 
The area below the peak is proportional to $2^{N/2}/{N \choose q}$. For $N=24$, it is given by $14.84$, $2.02$, $0.39$, and $0.03$ for $q=2$, $3$, $4$, and $6$, respectively.

\begin{figure}[t!]
	\includegraphics[width=8cm]{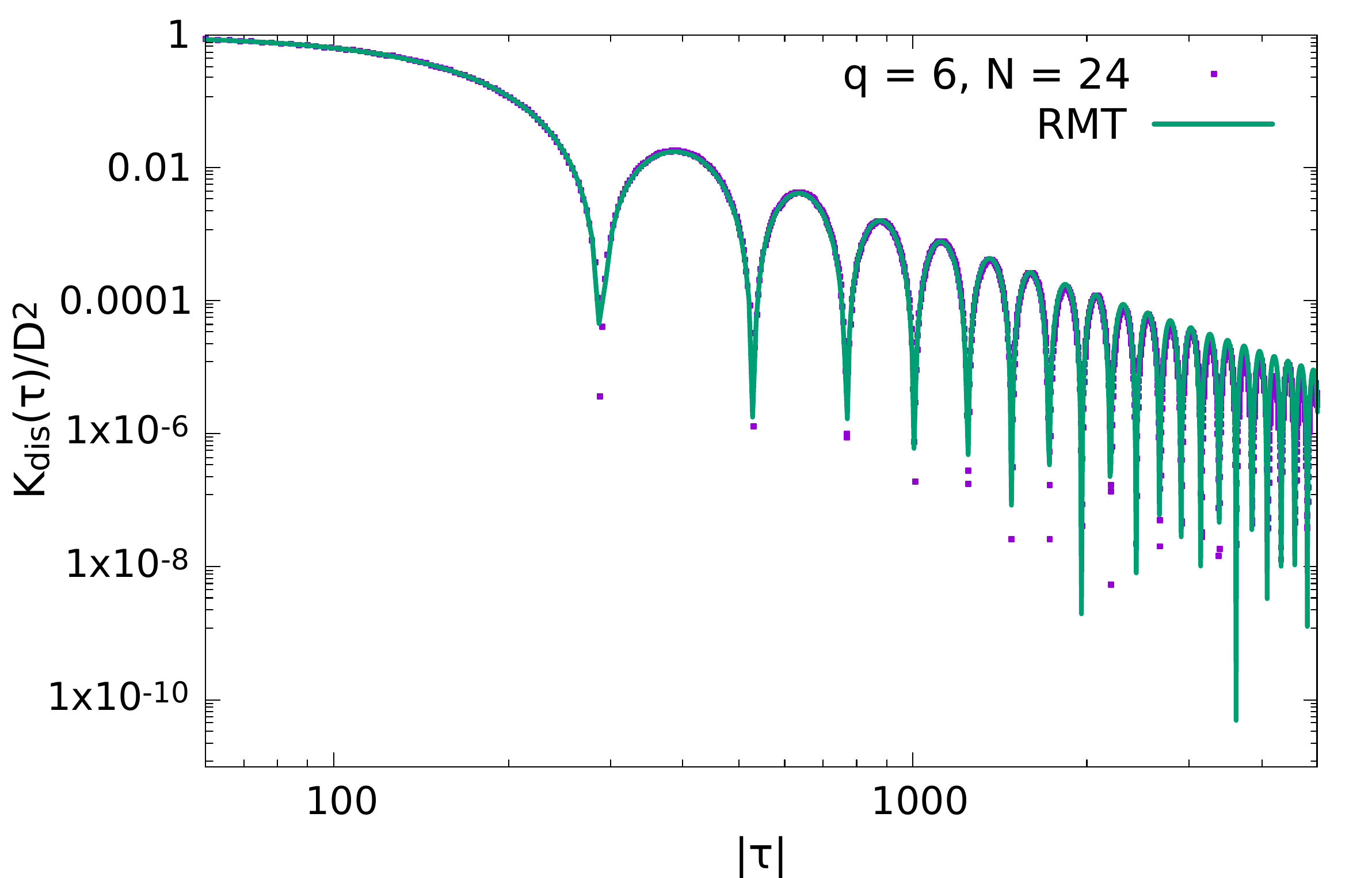}
	\includegraphics[width=8cm]{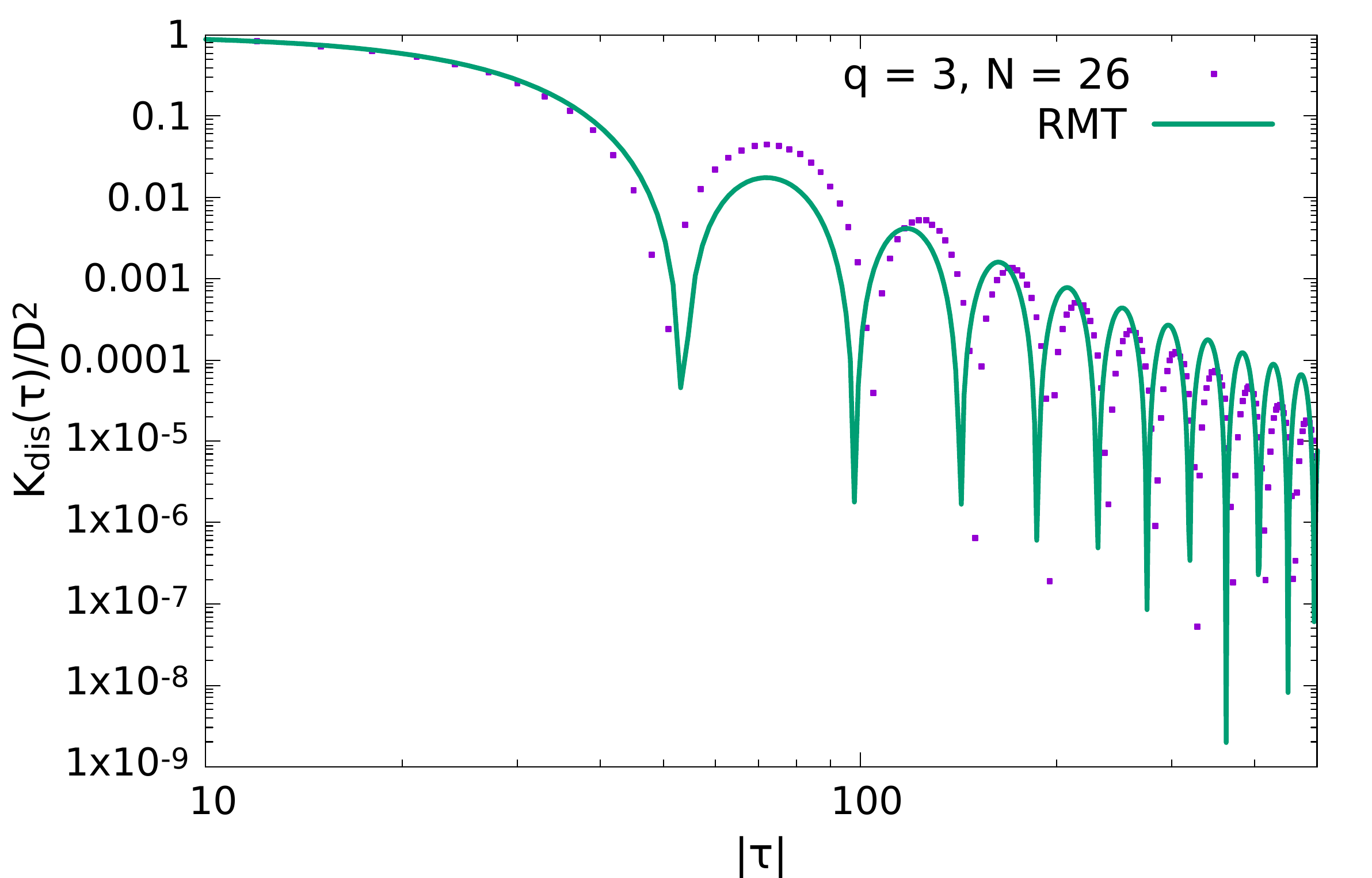}
	\caption{Disconnected part of the spectral form factor of the real parts of the eigenvalues (not unfolded), Eq.~(\ref{eq:dsff_dis}), normalized by the number of eigenvalues $D=2^{N/2-1}$ of the Hamiltonian. Solid curves are the random matrix prediction. Squares correspond to the $q=6$ (left) and $q=3$ (right) nHSYK model for different $N$. Although this is a nonuniversal observable related to the Fourier transform of the spectral density, for $q = 6$, we find excellent agreement with the random matrix prediction, $K_\mathrm{dis} (\tau) = 4D^2J_1(\tau)^2/\tau^2$ after an overall rescaling. However, for $q =3$, the agreement is only qualitative. This is not surprising, as the spectral
		density for $q = 3$ is not constant while for $q = 6$ it is already almost constant resulting in a semicircular spectral
		density of the real parts.
		We also expect that, for larger $N$, some deviations will be observed in the latter case because the spectral density is less uniform. 
	}
	\label{fig:dsffdis}
\end{figure}

The oscillatory behavior in the small-$\tau$ region for $q=3$, although not qualitatively different from $q=4$, has a much larger amplitude than in the $q=4$ case (see the right panel of
Fig.~\ref{fig:kt26q3}). This results in
a correlation hole that is shifted to a larger value of $\tau$.
On the other hand, for $q=6$, the amplitude of the oscillations is very small, and we barely observe
any deviation from the random matrix predictions.
For the SYK model with real couplings, it can be shown~\cite{erdos2014} that for $q \gg \sqrt{N}$ the SYK Hamiltonian resembles a random matrix with a semicircular spectral density.
For the nHSYK model, this corresponds to a constant level density inside the eigenvalue disk
so that the real parts of the eigenvalues are distributed according to a semicircle.
For $q = 6$, we are likely in this asymptotic region. Indeed, this is confirmed by
a comparison of the disconnected part of the spectral form factor, see Fig.~\ref{fig:dsffdis}, with the random matrix prediction $K_{\mathrm{dis}}(\tau) = 4D^2J_1^2(\tau)/\tau^2$, the square of the
Fourier transform of the semicircle law, where $J_1$ is a Bessel function. They are almost indistinguishable which explains why, for $q = 6$, the spectral density is very close
to that of the Ginibre ensemble. This also suggests that the spectral correlations
are very close to that of the Ginibre ensemble. 
In contrast, for $q = 3$, we observe larger deviations with respect to the semicircle law in the disconnected part. This is consistent with the fact that, by reducing $q$, the Hamiltonian is much sparser and, therefore, deviations from the RMT predictions should be more visible. Another issue is that there is a systematic difference between even $q$ and odd
$q$ related to cancellations that occur in the calculation of moments of
eigenvalues of the supercharge
\cite{Jia:2018ccl} which we expect to persist in the non-Hermitian case.

\subsubsection{Integrable behavior for $q=2$}

\begin{figure}[t!]
	\centerline{\includegraphics[width=8cm]{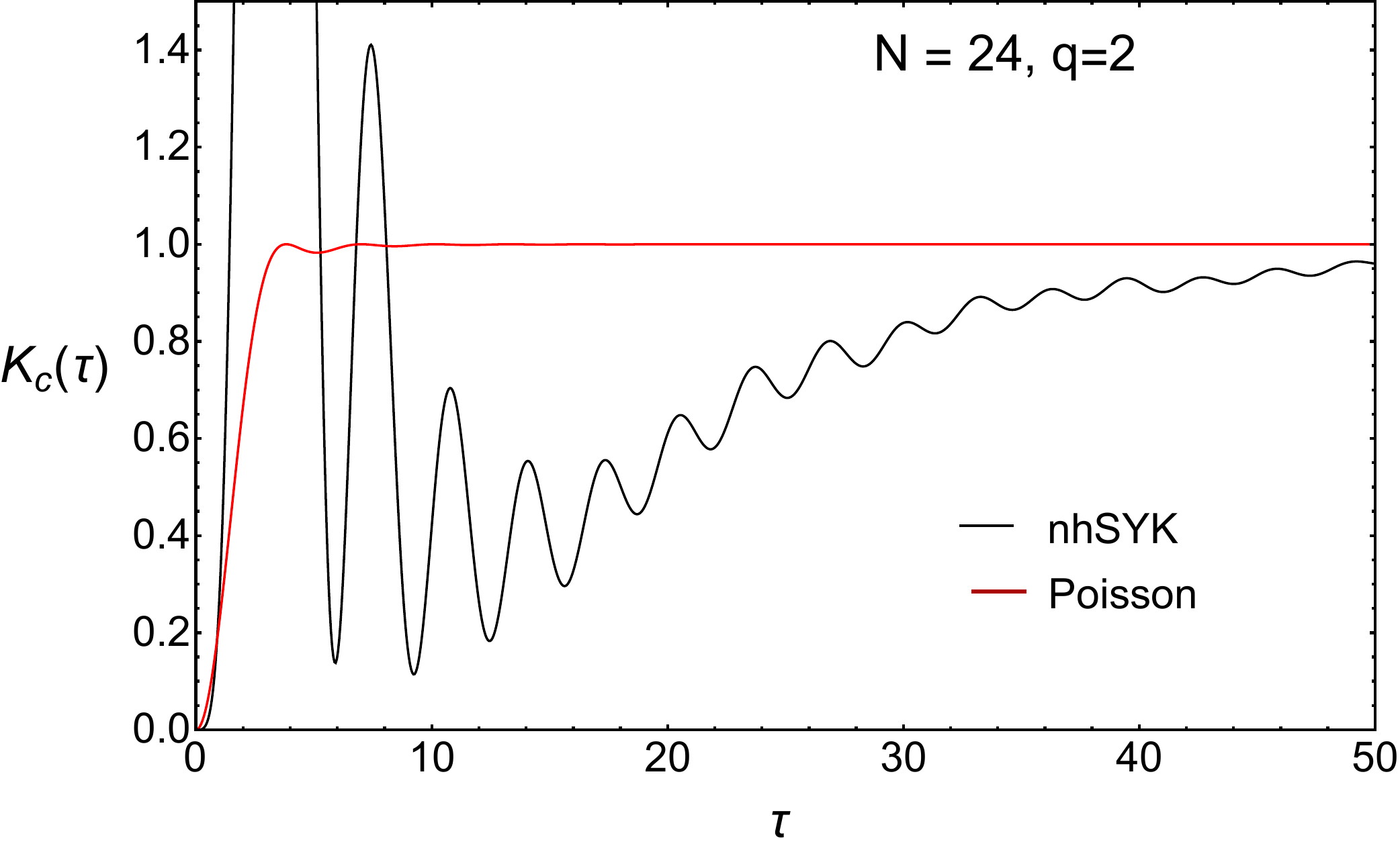}
		\includegraphics[width=8cm]{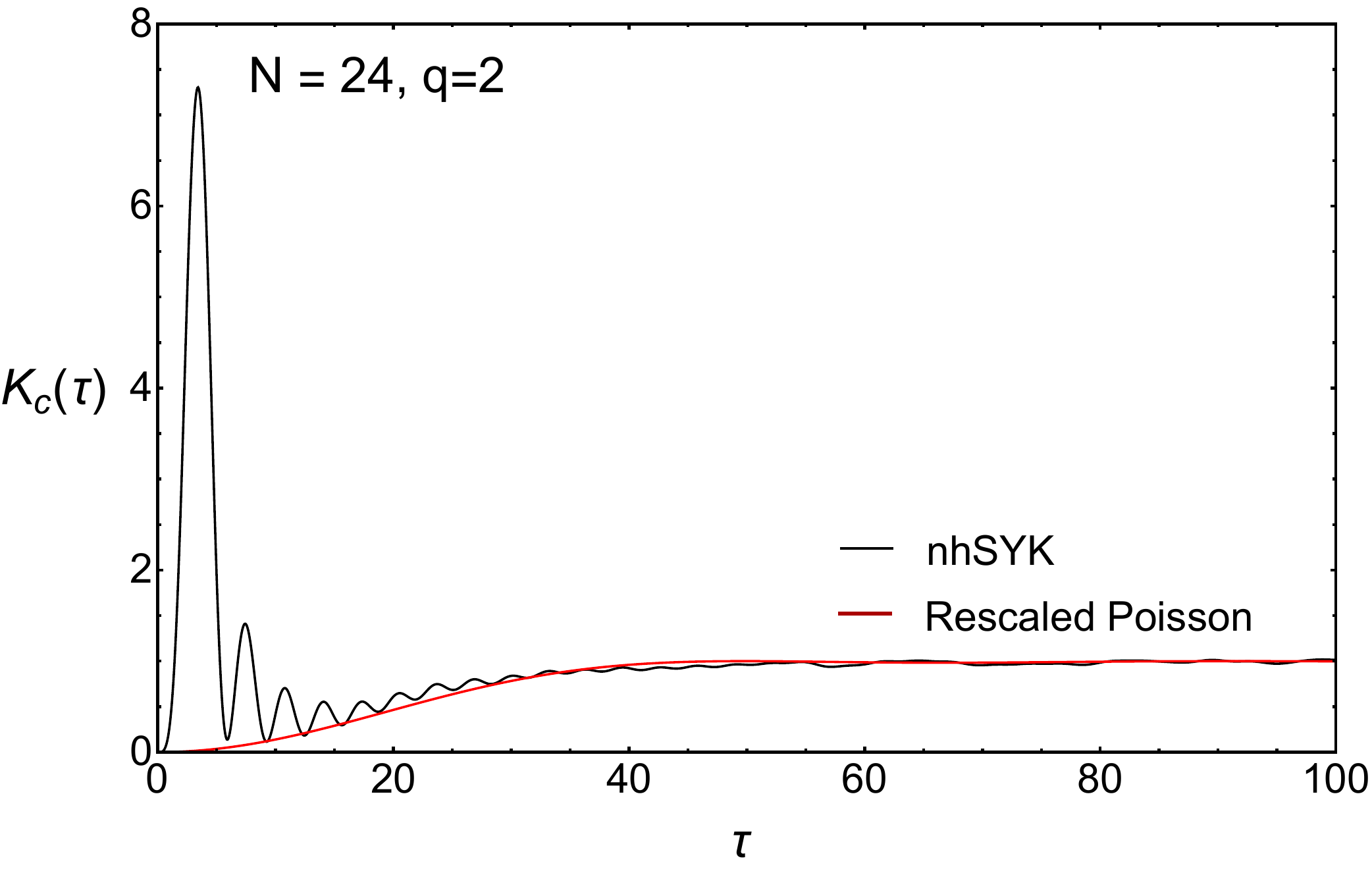}}
	\caption{The connected part of the spectral form factor of the real parts of the unfolded eigenvalues, Eq.~(\ref{form}), normalized by the number
		of eigenvalues $D\sim2^{N/2-1}$, of the $q = 2$ nHSYK model
		for $N=24$.
		The solid red curves show the result for Poisson statistics, Eq.~(\ref{eq:Poisson}), in the left panel, and the same expression with the horizontal axis rescaled by a factor of $\log D$ in the right panel. As was expected, the $q = 2$ SYK model shows correlations that are in between Poisson statistics and RMT statistics. The reason is that the integrable many-body spectrum is determined by a small number ($N/2$) of chaotic single-particle energies.
		The oscillations observed for small, but nonzero, $\tau$
		are due to collective fluctuations of the spectral density.
	}
	\label{fig:dsffq2}
\end{figure}

The $q=2$ SYK and nHSYK models are both integrable with all energy levels determined
by $N/2$ single-particle energies. In this case, we expect Poisson level statistics for sufficiently long times, but deviations from Poisson
statistics may be observed for shorter times.
Indeed,
as illustrated in 
Fig.~\ref{fig:dsffq2}, the spectral form factor
saturates to the Poisson limit, $K_c(\tau)= 1$, at a scale of order $\log D$,
which is 
much shorter than for $q>2$,
where the scale is determined by $\sqrt D$ (for $N=24$ the two scales are of
the same order of magnitude and our data cannot really distinguish between
the two). 
The analytical result for uncorrelated eigenvalues unfolded to
constant density inside the complex unit disk, given by Eq.~(\ref{eq:Poisson}),
saturates to Poisson statistics at $\tau = O(1)$ and
does not match the numerical result (see the solid red curve in Fig.~\ref{fig:dsffq2}, left).
A reasonable fit is obtained by replacing
$ \tau \to \tau / \log D $ (solid red curve in Fig.~\ref{fig:dsffq2}, right), but we have no rigorous argument for this substitution.

Physically, the saturation scale of the $q=2$ spectral form factor
is related to the fact that the model can be mapped onto free fermions with single-particle energies correlated according to RMT~\cite{cotler2016}.
The short-time dynamics,
controlled by the single-particle excitations, will be very different from that expected for a generic integrable system. However, for longer times
of the order $\log D$, multiparticle excitations
will reveal the generic integrable nature of the quantum dynamics.

As is the case for $q=3$ and $q=4$, we find oscillations for
small values of $\tau $ with the same period but with a larger amplitude.
These oscillations, which dominate the quadratic $\tau$-dependence,
are due to scale fluctuations of the average spectral density.
  
\subsection{Number variance of the nHSYK model}

We now turn to the analysis of the number variance of the real parts of the eigenvalues of the nHSYK model. We only
compute this observable for $q=2$ ($N=22$ and $N=26$), $q=3$ ($N=26$), and $q=4$ ($N=22$ and $N=26$), which
are in the Poisson, GinUE, and AII$^\dagger$ bulk universality classes, respectively (the $q=3$, $N=26$ case also has chiral symmetry and the full symmetry class is AII$^\dagger_-$).
For the GinUE, we can compare with the analytical 
expression \eref{nv-ann}, while in the case of class AII$^\dagger$ we have to rely on a numerical calculation of the spectrum of the corresponding random matrix ensemble. We assume that the number variance of AII$^\dagger$
still has the scaling behavior~(\ref{eq:univSigma}) obtained analytically
for the Ginibre ensemble,
\be
\Sigma^2(n) =\frac{\overline{ \rho}(E)}{\sqrt{D}}\, f\! \left(\frac{n \sqrt{D}}{\overline{ \rho}(E)}\right),
\ee
where $\bar \rho(E) $ is the eigenvalue density on the real axis, for some universal function $f$.

\subsubsection{$q=4$ and nonstationarity}

\begin{figure}[t!]
	\centerline{\includegraphics[width=8cm]{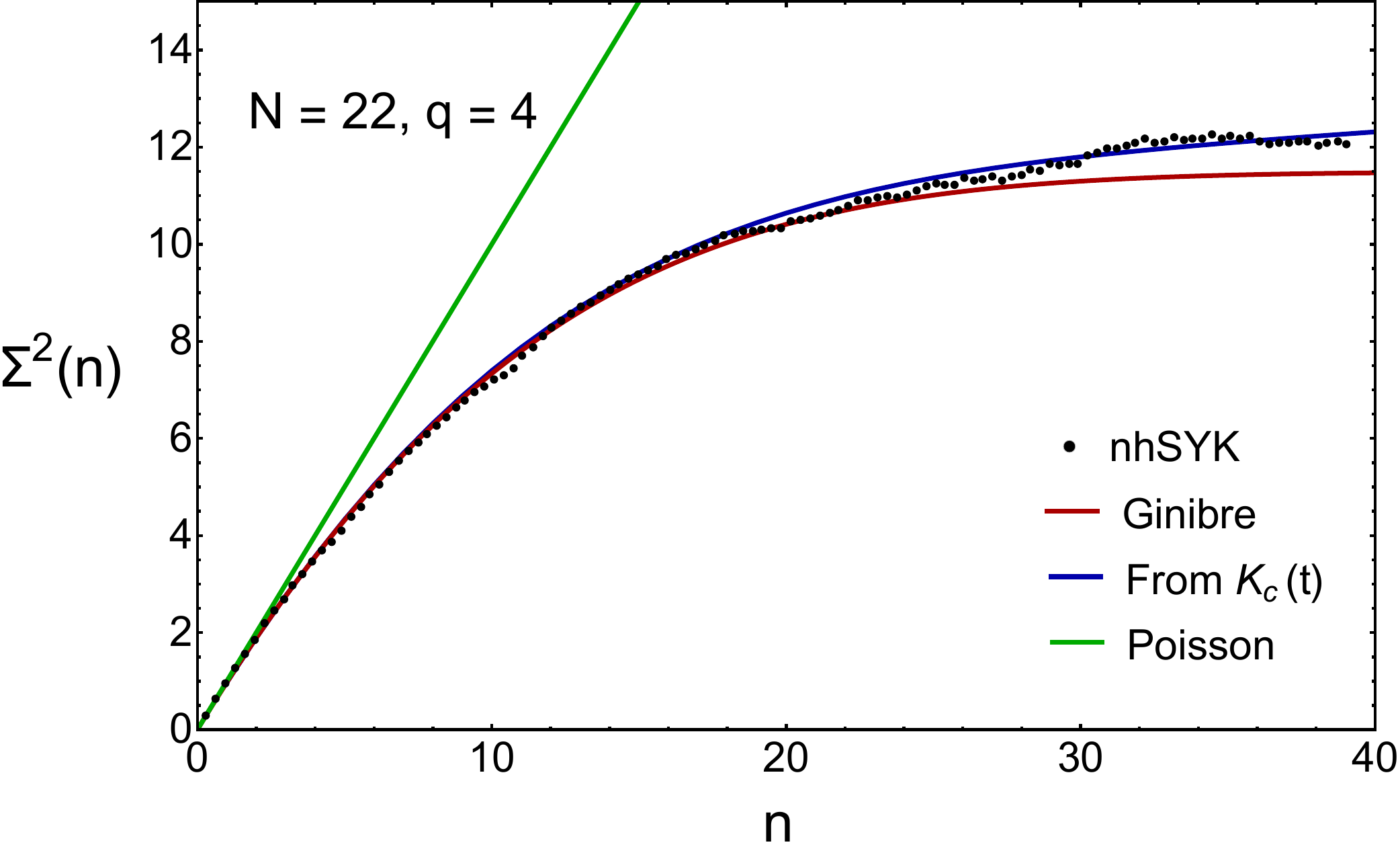}
		\includegraphics[width=8cm]{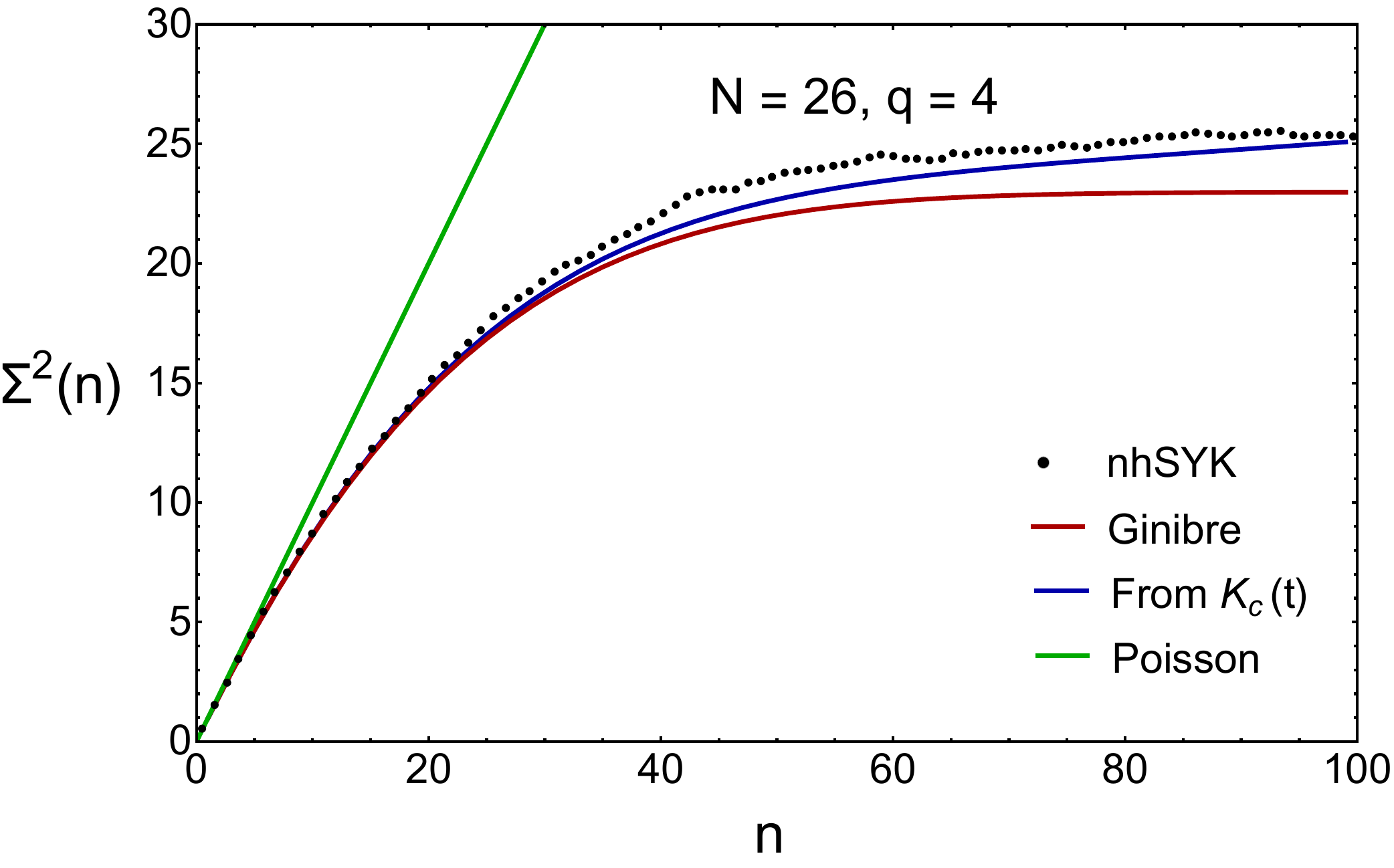}}
	\caption{The number variance of the real parts of the eigenvalues. The black dots are the result for the nHSYK model for $q =4$ and $N=22$ (left) and $N = 26$ (right). We find a good agreement with both the analytical result for the Ginibre ensemble [Eq.~(\ref{nv-ann}), red curve] and the result obtained by integrating the numerical spectral form factor for the nHSYK model [Eq.~(\ref{sigform}), blue curve].}
	\label{fig:nvq4}
\end{figure}

We first discuss the $q=4$ case.
In Fig.~\ref {fig:nvq4}, we plot the number variance of the real parts, $\Sigma^2(n)$,
versus the average number of levels, $n$, in an interval that is chosen to be symmetric around
zero. No unfolding is necessary this way---if we would have unfolded the real part of
the eigenvalues (times a phase factor) we would have obtained the same result. In order to suppress statistical fluctuations, we also average $\{ e^{i\theta} |z_k|\}$ over ten values of $\theta$ as
we did for the calculation of the spectral form factor.
The results are compared to the analytical result~\eref{nv-ann} for the Ginibre ensemble
(red curves) and the result obtained by integrating the
numerical spectral form factor for $\tau <\sqrt D$ using Eq.~\eref{sigform}
(blue curves) instead of using the analytical result of the spectral form factor all the
way to $\tau =0$.
For $N=22$, this correction explains the difference between
the number variance for the nHSYK model and the Ginibre ensemble, but for $N=26$
a discrepancy remains. One issue, as was discussed above, is that by integrating
the spectral form factor, we obtain the spectral average of the number variance, while in
Fig.~\ref{fig:nvq4} we show the number variance for intervals centered about zero energy.
The conclusion is that the number variance is not stationary, but this is also
the case for $N=22$ and the Ginibre ensemble. Apparently, the nonstationarity
of the real parts of the eigenvalues
for
$N=26$ is different from the nonstationarity for the Ginibre ensemble.

\begin{figure}[t!]
	\centerline {\includegraphics[width=8cm]{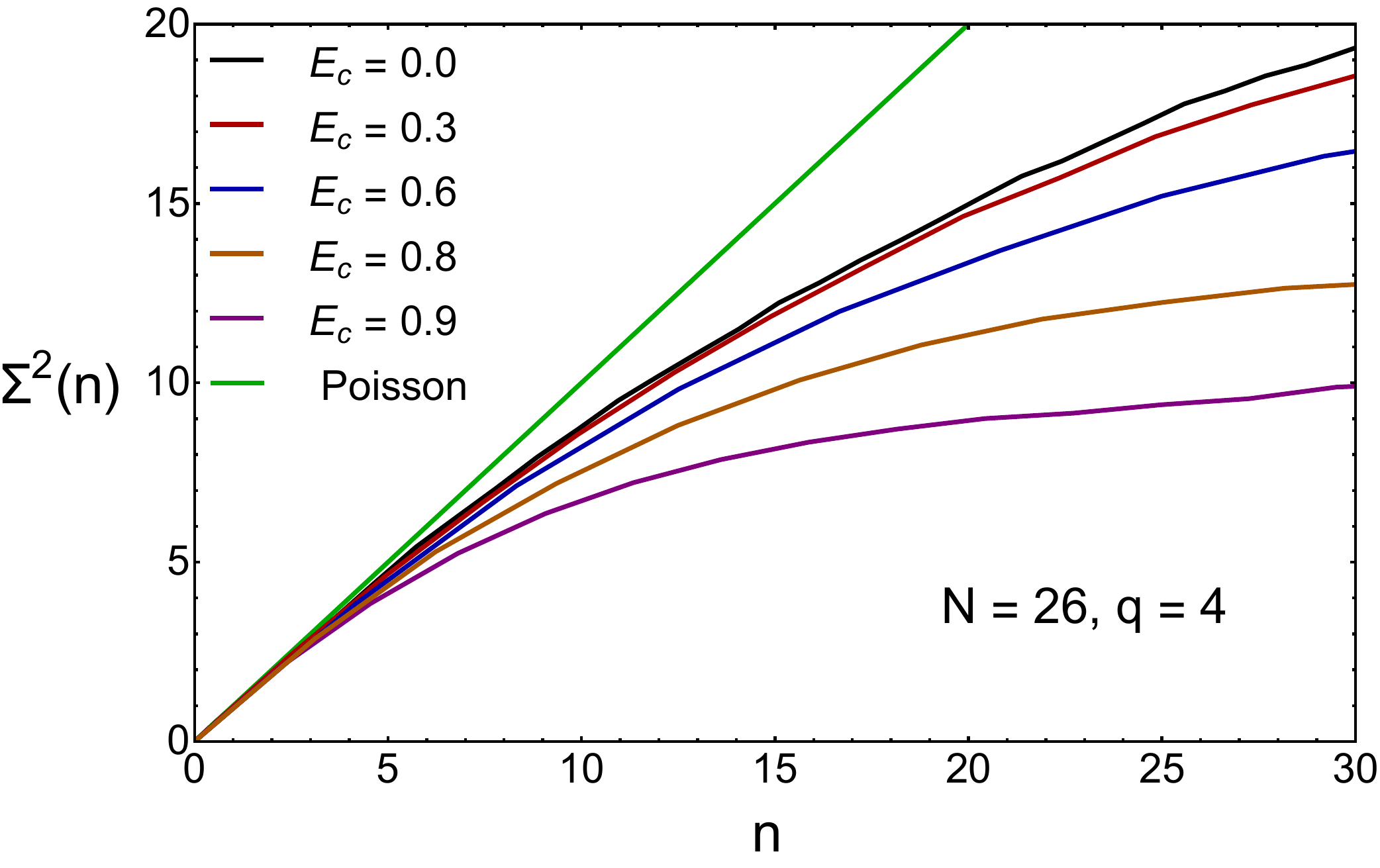}
		\includegraphics[width=8cm]{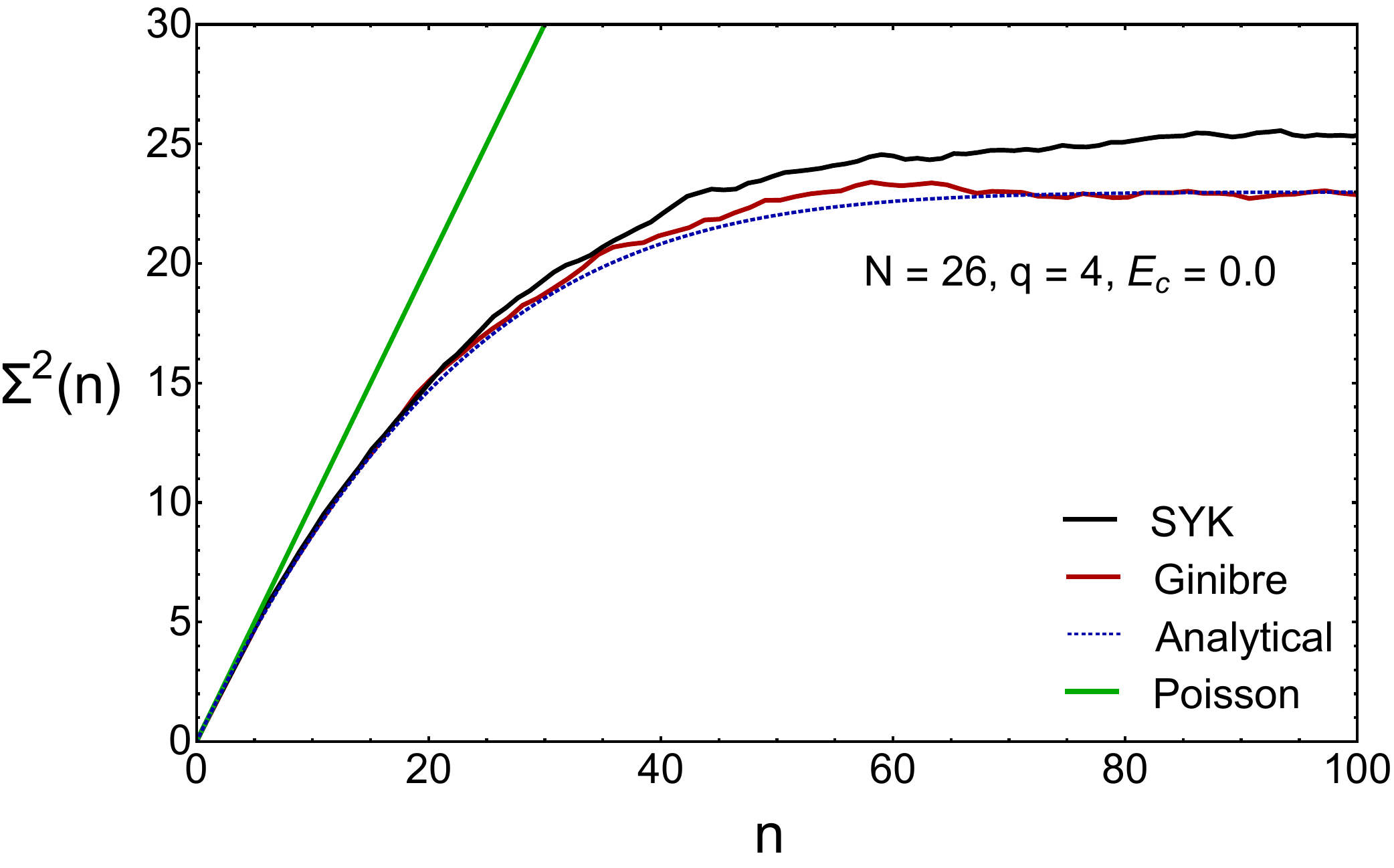}}
	\centerline {\includegraphics[width=8cm]{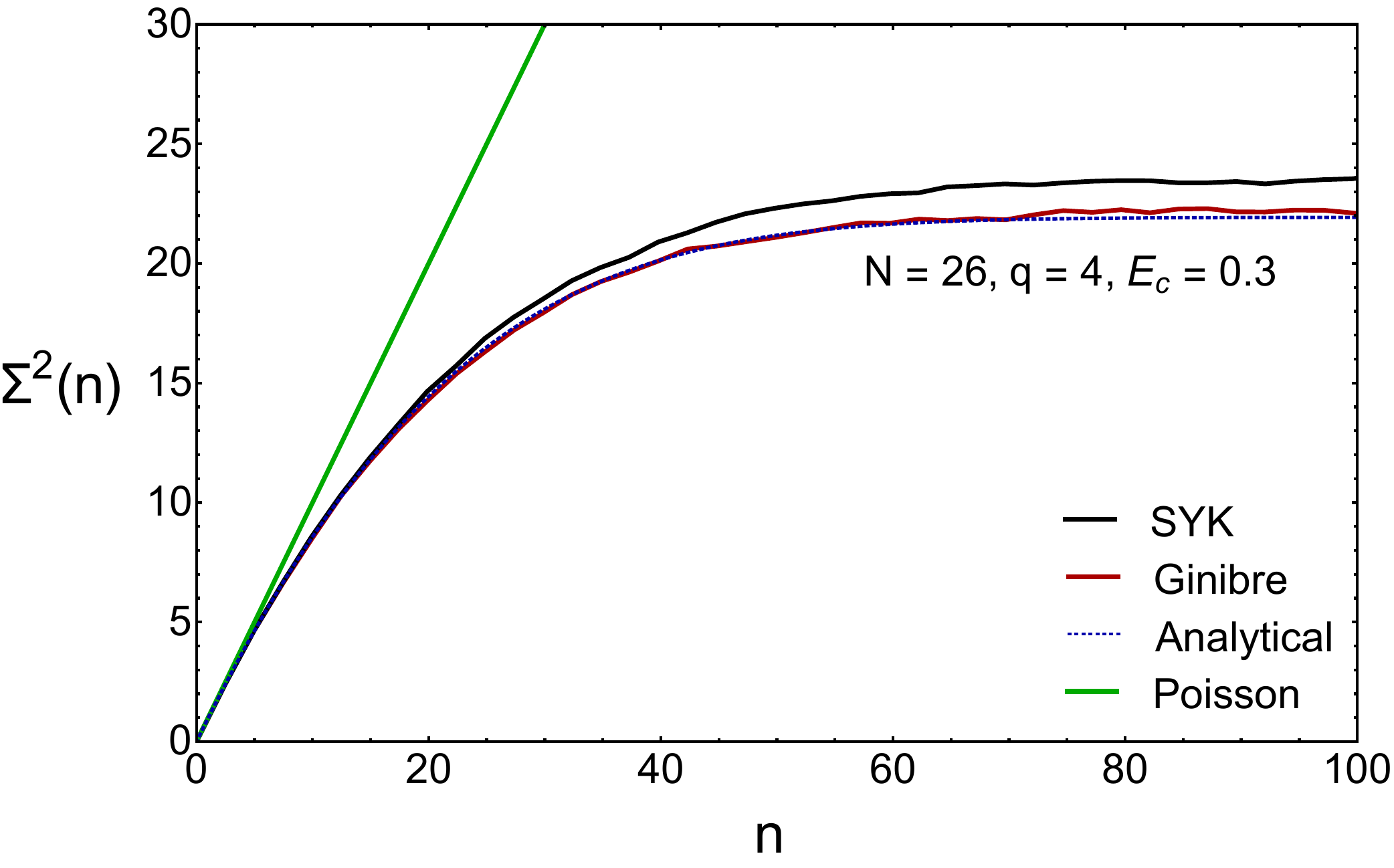}
		\includegraphics[width=8cm]{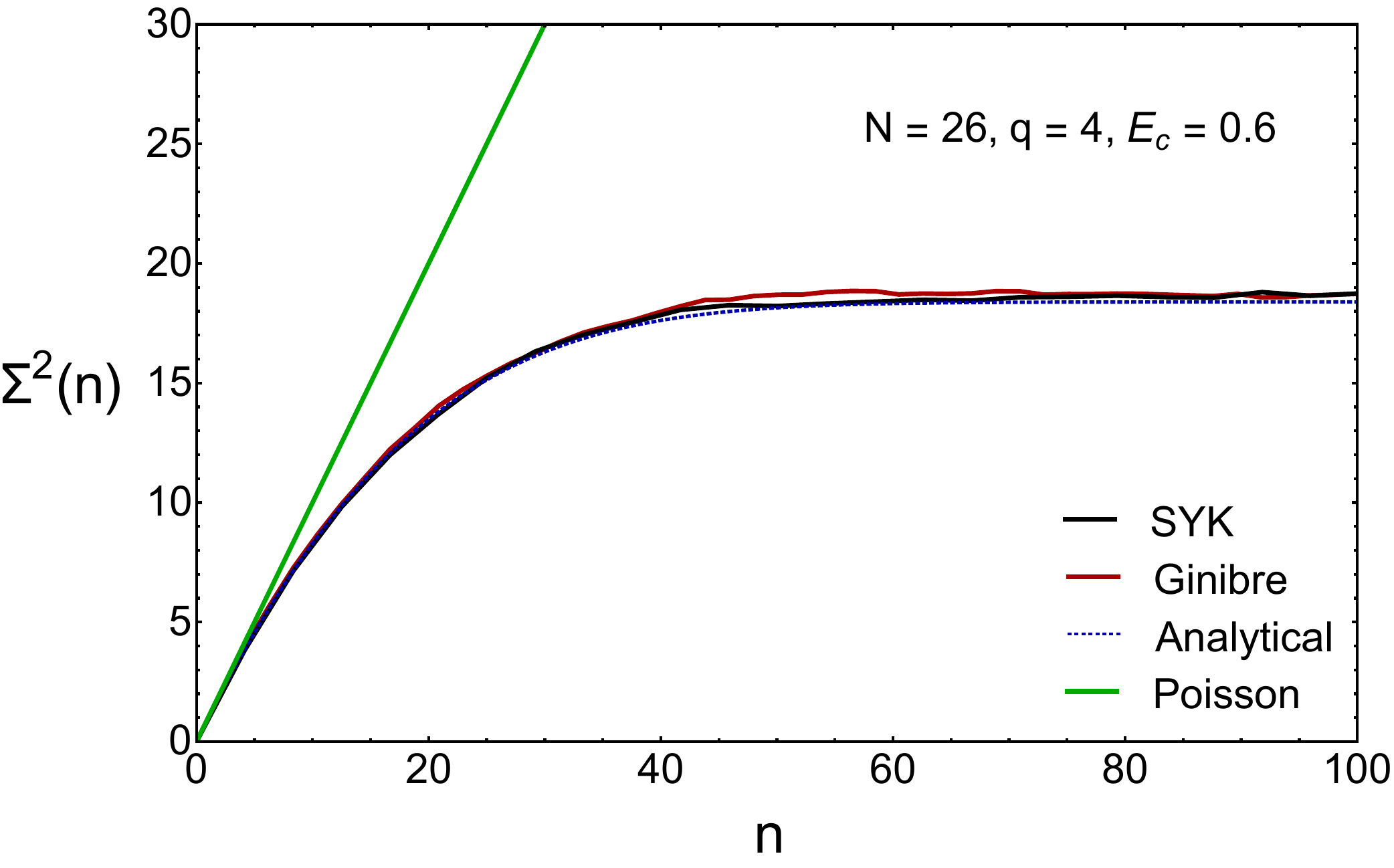}}
	\centerline {\includegraphics[width=8cm]{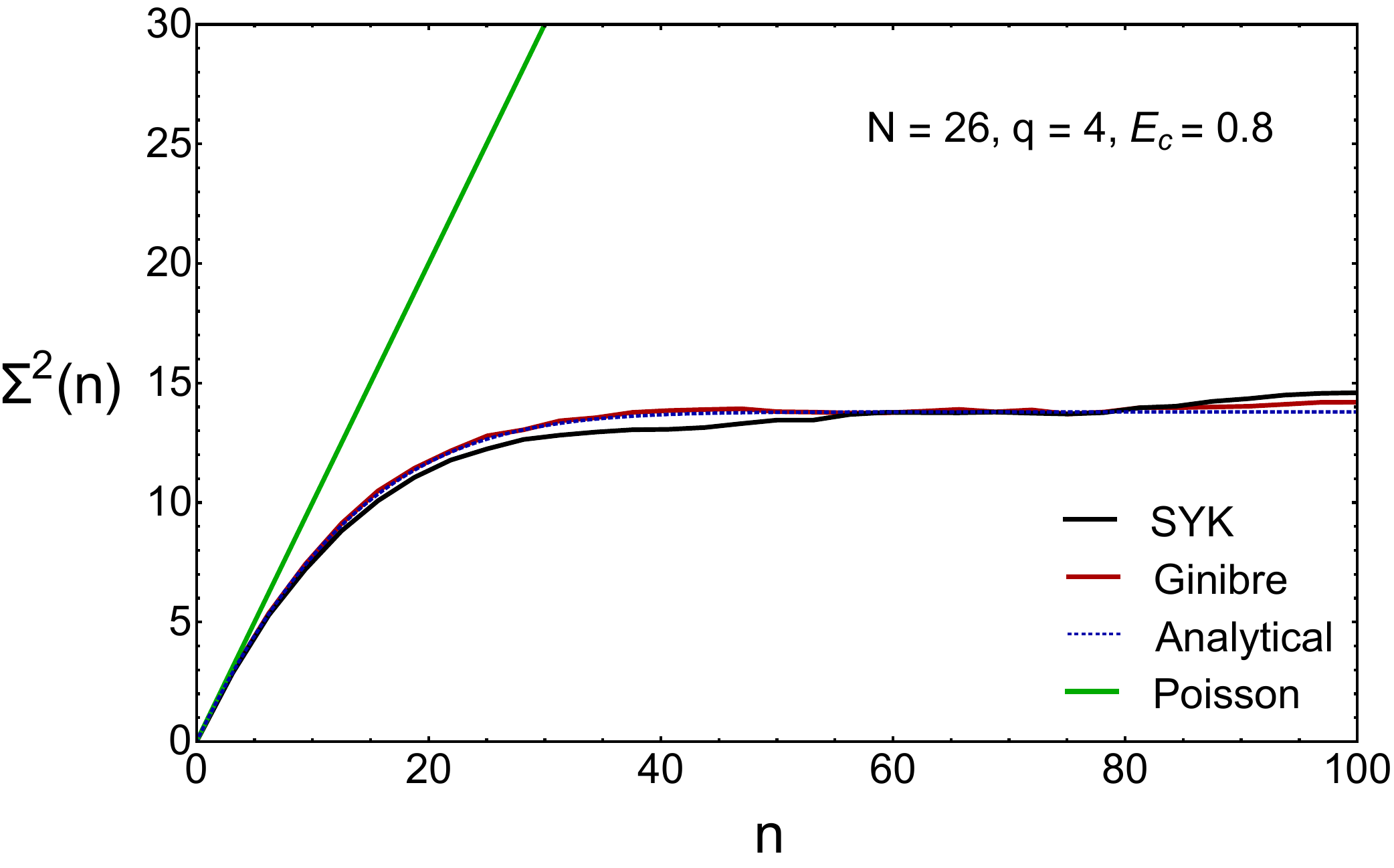}
		\includegraphics[width=8cm]{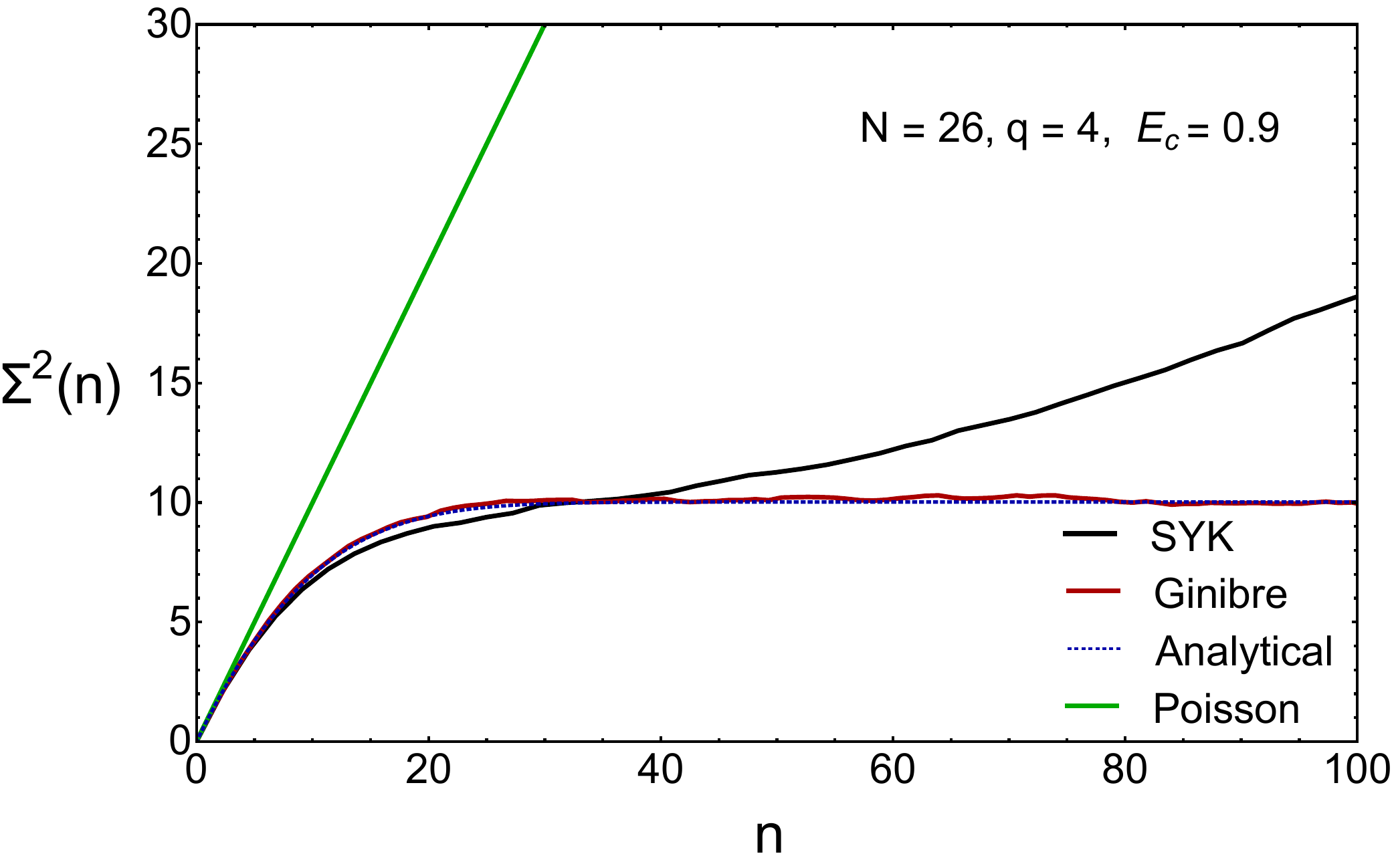}}
	\caption{The number variance of the real parts of the eigenvalues for the $q = 4$ and $N = 26$ nHSYK model. The upper left panel shows the number variance for intervals centered at different points of the spectrum: $E_c =0, \; 0.3, \; 0.6,\; 0.8$, and $ 0.9$. In the other panels, we compare the number variance for each of these values of $E_c$ (SYK) with Poisson, the numerical random matrix (Ginibre) result, and the analytical result. The latter is obtained by replacing $
		D \to D (1-E^2_c)$
		in Eq.~\eref{nv-ann}.  }
	\label{fig:stationarity}
\end{figure}

The nonstationarity at $N=26$ is illustrated in Fig.~\ref{fig:stationarity}, where
we show the number variance of the real parts of the eigenvalues
for an ensemble of 3000 $N=26, q=4$ nHSYK matrices
for intervals
centered at $E_c=0$, $0.3$, $0.6$, $0.8$, and $0.9$. In the upper left panel,
we show all nHSYK curves in one figure. This shows that the number variance
loses its stationarity as soon as it starts deviating from Poisson 
at a few level spacings. In the other panels, we
compare each of the curves with the numerically obtained
random matrix result for a Ginibre ensemble of 2048 $4096\times 4096$
matrices. For up to about $10$--$20$ level spacings, the Ginibre and nHSYK results are in agreement, but they start deviating for larger values
of $n$, in particular, if $E_c$ is close to the edge. 

Remarkably, for
$E_c = 0.6$ we find agreement between the nHSYK model and the Ginibre ensemble
for the entire range of $n$ we looked at. Since the number variance of the
nHSYK model ensemble overshoots the Ginibre number variance for $E_c>0.6 $ and undershoots it for $E_c <0.6$, its spectral average 
will agree with the random matrix prediction up to a larger value of $n$, i.e., to
$n \approx 50$.
This is the spectral average 
related to the spectral form factor of the real parts of the eigenvalues. We thus expect it will agree with RMT for $\tau > 4096/100 \approx 40$, which is
consistent with Fig.~\ref{fig:ktdel}. This calculation also shows that the
good agreement we find for the spectral form factor for $q = 4$ is due to the fact that the stationarity behavior of the nHSYK model and the Ginibre ensemble are similar.
We find that the analytical number variance away from the center of the spectrum resulting from the replacement $	D \to D (1-E^2_c)	$ in Eq.~\eref{nv-ann} is in excellent agreement with the numerical results.

\begin{figure}[t!]
	\centerline {\includegraphics[width=8cm]{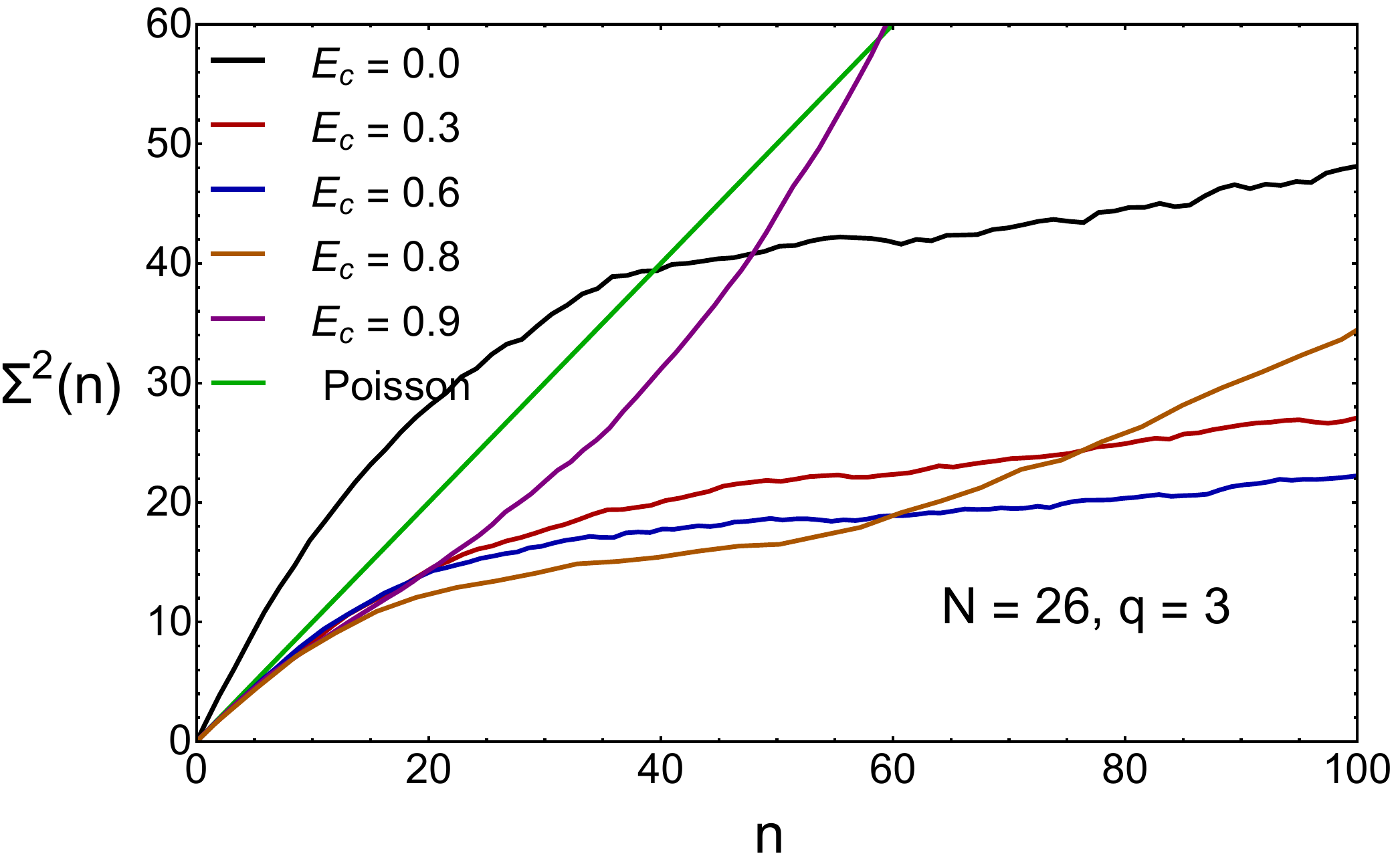}
		\includegraphics[width=8cm]{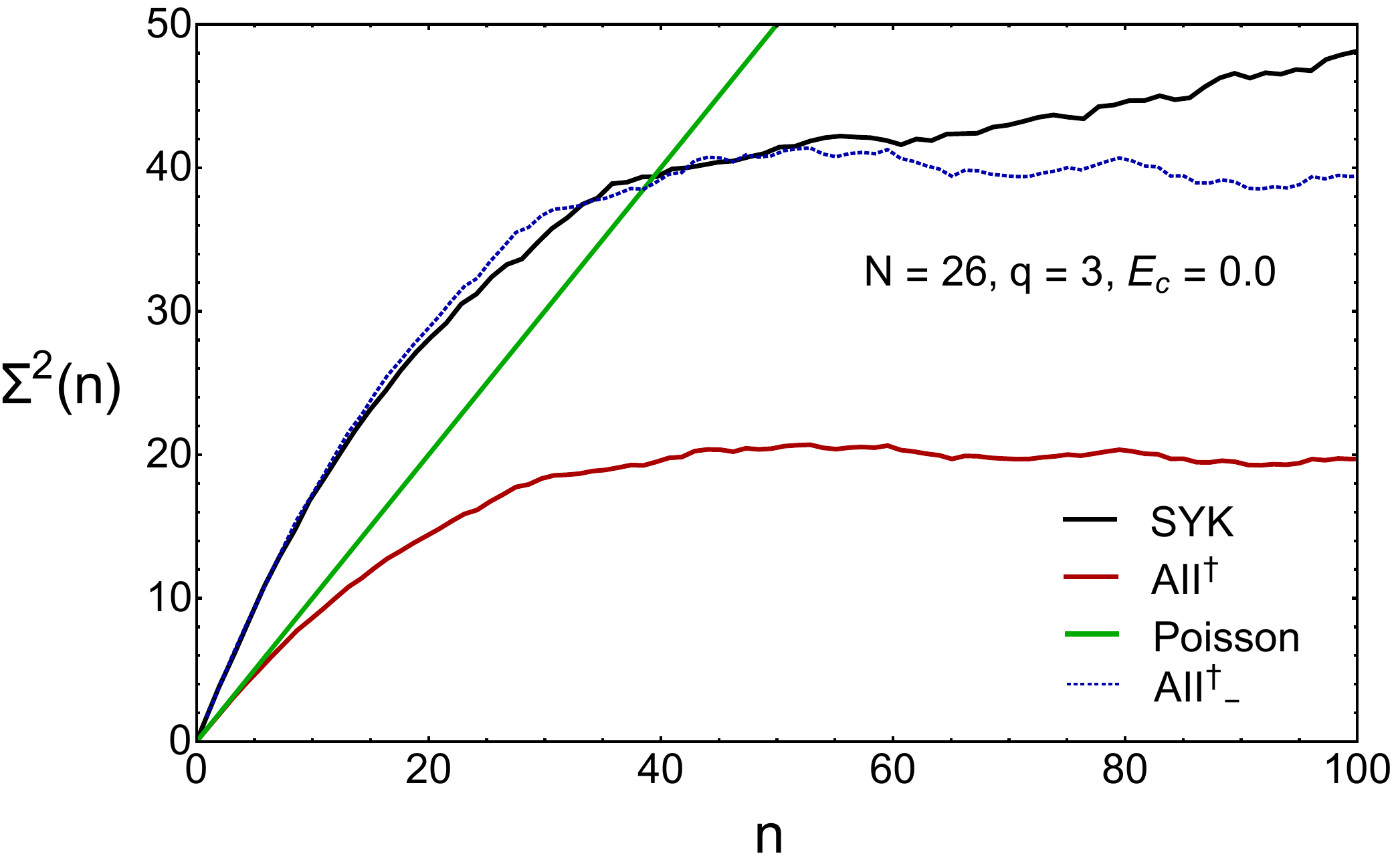}}
	\centerline {\includegraphics[width=8cm]{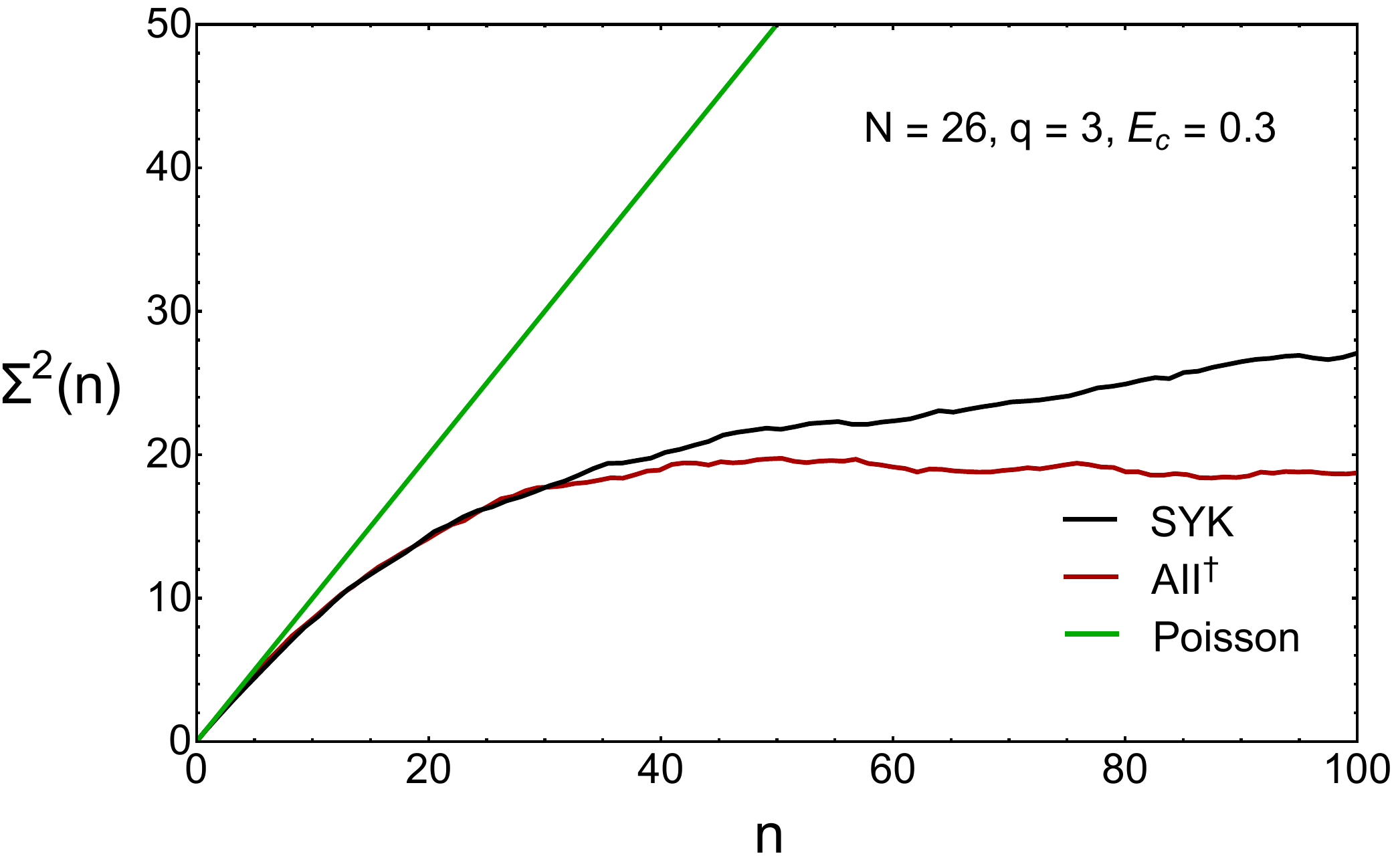}
		\includegraphics[width=8cm]{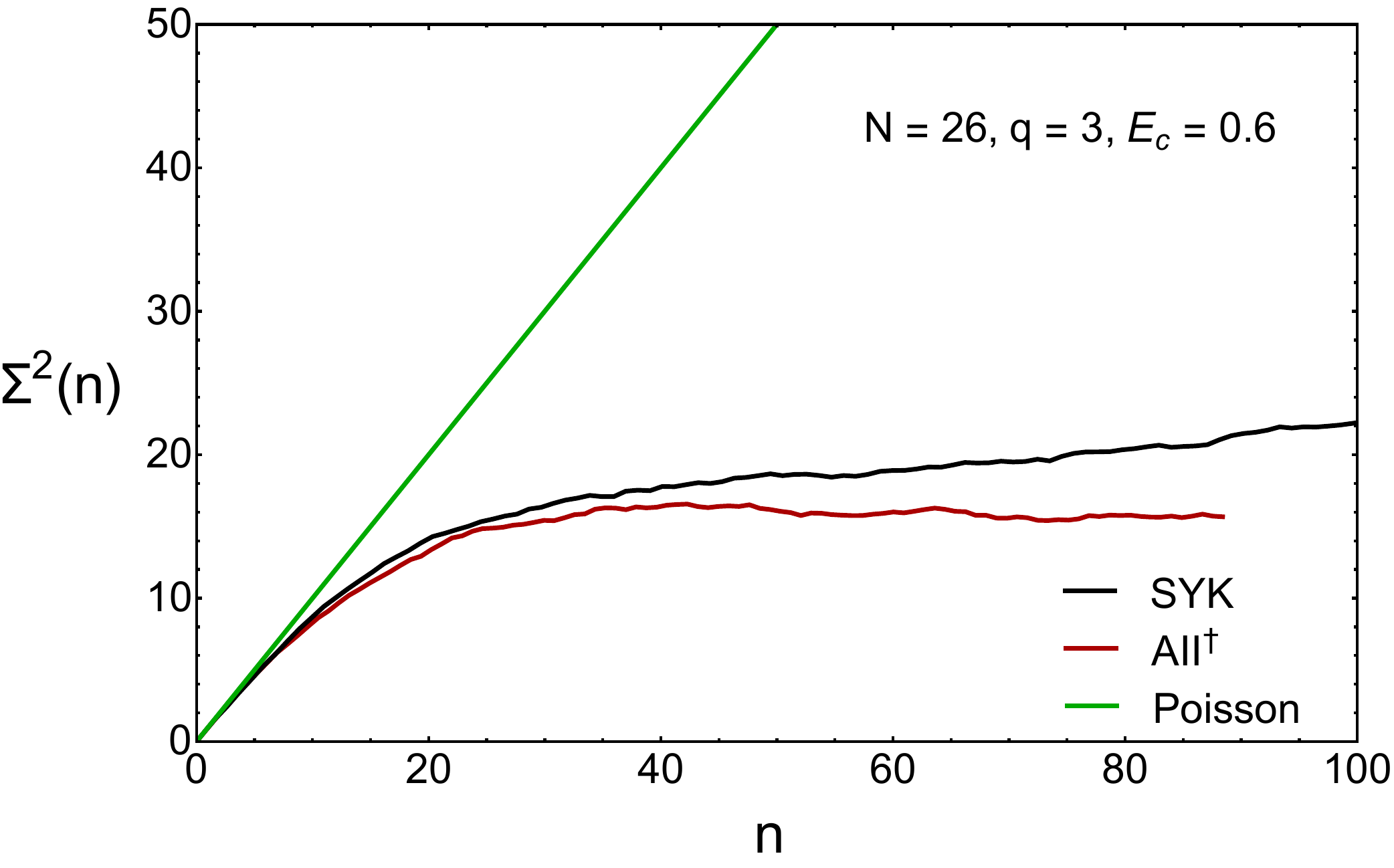}}
	\centerline {\includegraphics[width=8cm]{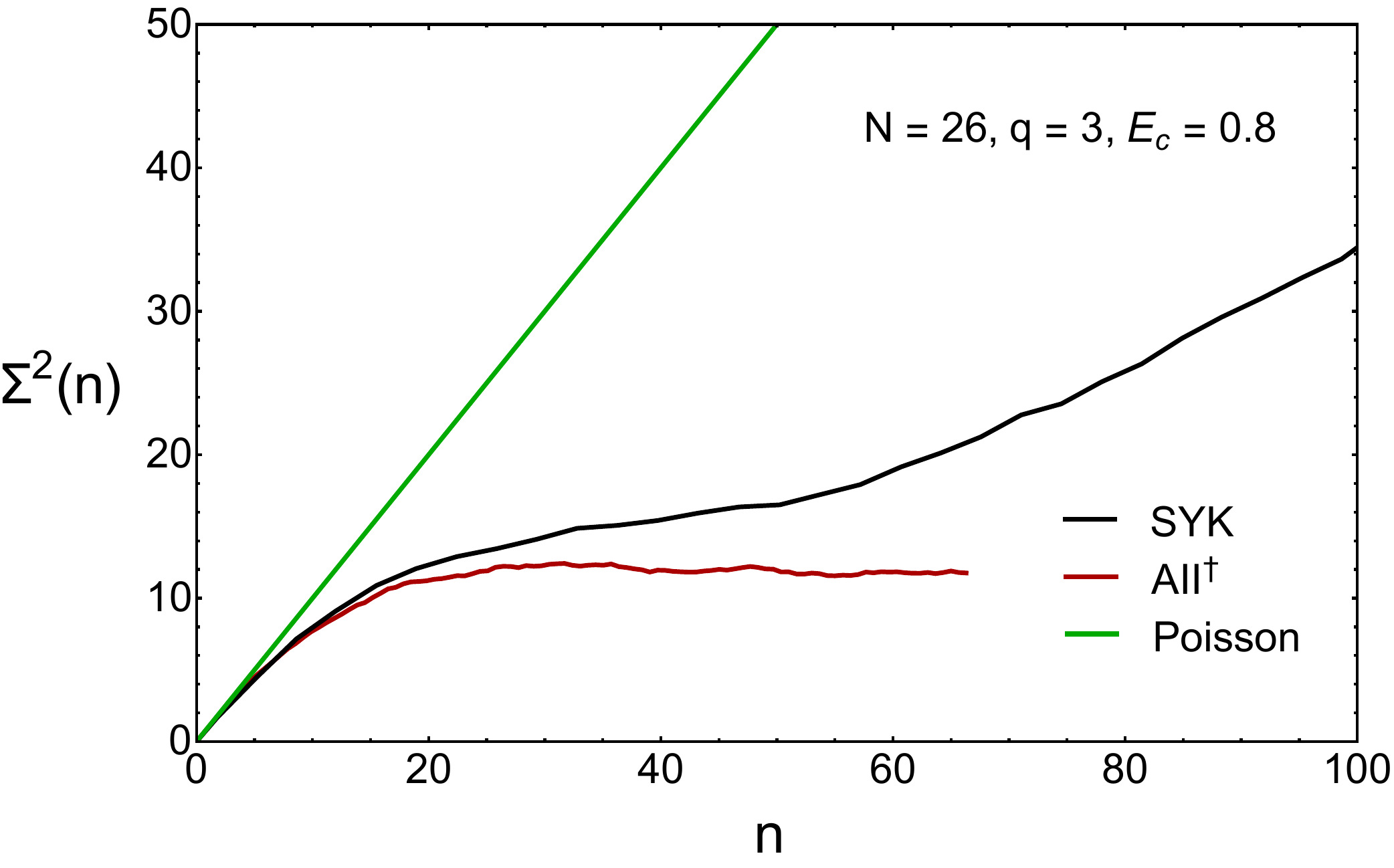}
		\includegraphics[width=8cm]{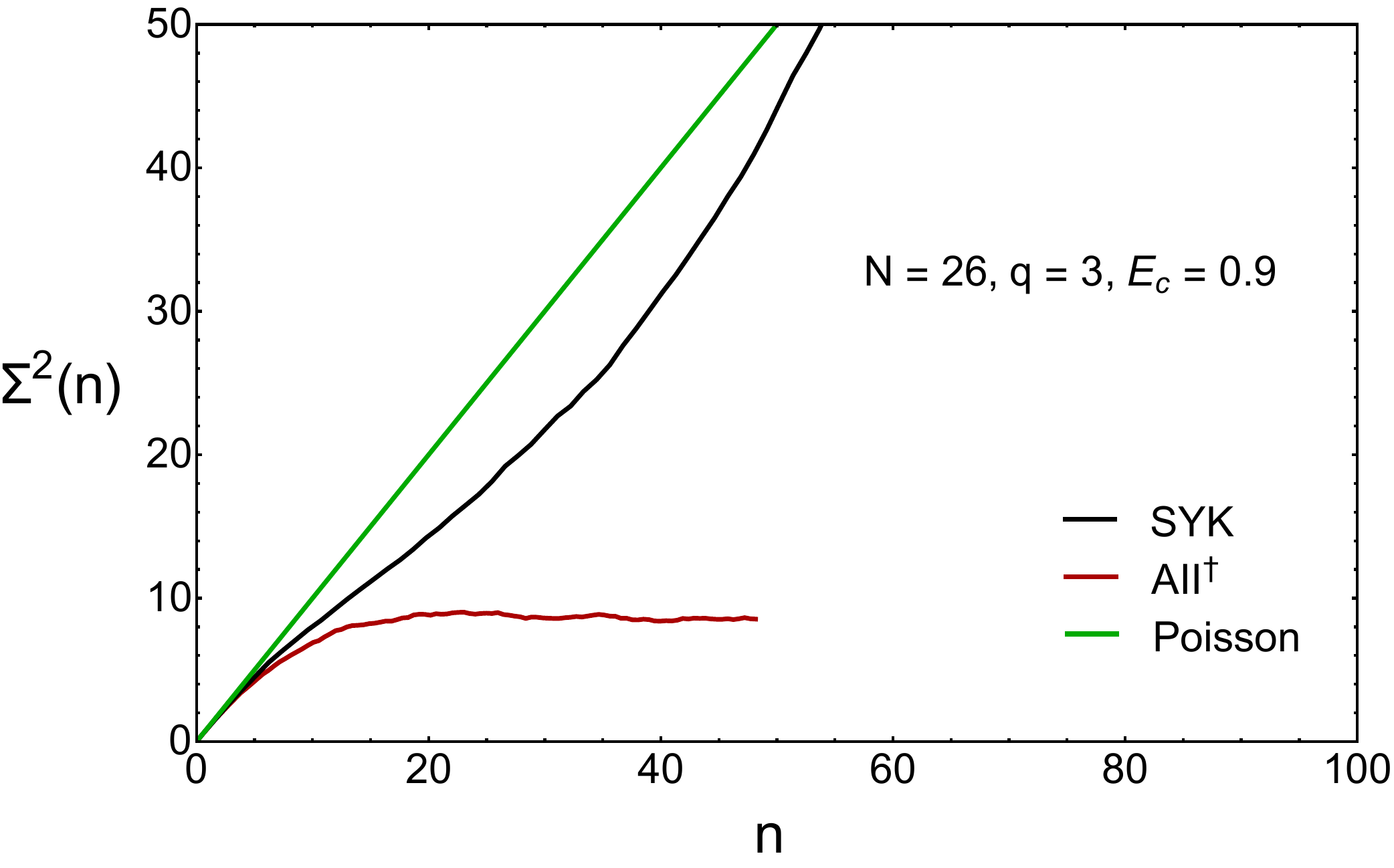}}
	\caption{The number variance of the real parts of the eigenvalues for the nHSYK model for $q = 3$ and $N=26$ for intervals centered at different points $E_c =0, \; 0.3, \; 0.6,\; 0.8$, and $ 0.9$. 
		In the upper left panel, we show all the nHSYK curves. In the rest of the panels, we compare each of these curves (labeled SYK) with the Poisson statistics and the result for the random matrix AII$^\dagger$ universality class with the same value of $E_c$.
	}
	\label{fig:nvN26q3}
\end{figure}

Although a good agreement has also been observed in other systems \cite{chan2021,ghosh2022,shivam2022}, we do not expect that the spectral form factor of the
real parts of the eigenvalues is a good observable for detecting
universal random matrix behavior in generic
quantum dissipative systems. The reason for that is that 
the statistical properties of
the real parts of the eigenvalues depend sensitively on the shape of the spectral density which is not universal. In addition, as seen in the previous section, the contribution due to
collective fluctuations may dominate the spectral form factor all the way up to
the Heisenberg time.

\begin{figure}[t!]
	\centerline{\includegraphics[width=8cm]{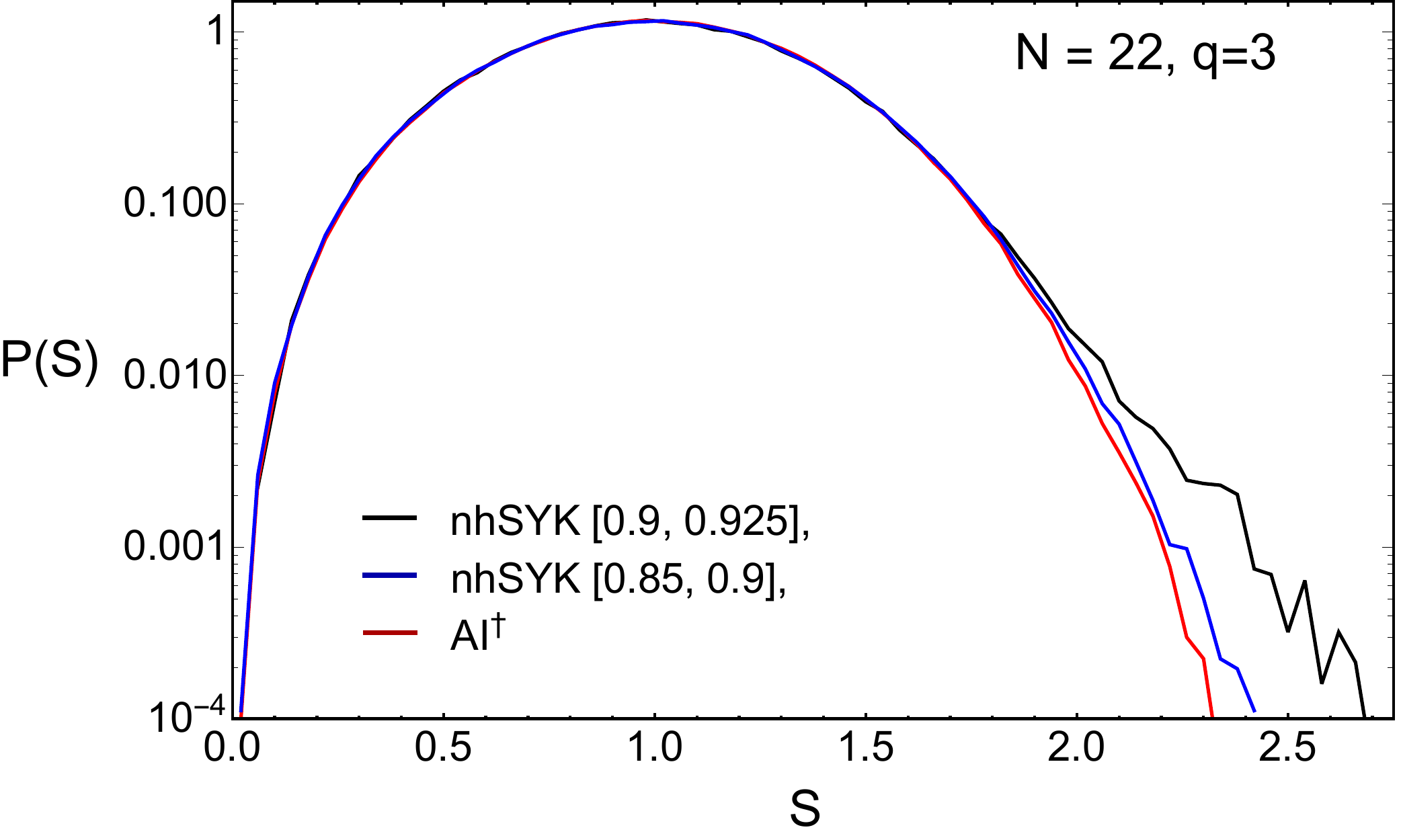}}
	\caption{Log plot of the nearest neighbor spacing distribution for $N = 22$ and $q = 3$. We compare the level spacing distribution for intervals $[0.85, 0.9]$ and $[0.9,0.925]$ to the random matrix result for the ensemble AI$^\dagger$.
		The discrepancy in the tail decreases rapidly as we move further into the bulk of the spectrum. Note that $72$ percent of the eigenvalues
		are larger than $0.85$ in absolute value.}
	\label{fig:ps}
\end{figure}

\subsubsection{$q=3$ and the effect of chiral symmetry}

We turn now to the $q =3$ case. In Fig.~\ref{fig:nvN26q3}, we show the number variance for intervals centered
at $E_c=0$, $0.3$, $0.5$, $0.8$, and $0.9$ (black curves), and compare it to
the number variance of the AII$^\dagger$ random matrix ensemble rescaled to match the number of eigenvalues.
In the upper left figure, we compare the curves with different values of $E_c$. We note that the number
variance for $E_c =0$ is a factor 2 larger than the random matrix result for other values of $E_c$
(see blue and red curves). This is due to the chiral symmetry
of the AII$^\dagger_-$ symmetry class
which doubles the number variance only for $E_c =0$. 
The agreement with the random matrix
result decreases rapidly for increasing $E_c$ and the number variance is close to Poisson for
$E_c=0.9$. 

The same behavior is also visible in the nearest neighbor spacing distribution,
$P(S)$ where $S$ is the absolute value of the distance between neighboring eigenvalues in units of the mean level spacing. It is
illustrated by the log plot in Fig.~\ref{fig:ps}, where we compare the
spacing distribution of the unfolded nHSYK eigenvalues for $E_k \in[0.85, 0.9]$ (blue curve), $E_k \in[0.9, 0.925]$ (black curve), and the corresponding random matrix result for AI$^\dagger$ (red curve).
We find excellent agreement for $S < 1.75$, but for larger spacings the distribution becomes exponential (characteristic of Poisson statistics)
instead of Gaussian (characteristic of random matrix statistics) as we move close to the edge. Note that the average
spacing is $0.024$ so that there are no edge effects for the interval
$[0.9, 0.925]$.

\subsubsection{$q=2$ and deviations from Poisson statistics}

\begin{figure}[t!]
	\centerline{\includegraphics[width=8cm]{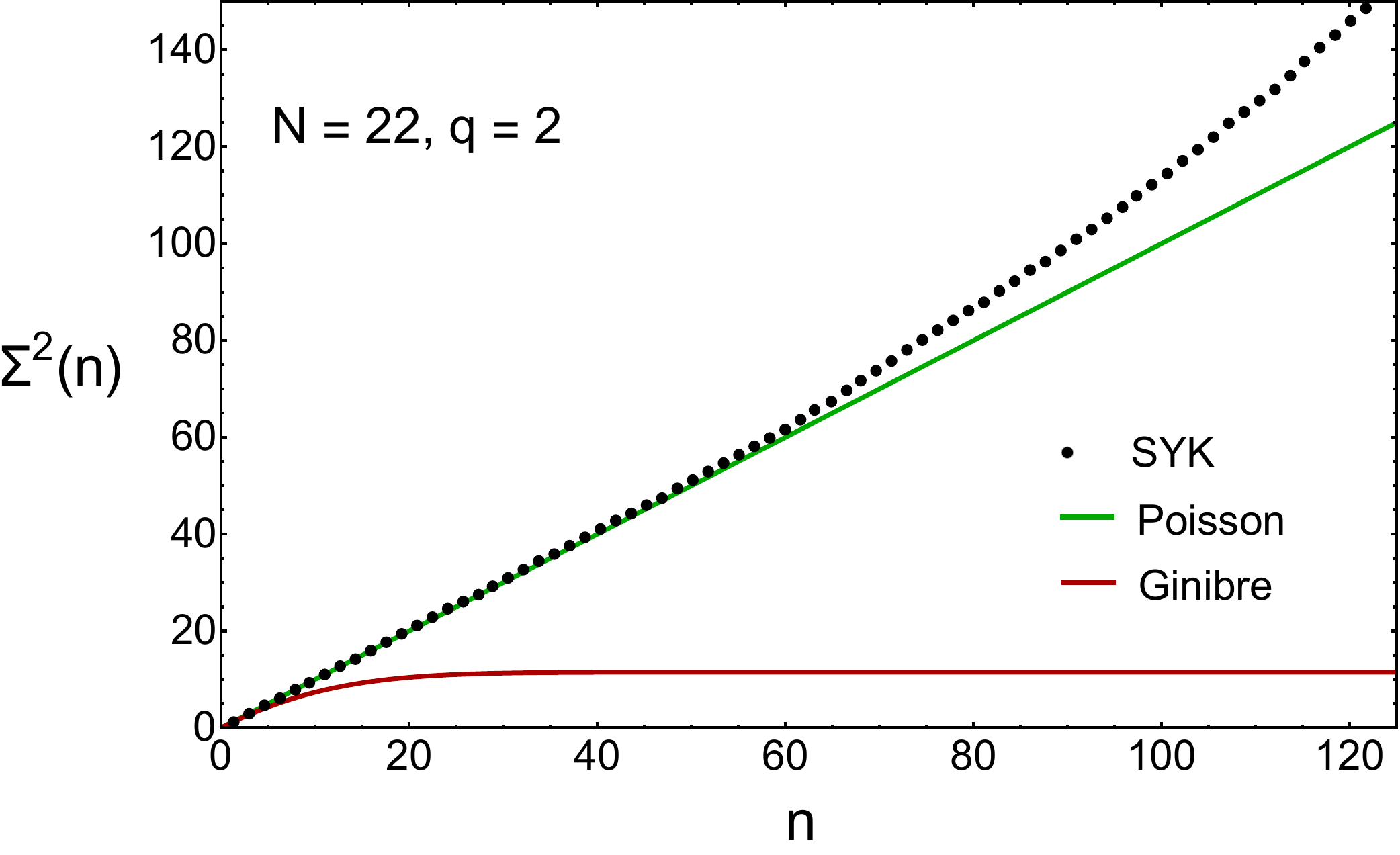}
		\includegraphics[width=8cm]{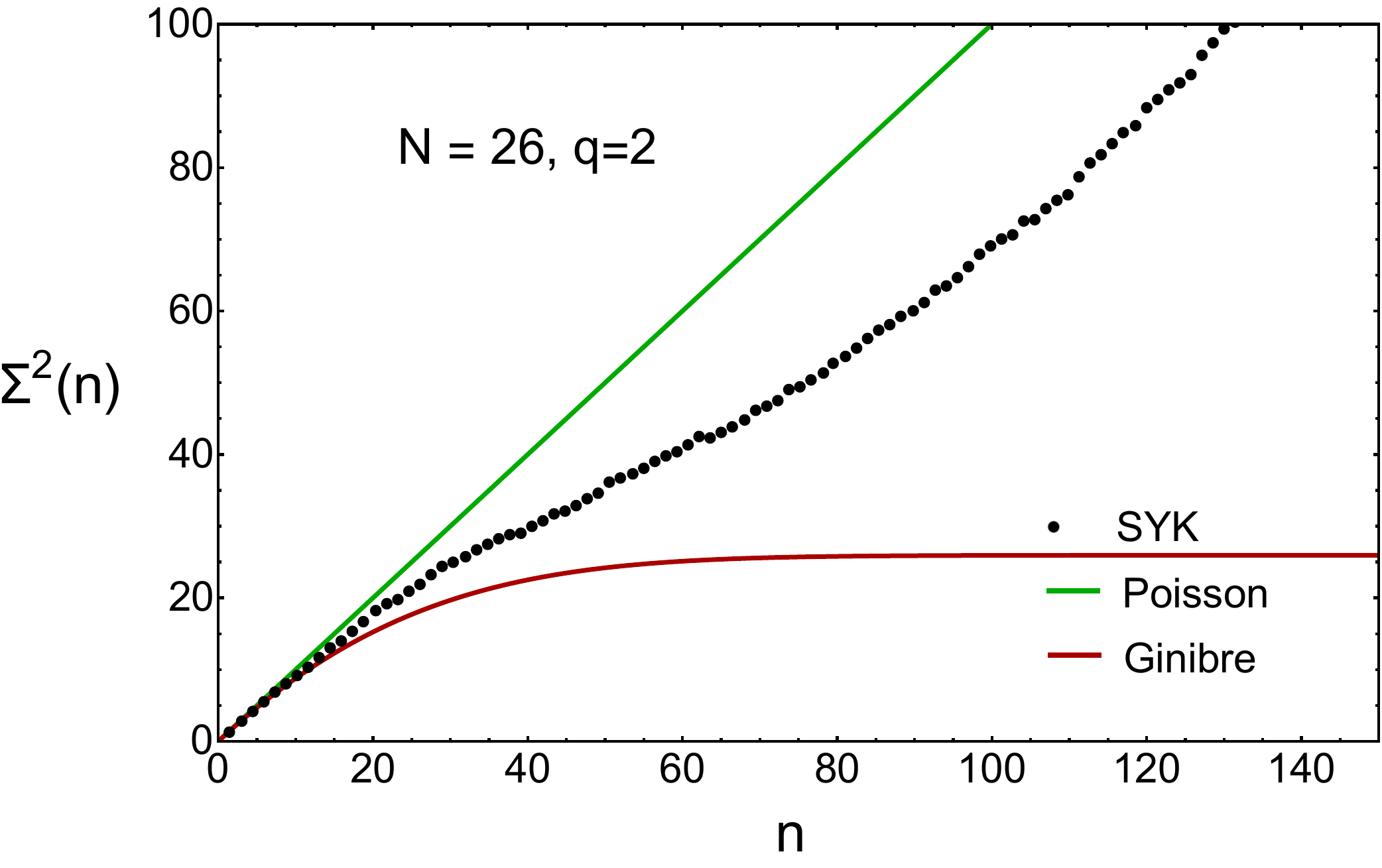}}
	\caption{The number variance of the real part of the eigenvalues of the nHSYK model for $q = 2$ and
		$N = 22$ (left) or $N = 26$ (right). The numerical data are given by the black dots,
		the Poisson result by the green line and the analytical RMT result \eref{nv-ann} by the red curve. The results are for an ensemble of $1000$ realizations averaged over $10$ rays through the origin.}
	\label{fig:nvq2}
\end{figure}

For $q=2$, for which the nHSYK model is integrable, we still observe large deviations from Poisson
(green line) also for smaller values of $N$, see Fig.~\ref{fig:nvq2}. Although there are substantial statistical fluctuations for $n>50$, we can see the onset of the quadratic
behavior due to the peak in the spectral form factor near $\tau = 0$. For comparison,
we also give the analytical result for the Ginibre ensemble evaluated for $D = 2^{N/2}/2$.

\subsection{Local spectral form factor and stationarity}

\begin{figure}[t!]
 	\centerline{\includegraphics[width=8cm]{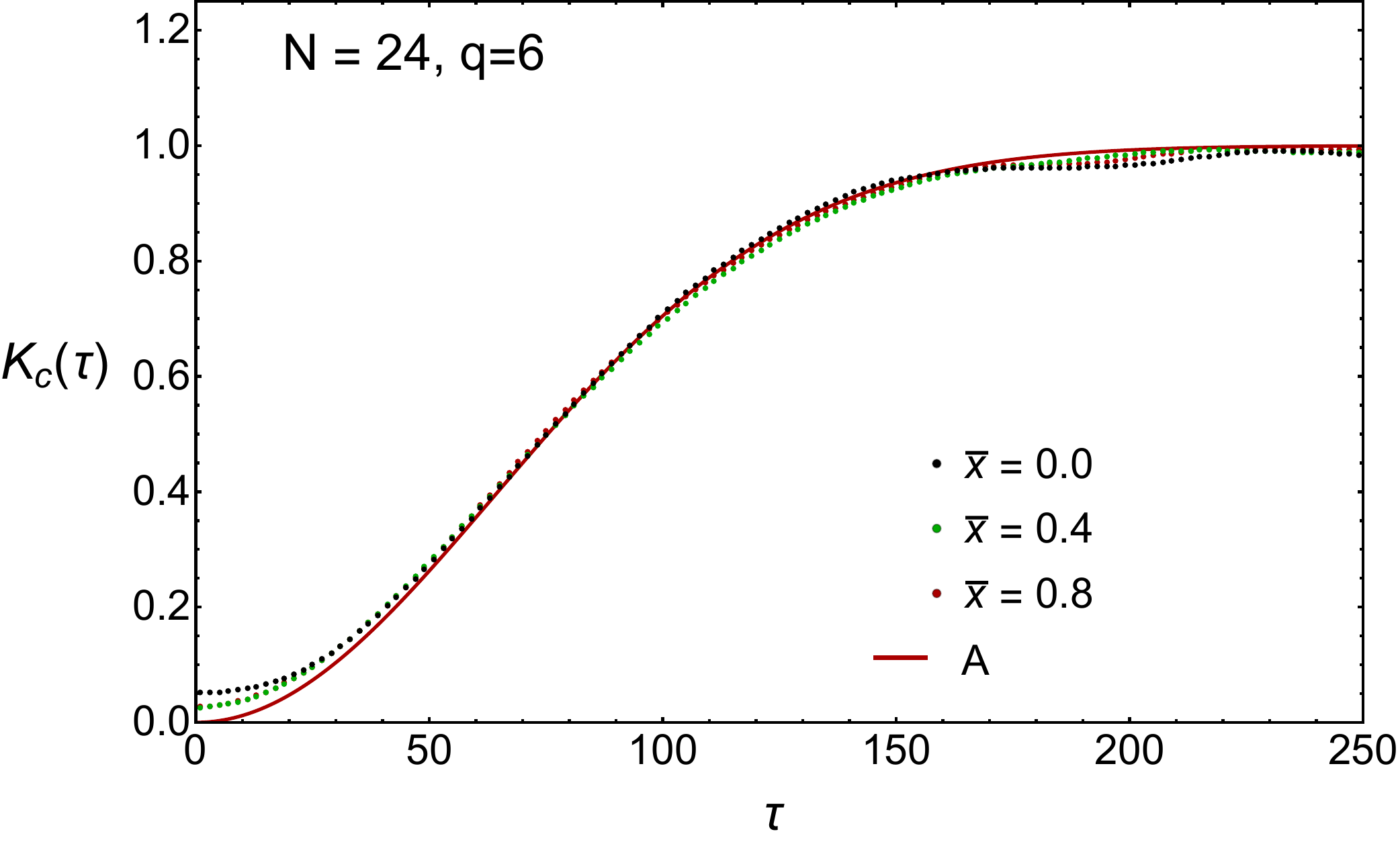}
 		\includegraphics[width=8cm]{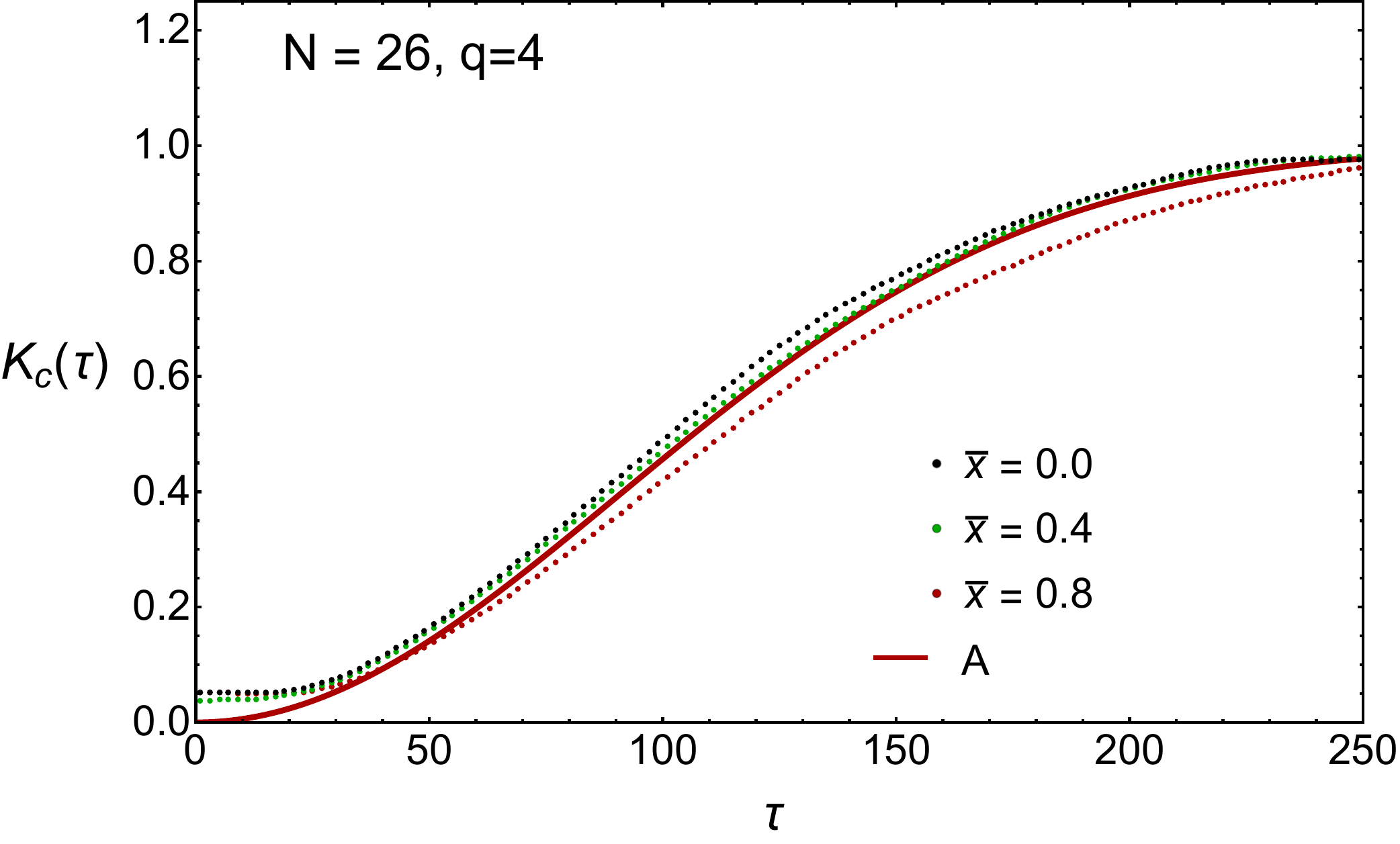}
 	}
 	\caption{The spectral form factor of the real parts of the eigenvalues for the
 		nHSYK model with 
 		$N=24$, $q=6$ (left) and $N=26$, $q=4$ (right). The nHSYK results are given
 		for $\bar x =0 $ (black dots), $\bar x =0.4  $ (green dots) and $\bar x=0.8$
 		(red dots). The solid red curve shows the result for the GinUE
 		(class A).} 
 	\label{fig:form-loc}
 \end{figure}

In this and the next section, we study the stationarity of the local spectral
form factor and the relation with the number variance in the context of the nHSYK model. In Fig.~\ref{fig:form-loc},
we show the local spectral form factor of the real parts of the eigenvalues, Eq.~\eref{kt-loc}, for $N=24$ and $q=6$ (left) and $N=26$ and $q=4$ (right). The local spectral form factor is calculated at $ \bar x = 0$, $ \bar x = 0.4$ and $ \bar x = 0.8$.
The width of the Gaussian cutoff is $w= 0.05$. As expected, the oscillations for small times are absent.
Contrary to the number variance though, we observe only a weak dependence of the local spectral form factor on $\bar x$. This is in agreement with the stationarity of the
local spectral form factor shown in Sec.~\ref{sec:local-form}.
We conclude that the local spectral form factor overcomes the shortcomings of the global spectral form factor (nonstationarity and a diverging nonuniversal collective-excitation peak) and is, therefore, a good diagnostic of non-Hermitian quantum chaos.

More surprising is the fact that, for $ q=6$, for which the spectrum has a reflection symmetry across the origin, the spectral form factor in the center of the spectrum is the same as away
from the center.
The explanation is as follows. The spectral form factor can be written as:
\be
K^\mathrm{loc}_c(\bar x,\tau) = &\frac 1{\cal N} \left \langle \sum_k \cos (x_k\tau)
e^{-\frac{(x_k-\bar x)^2}{2 w^2}}
 \sum_l \cos (x_l \tau) e^{-\frac{(x_l-\bar x)^2}{2 w^2}} \right \rangle_c 
  \nn \\
  +&\frac 1{\cal N} \left \langle \sum_k \sin (x_k \tau) e^{-\frac{(x_k-\bar x)^2}{2 w^2}} 
 \sum_l \sin (x_l \tau) e^{-\frac{(x_l-\bar x)^2}{2 w^2}} \right \rangle_c.
\ee
Away from the center of the spectrum, the contribution of both terms is equal.
However, for $\bar x=0$, the term containing the sine functions vanishes
because the eigenvalues occur in pairs $\pm x_k$.
In the cosine term, we can restrict the sum to the positive real parts only at the expense of
an overall factor of four. For a finite correlation length, we then only sum over half the
total number of eigenvalues so that the cosine term in the reflection-symmetric
case is twice as large as without this symmetry. Since the sine term does not
contribute in the reflection-symmetric case, we find that the spectral form factor is
the same in both cases.
 We thus conclude that chiral symmetry does not affect the local spectral form factor.
 
\subsection{Local number variance}

\begin{figure}[t!]
	\centerline{\includegraphics[width=8cm]{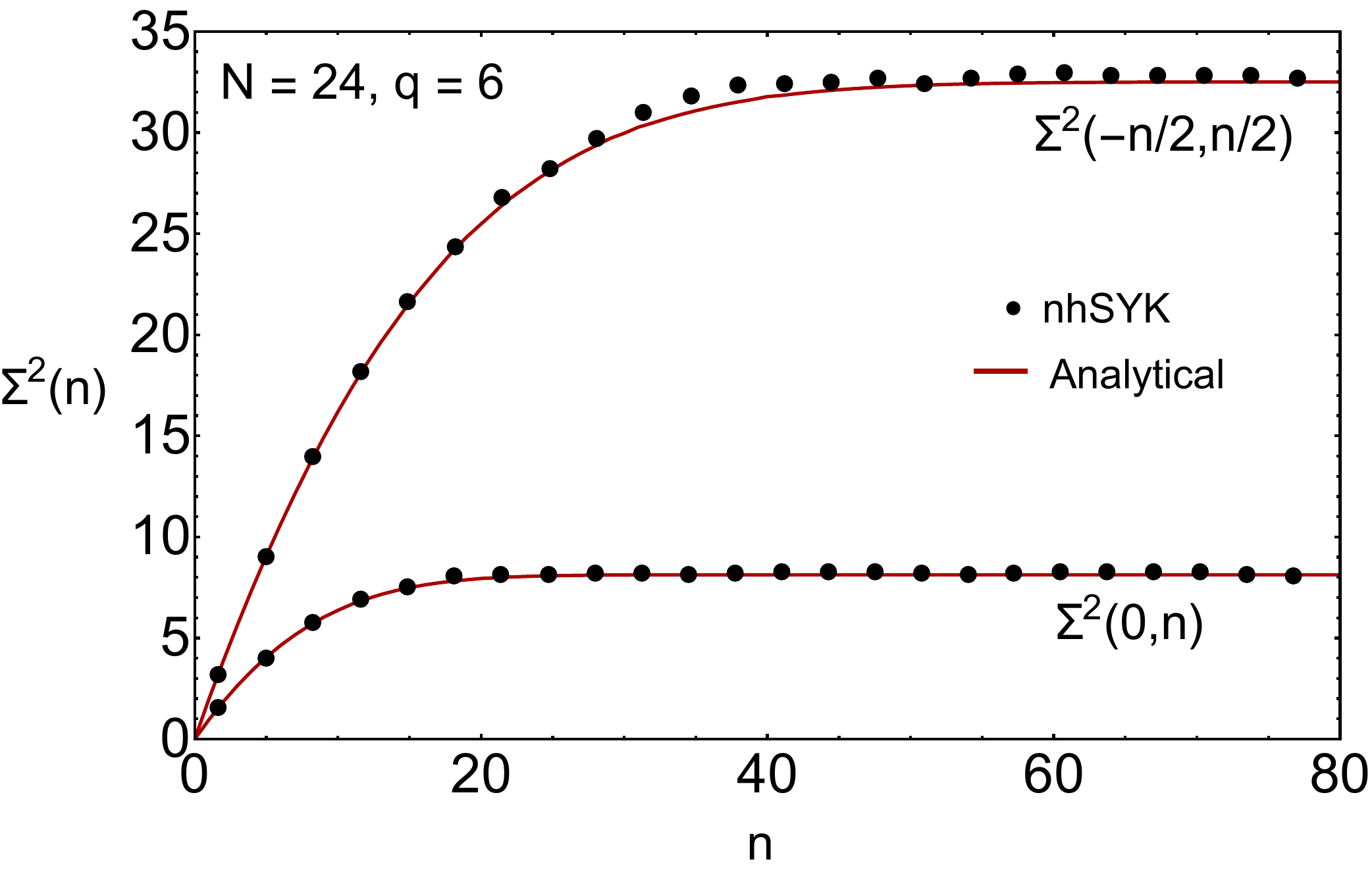}
		\includegraphics[width=8cm]{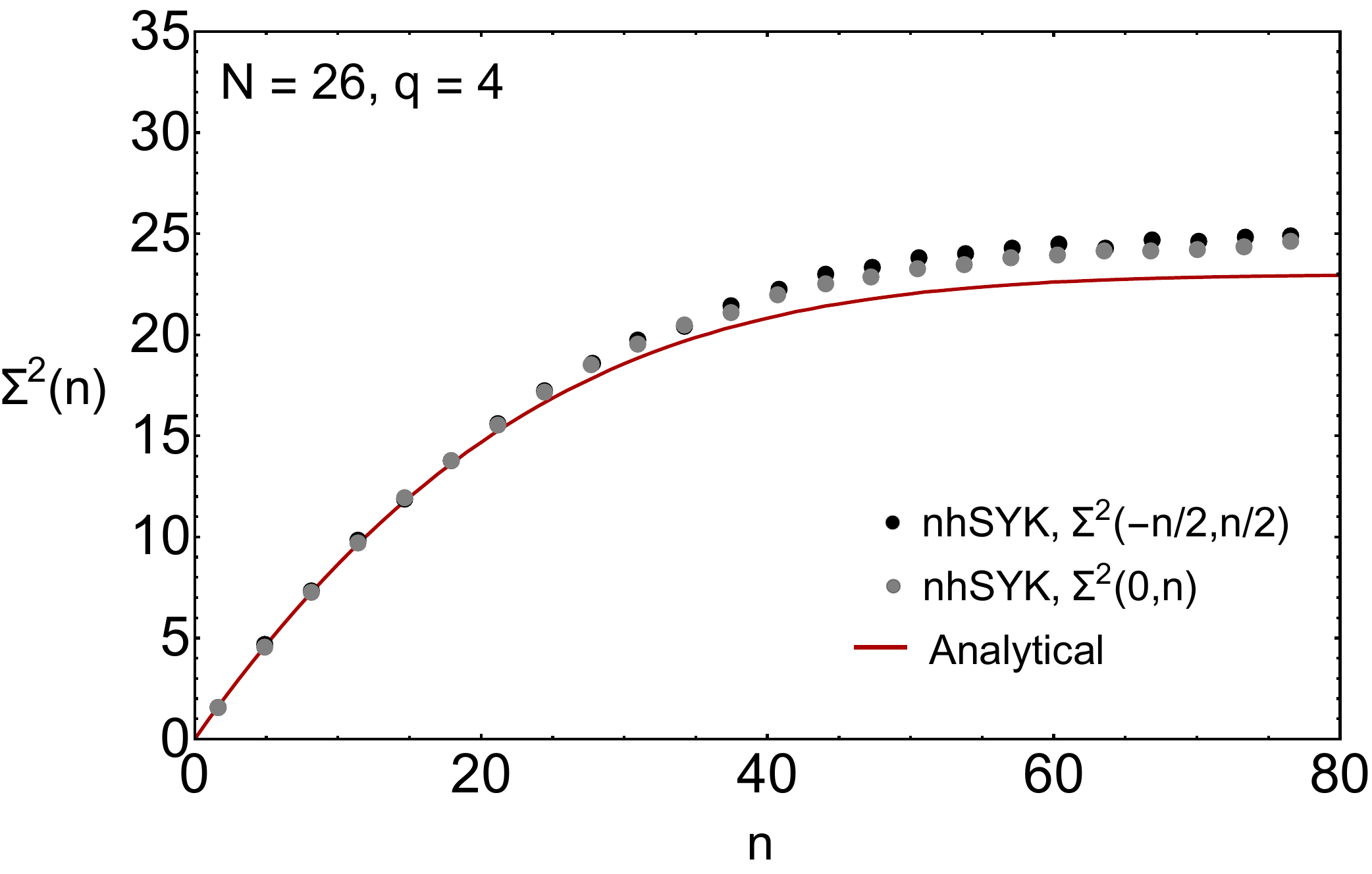}
	}
	\caption{The number variance of the real parts of the eigenvalues for the
		nHSYK model with 
		$N=24$, $q=6$ (left) and $N=26$, $q=4$ (right). The nHSYK results are given
		for the intervals $[0,n]$ and $[-n/2, n/2]$. The
		analytical results are given by $\Sigma^2(0,n)=\frac 12 \Sigma^2_A(n)$
		and $\Sigma^2(-n/2,n/2)=2 \Sigma^2_A(n)$ in the left panel and by
		$ \Sigma^2_A(n)$ in the right panel.}
	\label{fig:nv-chir}
\end{figure}

Contrary to the spectral form factor, the number variance for an interval
centered about zero is quite different from the number variance in the
bulk of the spectrum in the case when the spectrum is reflection symmetric
(see Fig.~\ref{fig:nv-chir}, left), but they are the same when there is no
reflection symmetry (see Fig.~\ref{fig:nv-chir}, right). The analytical result
in the right panel of this figure is just the result for class A (GinUE) evaluated
at the center of the spectrum (red curve). In the case of chiral symmetry in the left panel, the
analytical result is obtained by the following approximation. For the interval
$[0,n]$ the number variance is half the number variance of class A, $\Sigma^2_A$, because
the eigenvalues can only fluctuate in and out of the interval at $n$, but
$\Sigma_A$ has to be evaluated at $2n$,
\be
\Sigma^2(0,n) = \frac 12 \Sigma^2_A(2 n).
\ee
For the internal $[-n/2, n/2]$ we have that
\be
\Sigma^2\left(-\frac n2, \frac n2\right) = 4\Sigma^2(0,n/2) = 2\Sigma^2_A(0,n).
\ee

\begin{figure}[t!]
	\centerline{\includegraphics[width=8cm]{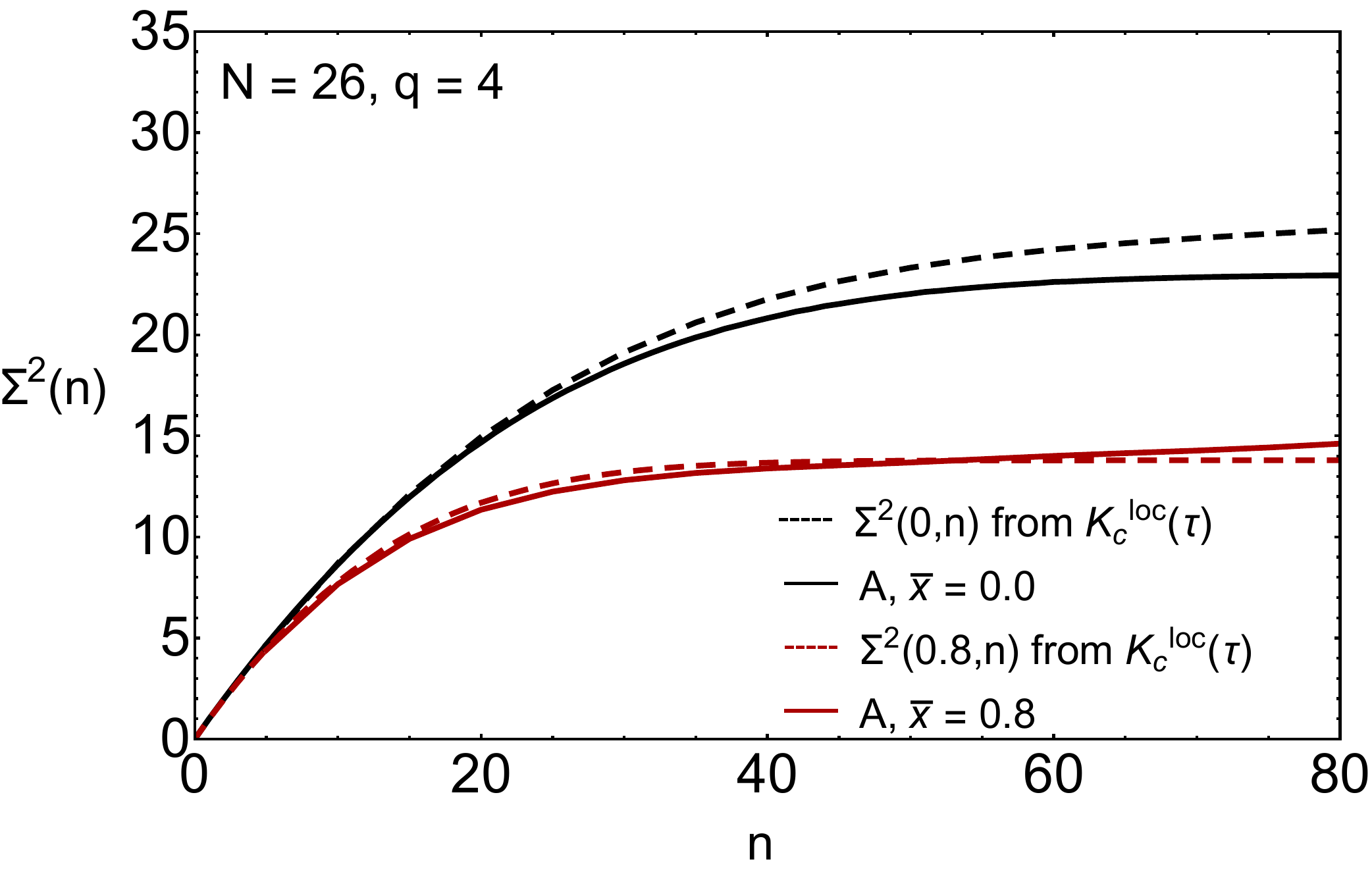}
	}
	\caption{Comparison of the directly calculated number variance (solid curves)
		to the number variance calculated from the spectral form factor of
		Fig.~\ref{fig:form-loc} (right)
		using Eq.~\eref{sigform}. Results are given for $\bar x =0.0$ and $\bar x = 0.8$. }
	\label{fig:concis}
\end{figure}

Finally, to show consistency between the spectral form factor and the number variance
we calculate the number variance from the spectral form factor using Eq.~\eref{sigform}. This relation assumes translational invariance and is not
applicable in the center of the spectrum when the spectrum is reflection
symmetric, and we can only give results for the $q=4$ case, see Fig.~\ref{fig:concis}. In this figure, we compare the direct evaluation of
the number variance to the result obtained from the local spectral form factor
at $\bar x =0$ and $\bar x = 0.8$ (dashed curves). The agreement between
the two shows that the nonstationarity of the number variance is
``kinematical'' and can be eliminated by a proper rescaling, see
Eq.~\eref{double-scaling}.

\section{Conclusions and outlook} 
In this paper, we have studied long-range correlations of the non-Hermitian
SYK model by means of the number variance and the spectral form factor
of the real parts of the eigenvalues with
results for the Ginibre or Ginibre-like ensembles as a benchmark.
To eliminate unfolding ambiguities we have only
considered non-Hermitian SYK models with a radially symmetric spectrum.
A feature of spectral correlations of the real parts of the eigenvalues
is that eigenvalues that are many level spacings apart, and are essentially
uncorrelated, can have real parts that are close. This results in Poisson
statistics already after a timescale of $\sim \sqrt D$.
The early onset of Poisson statistics has the consequence that the
collective spectral fluctuations can no longer be separated from universal
eigenvalue fluctuations. 

For small times, the spectral form factor deviates from the Ginibre result and shows an oscillatory behavior with a period conjugate to the overall width of the spectrum and an amplitude decreasing with time 
that is very sensitive to the number $q$ of interacting Majoranas, a parameter that controls the fraction of
independent matrix elements of the Hamiltonian in Fock space. The area below the peak is proportional to $2^{N/2} / \binom{N}{q}$ and decreases
rapidly going from $q=2$ to $q=6$. Averaging over the oscillations, the small-time behavior of the spectral form factor is similar to that of 
Hermitian systems with a correlation hole. 
Although this is a typical feature of strongly interacting quantum chaotic systems, in this case, it is caused by collective ensemble fluctuations rather than by the nonuniversal dynamics at that timescale. 
Therefore, the nHSYK model describes both generic features of the universal quantum ergodic state reached around the Heisenberg time and nonuniversal, but still rather generic, properties of quantum interacting systems
in its approach to ergodicity.

Having said that, we note that the spectral correlations of the real parts of the eigenvalues
are not stationary and show deviations from
the Ginibre ensemble that depend on the region of the spectrum that is
considered. Since the corresponding spectral form factor is an average over the complete
spectrum, we expect that, in general, it is not universal with a result that depends on the nonstationarity of the spectral correlations. Remarkably, in the nHSYK model, deviations from stationary compared to those of the Ginibre ensemble seem to average out, resulting in a much better agreement than could be expected.
At this point, we do not have a good understanding of this remarkable
coincidence, but we hope to further explore this in future work.
Most likely it is related to the stationarity of the local spectral form factor introduced
in this paper.
In any case, our results also point to intrinsic limitations of the global observables that we have investigated to describe dynamical features. Their local counterparts, on the other hand, overcome these shortcomings and could prove an effective diagnostic of non-Hermitian quantum chaos.

Other related problems that are worthwhile to pursue, and seem within reach, are 
to find analytical results for the spectral density of the nHSYK model and also to compute analytically the spectral form factor, and number variance, for other universality classes, such as AI$^\dagger$ and AII$^\dagger$.

\appendix

\section{\uppercase{Calculation of the spectral form factor of the Ginibre ensemble}}
\label{app:DSFF_GinUE}

In this appendix, we calculate the form factor of the real parts of the eigenvalues of the Ginibre ensemble. It is given by
\be
K_c(\tau) = \frac 1D\int d^2z_1 dz^2_2 e^{i\tau {\rm Re} (z_1-z_2)} \rho_{2,c}(z_1,z_2)
\ee
with the connected two-point correlation function given by \cite{ginibre1965,mehta2004}
\be
\rho_{2, c}(z_1,z_2) = \bar \rho(z_1)\delta^2(z_1-z_2)- |K(z_1,z_2)|^2,
\ee
and kernel given by
\be
K(z_1,z_2) =\frac D\pi e^{-\frac D2 (|z_1|^2 +|z_2|^2)}\sum_{k=0}^{D-1}\frac{(D z_1 z_2^*)^k}{k!}.
\ee
This results in
\be
K_c(\tau) =1 - \frac{D}{\pi^2}\sum_{p=0}^{D-1}\sum_{s=0}^{D-1} \frac{D^{p+s}}{p! s!} I(\tau,p)I^*(\tau,s),
\ee
with
\be 
I(\tau,p,s) = \int d^2z e^{i\tau (z+z^*)/2} z^p z^{* s} e^{-\frac D 2 |z|^2}.
\ee
This integral can be evaluated using polar coordinates
\be
I(\tau,p,s)& = &\int_0^\infty rdr \int_{-\pi}^\pi d\phi e^{irt \cos\phi} r^{p+s}e^{i\phi(p-s)} e^{-D r^2}
\nn\\
&=& 2\pi(-i)^{|p-s|} \int_0^\infty rdr J_{|p-s|}(r\tau)  r^{p+s} e^{-D r^2}.
\ee
The radial integral is evaluated as a hypergeometric function, resulting in
\be
I(\tau,p,s)& = & i^{|p-s|} 2\pi  \frac {\tau^{|p-s|}}{ 2^{|p-s|+1}D^{ {\rm max}(p,s) +1}}
\frac{ {\rm max}(p,s)!}{|p-s|!} ~_1F_1({\rm max}(p,s)+1,|p-s|+1,\frac {-\tau^2}{4D}).
\ee
Our final expression for the spectral form factor is given by
\be
K_c(\tau)=1- \frac 1D\sum_{p=0}^{D-1} \sum_{s=0}^{D-1}
\frac{D^{p+s}}{p! s!} \left[\frac{ {\rm max}(p,s)!}{|p-s|!} \frac {\tau^{|p-s|}}{ 2^{|p-s|}D^{ {\rm max}(p,s) }} ~_1F_1\left ({\rm max}(p,s)+1,|p-s|+1,\frac {-\tau^2}{4D} \right ) \right    ]^2.\nn\\
\label{kcgin}
\ee
This result differs from the expression quoted in Ref.~\cite{chan2021} by the factor $2^{-|p-s|}$.
Asymptotically, for large $D$, it simplifies to
\be
K_c(\tau) \sim 1-e^{-t^2/4D},
\label{klarge}
\ee
which is in agreement with Ref.~\cite{chan2021}. We note that the large-$D$ limit of
the spectral form factor \eref{kcgin} does not commute with the $\tau\to 0 $ limit. The
Taylor expansion of Eq.~\eref{kcgin} gives
\be
K_c(\tau) = \frac {\tau^2}{2D}- \frac {D+3}{32 D^2}\tau^4 + O(\tau^6),
\ee
while from the large $D$ result \eref{klarge} we obtain
\be
K_c(\tau) = \frac {\tau^2}{4D}-\frac {\tau^4}{32 D^2} + O(\tau^6).
\ee

%\newpage
\acknowledgments{
	AMGG was partially supported by NSFC Grant No.\ 11874259 (AMG), the National Key R$\&$D Program of China (Project ID: 2019YFA0308603), and a Shanghai talent program. LS acknowledges support from Funda\c{c}\~ao para a Ci\^encia e a Tecnologia (FCT-Portugal) through grant No.\ SFRH/BD/147477/2019. This project was funded within the QuantERA II Programme that has received funding from the European Union’s Horizon 2020 research and innovation programme under Grant Agreement No 101017733 (LS). This work was also supported by U.S. DOE Grant No. DE-FAG-88FR40388 (JJMV).
}

\bibliography{23_02_07-lrnhsyk-lucas.bbl}

\end{document}